\documentclass{aa}
\usepackage{graphicx}
\usepackage{txfonts}
\usepackage{lscape,url}
\usepackage{natbib}
\usepackage{color}
\bibpunct{(}{)}{;}{a}{}{,} 
\newcommand\kms{{\rm\,km\,s^{-1}}}

\newcommand\zmax{z_{\rm max}}
\newcommand\rmean{R_{\rm mean}}
\newcommand\ulsr{U_{\rm LSR}}
\newcommand\vlsr{V_{\rm LSR}}
\newcommand\wlsr{W_{\rm LSR}}
\newcommand\uw{(U^2+W^2)^{1/2}}

\newcommand\teff{T_{\rm eff}}

\begin{document}

\title{Exploring the Milky Way stellar disk\thanks{
This paper includes data gathered with the 6.5 meter Magellan 
Telescopes located at the Las Campanas Observatory, Chile; 
the Nordic Optical Telescope (NOT) on La Palma, Spain; 
the Very Large Telescope (VLT) at the European Southern Observatory 
(ESO) on Paranal, Chile (ESO Proposal ID 69.B-0277 and 72.B-0179); 
the ESO\,1.5-m, 2.2-m, and 3.6-m telescopes on 
La Silla, Chile (ESO Proposal ID 65.L-0019, 67.B-0108, 76.B-0416, 82.B-0610); 
and data from the UVES Paranal Observatory Project 
(ESO DDT Program ID 266.D-5655).
Tables~\ref{tab:rejected}, \ref{tab:atomdata}, and \ref{tab:parameters},   
are only available in electronic form at the CDS
via anonymous ftp to {\tt cdsarc.u-strasbg.fr (130.79.128.5)} or via 
{\tt http://cdsweb.u-strasbg.fr/cgi-bin/qcat?J/A+A/XXX/AXX}.}
}

\subtitle{
A detailed elemental abundance study of 714
F and G dwarf stars\\ in the Solar neighbourhood}
\titlerunning{714 dwarf stars in the Solar neighbourhood}
\author{
T. Bensby\inst{1}
\and
S. Feltzing\inst{1}
\and
M.S. Oey\inst{2}
}

\institute{
Lund Observatory, Department of Astronomy and Theoretical physics, 
Box 43, SE-221\,00 Lund, Sweden
\and
Department of Astronomy, University of Michigan, Ann Arbor, 
MI 48109-1042, USA
}

\date{Received 9 September 2013 / Accepted XX Xxxx 201X}

\offprints{T.~Bensby \email{tbensby@astro.lu.se}}
\abstract{}
{
The aim of this paper is to explore and map the age and abundance structure
of the stars in the nearby Galactic disk. 
 }
{
We have conducted a high-resolution 
spectroscopic study of 714 F and G dwarf and subgiant stars in the Solar neighbourhood. 
The star sample has been kinematically selected to trace the Galactic
thin and thick disks to their extremes, the metal-rich stellar halo, 
sub-structures in velocity space such as the Hercules stream and 
the Arcturus moving group, as well as stars that cannot (kinematically) be 
associated with either the thin disk or the thick disk. The determination of 
stellar parameters and elemental abundances is based on a 
standard 1-D LTE analysis using equivalent width measurements in 
high-resolution ($R=40\,000-110\,000$) and high
signal-to-noise ($S/N=150-300$) spectra obtained with 
FEROS on the ESO\,1.5-m and 2.2-m telescopes, SOFIN and FIES on the Nordic Optical 
Telescope, UVES on the ESO Very Large Telescope, 
HARPS on the ESO\,3.6-m telescope, and MIKE on the 
Magellan Clay telescope. NLTE corrections for individual \ion{Fe}{i} 
lines were employed in every step of the analysis.
 }
{
We present stellar parameters, stellar ages, kinematical 
parameters, orbital parameters, and detailed elemental abundances for O, Na, 
Mg, Al, Si, Ca, Ti, Cr, Fe, Ni, Zn, Y, and Ba for 714 nearby F and G dwarf
stars. Our data show that there is an old 
and $\alpha$-enhanced disk population, and a younger and less 
$\alpha$-enhanced disk population.
While they  overlap greatly in metallicity between 
$\rm -0.7<[Fe/H]\lesssim+0.1$, they show a bimodal distribution in 
$\rm [\alpha/Fe]$. This bimodality becomes even clearer if stars 
where stellar parameters and abundances show larger uncertainties 
($\teff\lesssim5400$\,K) are discarded, 
showing that it is important to constrain the data set to a narrow
range in the stellar parameters if small differences between 
stellar populations
are to be revealed. We furthermore find that the $\alpha$-enhanced 
population has orbital parameters placing the stellar birthplaces in the
inner Galactic disk while the low-$\alpha$ stars mainly come from the
outer Galactic disk, fully consistent with the recent claims
of a short scale-length for the $\alpha$-enhanced Galactic thick disk.
We have also investigated the properties of the Hercules stream and the
Arcturus moving group and find that neither of them present 
chemical or age signatures that could point to that they are
disrupted clusters or extragalactic accretion remnants from 
ancient merger events. Instead, they are most likely dynamical features originating
within the Galaxy. We furthermore have discovered that a standard
1-D, LTE analysis, utilising ionisation and excitation balance of \ion{Fe}{i}
and \ion{Fe}{ii} lines produces a flat lower main sequence. As the exact cause
for this effect is unclear we chose to apply an empirical correction.
Turn-off, and more evolved, stars, appears to be un-affected.
 }
{}

   \maketitle

\section{Introduction}

How galaxies form and evolve is a vast subject that has in the last 
decades rapidly developed into one of the most exciting areas in 
contemporary astrophysics. The goal has been to unveil the mysteries 
of the formation, assembly and chemical history of galaxies, and our 
own galaxy, the Milky Way, in particular. 
As the Milky Way currently is the only 
galaxy whose stellar populations can be studied in great detail with high-resolution 
spectrographs, and may serve as a ``benchmark galaxy'' for extra-galactic 
studies, it is essential to establish the properties 
of the different Milky Way stellar populations.

Major pieces to the 
puzzle of galaxy formation are held by the atmospheres of stars 
which may remain intact over 
time and act as time capsules showing the mixture of chemical elements 
that were present in the gas cloud out of which the stars formed 
billions of years ago \citep[e.g.,][]{lambert1989,freeman2002}. 
F and G dwarf stars are especially reliable tracers
as their expected lifetimes on the main sequence, burning hydrogen to 
helium in their centres, are similar to, or possibly even longer than, the current age of the 
Galaxy. For instance, a solar-type star will spend around 10\,Gyr on the main 
sequence \citep[e.g.,][]{sackmann1993}. During this time its atmosphere is
untouched by internal nuclear processes.
By obtaining high-resolution spectra of such stars it is possible to 
determine their detailed chemical compositions and
ages, which allow us to trace the histories of
different stellar populations.
In the last 20 years, several studies have aimed at characterising 
the Galactic stellar disk using nearby F and G dwarf stars
\citep[e.g.,][]{edvardsson1993,feltzing1998,fuhrmann1998,fuhrmann2000unpubl,fuhrmann2004,fuhrmann2008,fuhrmann2011,prochaska2000,gratton2000,chen2000,mashonkina2001,tautvaisiene2001,trevisan2011,bensby2003,bensby2004,bensby2005,bensby2007letter2,bensby2006,feltzing2007,soubiran2003,reddy2003,reddy2006}.
The evidence from these high-resolution spectroscopic
studies have so far shown that the Milky Way appears to contain 
two disk populations, 
with different chemical and age properties, indicating different 
origins and different chemical histories. 

\begin{figure}
\resizebox{\hsize}{!}{
\includegraphics{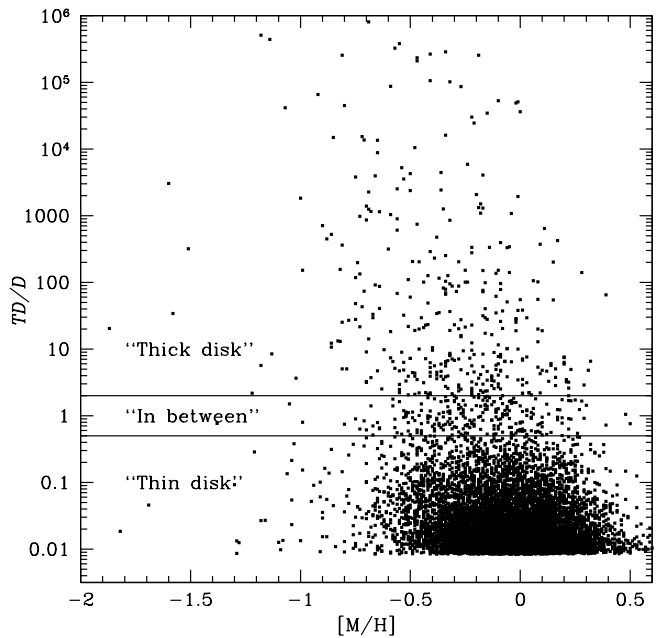}}
\caption{The kinematical thick disk-to-thin disk probability
ratio ($TD/D$) versus metallicity for the $\sim 14000$ stars in the 
GCS. Stars with $TD/D>2$ are, to a first approximation, classified as potential
thick disk stars, and stars with $TD/D<0.5$ are, to a first approximation,
classified as potential thin disk stars. Stars with probability 
ratios between these two limits are here classified as 
``in-between stars''. Note that all metallicities, [M/H],
are from the Str\"omgren calibration by 
\cite{casagrande2011}.
\label{fig:selection}
}
\end{figure}

However, after more than two decades of observational efforts, we are still 
lacking much information about the complex abundance structure of the 
Galactic stellar disk. For instance,
the Geneva-Copenhagen Survey (hereafter GCS) by \cite{nordstrom2004}
contains approximately 14\,000 dwarf stars in the Solar 
neighbourhood, all of which have full three-dimensional kinematic 
information available, as well as ages and metallicities estimated from
Str\"omgren photometry. It is evident from the GCS
data that there are substantial kinematical sub-structures present
in the Solar neighbourhood that can be associated with various stellar 
streams and moving groups \citep[e.g.,][]{nordstrom2004,navarro2004,
famaey2005,soubiran2005,arifyanto2006,helmi2006}. These 
kinematical substructures, seen in the immediate Solar neighbourhood, 
have recently been confirmed to persist to distances of at least 1\,kpc from the 
Sun, although with slightly shifted velocity components \citep{antoja2012}.
It is unclear whether such structures are of Galactic or extragalactic origin. 
The GCS also contains many stars with typical thick-disk kinematics, and 
with very  high metallicities, well above solar (cf. Figs.~\ref{fig:selection} 
and \ref{fig:uvwfeh}). The question is whether these stars are true thick disk 
stars.  It is also unclear what the lowest metallicities are in the thin disk,
and whether the thin and thick disks show distinct abundance trends.

In addition, recent studies of the SDSS Segue G and K dwarf stellar 
sample by \cite{abazajian2009,yanni2009} of more than 5000 stars at 
larger distances add a new dimension to this discussion. From this 
data (but treated in different ways) \cite{bovy2012} finds that there 
is no distinct thick disk, whilst  \cite{lee2011} and \cite{liu2012}
find two or perhaps even three components in the stellar disk. 
Furthermore, other recent studies actually show that many, if not all, 
edge-on spiral galaxies appear to host dual disk systems 
\citep[e.g.,][]{yoachim2006,comeron2011}.

Distinct and different multiple stellar disks are
an important component in galaxy formation models, 
and the signature of a unique thick disk in such models depends 
on the formation scenario. For example, 
if radial migration is the responsible mechanism,
then it is a continuous process and the result could very well
be that the thick and thin disks form a smooth transition. On the
other hand, if the formation of the thick disk is fast, e.g., through
kinematical heating of an old disk due to an ancient merger event,
it is more likely that the two disks are distinct components in 
chemistry and phase-space \citep{minchev2012}. It is therefore extra 
important that the dichotomy of the Milky Way stellar disk is 
well-understood, helping us to better understand galaxy formation in 
general.

On larger scales, there are several ongoing and upcoming large 
spectroscopic surveys that 
will probe the abundance structure of the Milky Way and its stellar
populations on much larger scales. Examples are the 
SDSS Segue \citep{yanni2009}, APOGEE \citep{allendeprieto2008},
the Gaia-ESO Survey \citep{gilmore2012}, the GALAH survey 
\citep[e.g.,][]{zucker2012} which together will gather spectra and 
determine stellar parameters and chemical abundances for several 
hundreds of thousands of dwarf and red giant stars in the
Galactic thin disk, thick disk, stellar halo, and bulge. 
However, these surveys are based
on low- or medium-resolution spectra that often have very limited 
wavelength coverages and sometimes lower signal-to-noise ratios. 
Hence, they will need to anchor their results to studies that present 
detailed elemental abundances
that have been homogeneously determined from high-resolution and high
signal-to-noise spectra. 

The stellar sample presented in this study aims at mapping and exploring 
the age and abundance structure of the Milky Way stellar disk
in a consistent and homogeneous way based on high-resolution and
high signal-to-noise spectra of nearby F and G dwarf stars.
In this paper we describe the star sample and the 
elemental abundance analysis, as well as presenting the observed 
properties of the Galactic disk.
In particular, the extent and variation of elemental abundances and stellar ages
with galactocentric radius is explored.

First results based on the current sample have been published in 
\cite{bensby2007letter,bensby2007letter2,feltzing2008uppsala,bensby2010rio}, 
and the sample has also been part of the recalibration of the Geneva-Copenhagen
Survey \citep{casagrande2010}, characterisation of planet signatures
in solar-type stars \citep{ramirez2010}, and most recently in the chemical
tagging experiment by \cite{mitschang2013}.
Further investigations into the dichotomy of the Galactic stellar disk
are conducted in a parallel paper (Feltzing et al.~2013, in preparation),
while work on odd iron peak elements will be presented in
Battastini \& Bensby (in preparation), and results for a larger range of 
$r-$ and $s-$process elements in Battistini et al.,~(in preparation).

\section{Sample selection} \label{sec:sample}

\begin{figure}
\resizebox{\hsize}{!}{
\includegraphics[bb=18 155 592 395,clip]{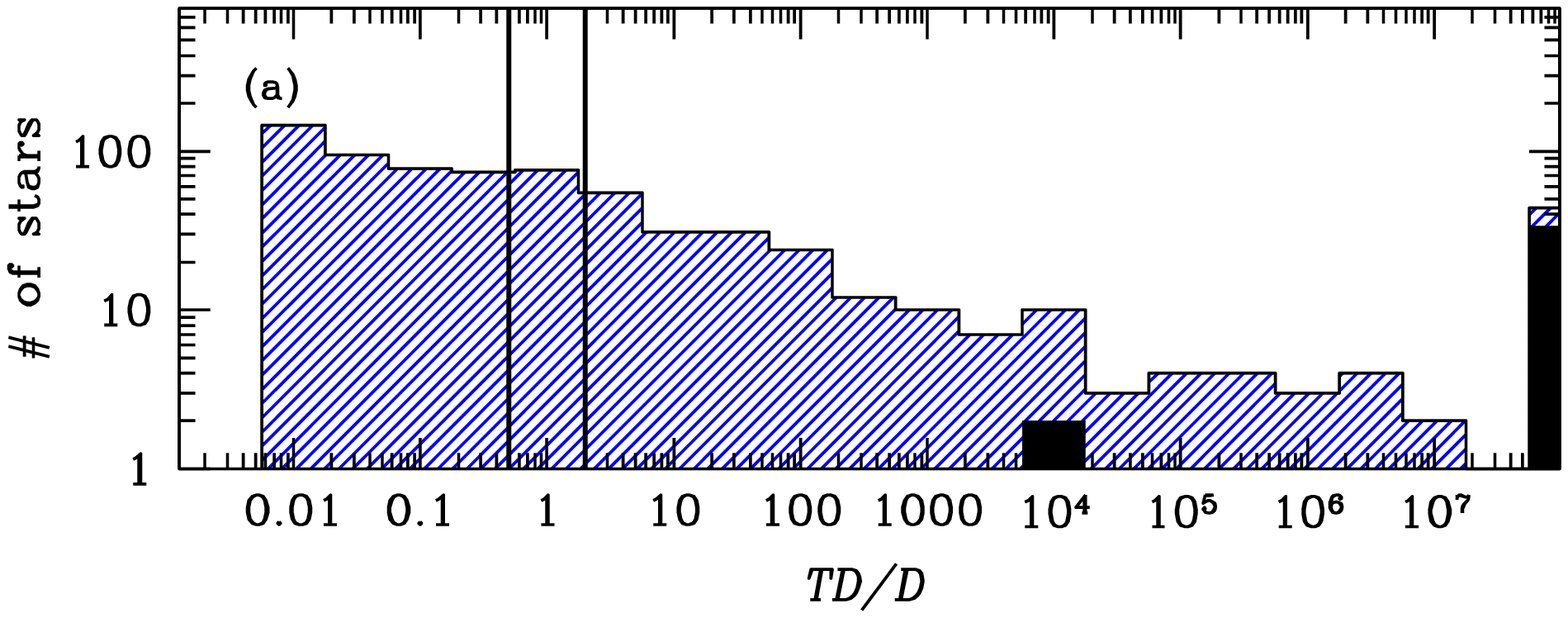}}
\resizebox{\hsize}{!}{
\includegraphics[bb=18 144 592 700,clip]{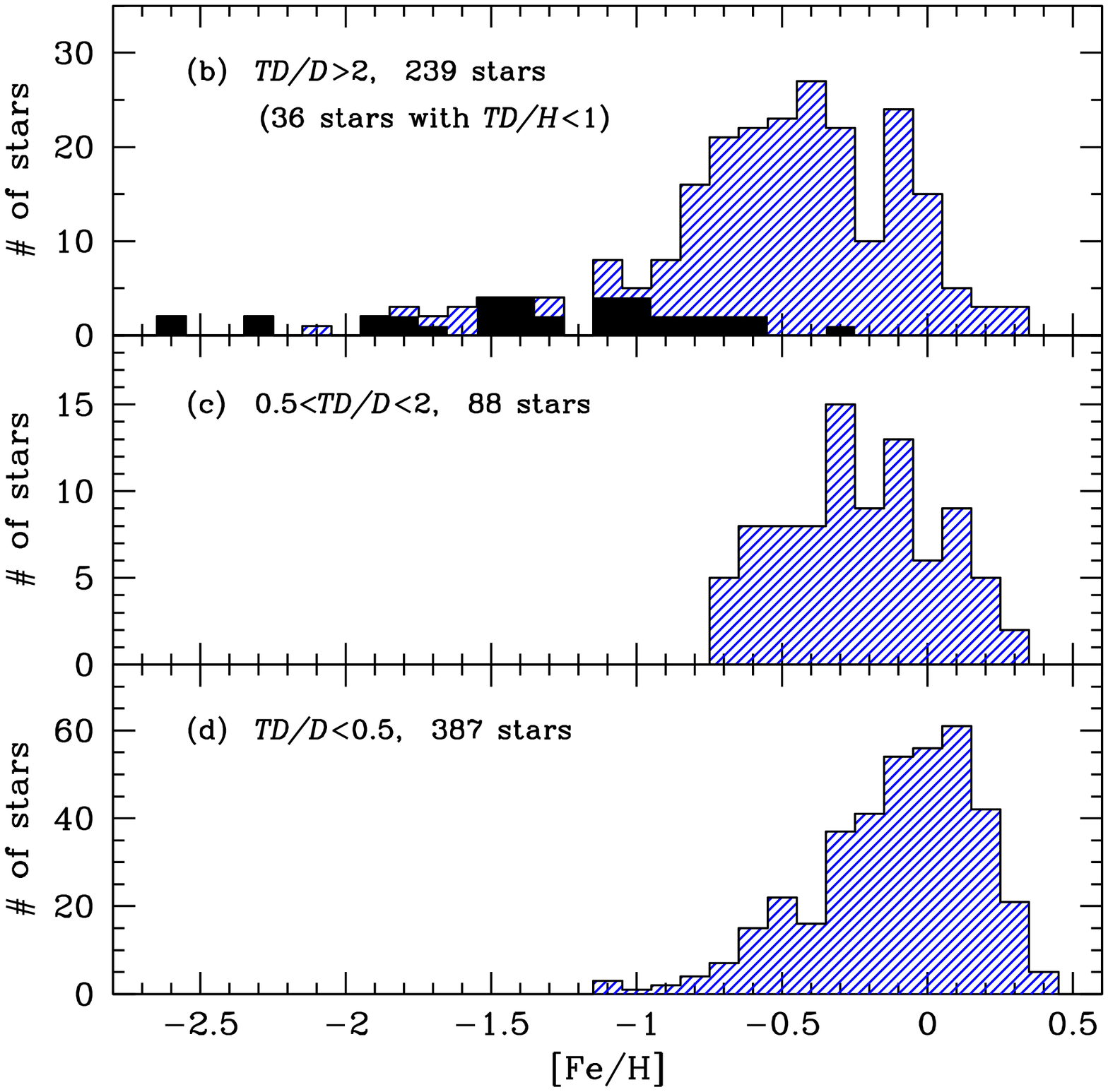}}
\caption{(a) The $TD/D$ distribution of our sample of 714 stars.
The solid vertical lines mark the $TD/D=0.5$ and $TD/D=2$ ratios.
Panels (b)--(d) show the metallicity distributions of 239 potential
thick disk stars with $TD/D>2$, out of which 36 stars have $TD/H<1$, i.e.
most likely halo stars; 88 stars with kinematics
``in between''; and the 387 potential thin disk stars with
$TD/D<0.5$. In (a) and (b) the likely halo stars ($TD/H<1$) are marked
by solid black histograms. The metallicities are from our spectroscopic analysis.
\label{fig:tddhist}
}
\end{figure}

\begin{figure}
\resizebox{\hsize}{!}{
  \includegraphics[bb=18 200 592 525,clip]{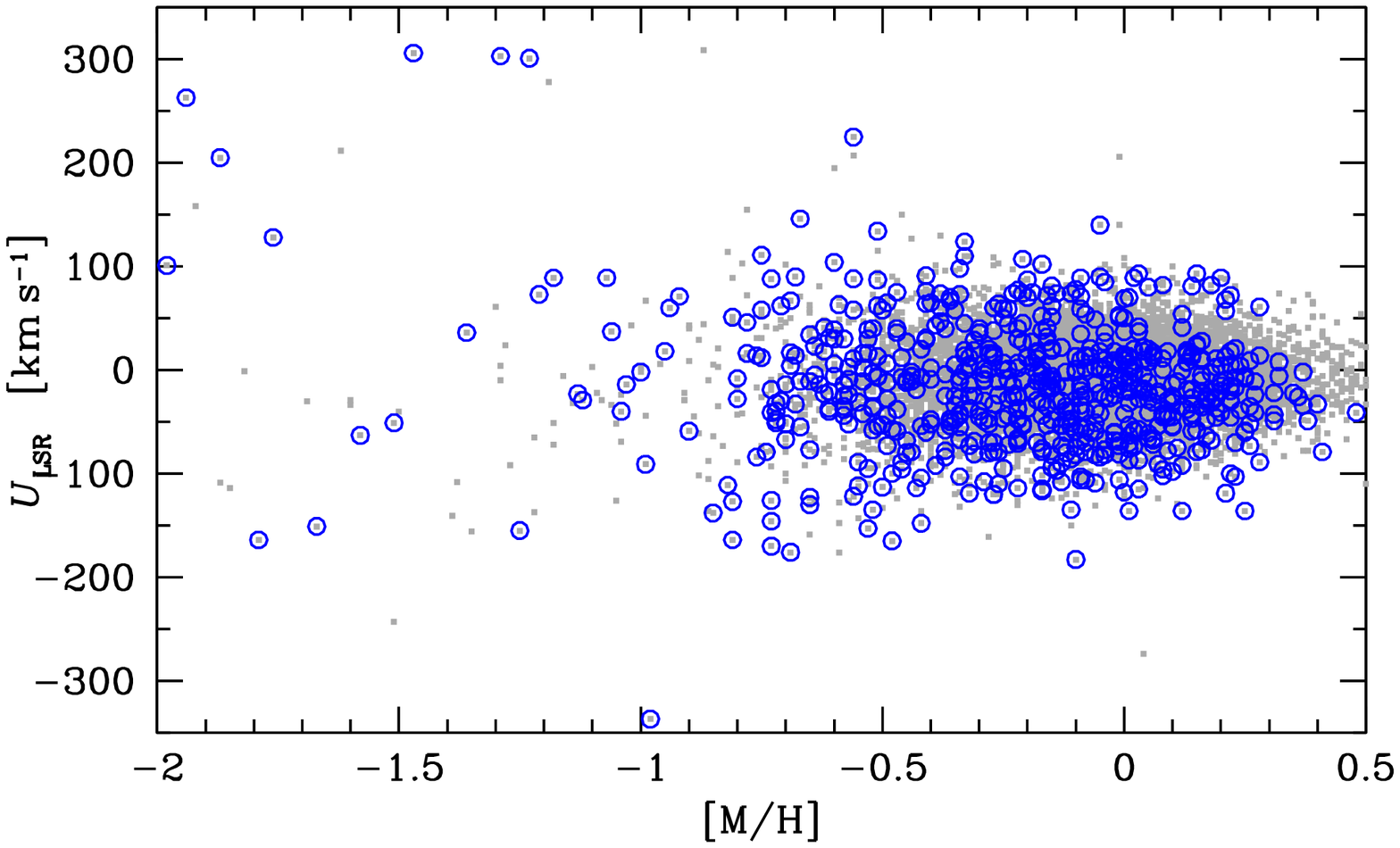}}
\resizebox{\hsize}{!}{
  \includegraphics[bb=18 200 592 500,clip]{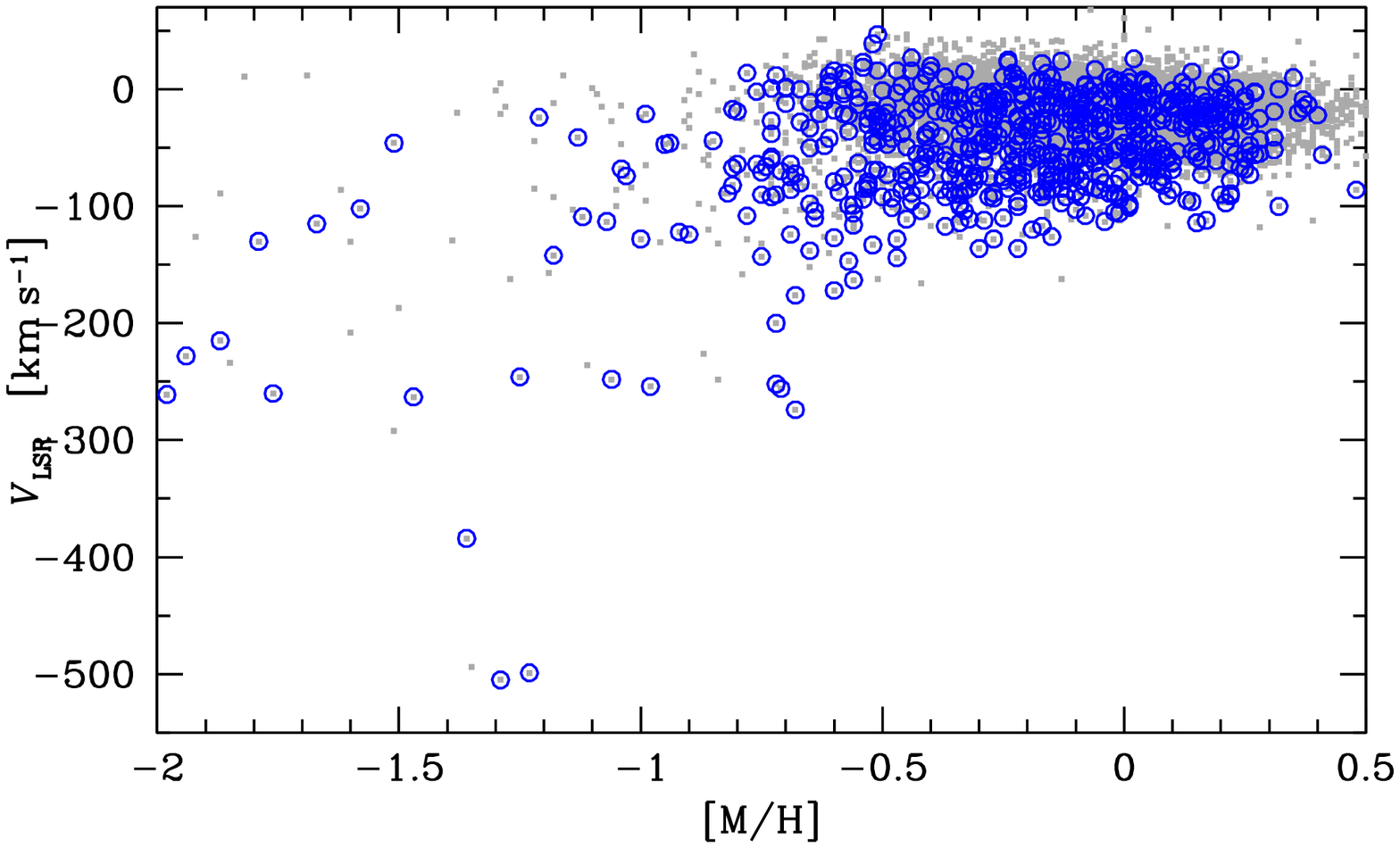}}
\resizebox{\hsize}{!}{
  \includegraphics[bb=18 144 592 500,clip]{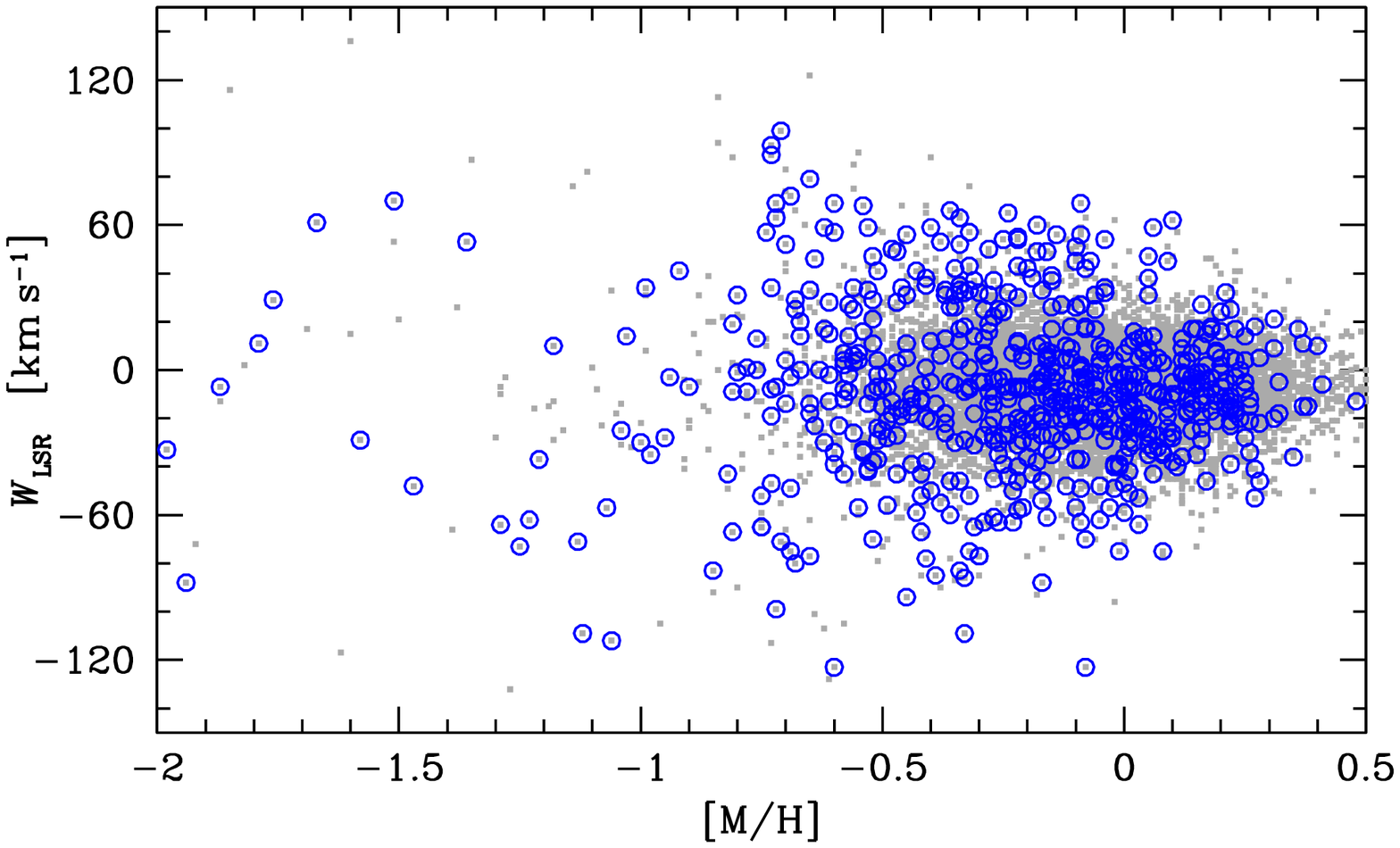}}
\caption{690 out of 714 stars in our sample are also present
in the GCS. The figure shows $\ulsr,\,\vlsr,\,\wlsr$ velocities versus 
[M/H] for the  $\sim 14000$ stars in the GCS (grey dots),
and open circles show our stars (those with
$\rm [M/H]>-2$). Note that all metallicities, [M/H],
are from the Str\"omgren calibration by 
\cite{casagrande2011}.
\label{fig:uvwfeh}
}
\end{figure}

The star sample presented here results from the joint effort
of several observing campaigns with different aims.
In particular, we wanted to trace the
metal-poor limit of the thin disk, the metal-rich limit of the thick disk, 
the metal-poor limit of the thick disk, the metal-rich limit of the stellar
halo, structures in velocity space such as the Hercules stream and 
the Arcturus moving group, and stars that have kinematical properties
placing them in between those of the thin and thick disks. Hence, our selection 
function is very complex and the sample
should not be used to determine the distributions of their properties such as velocity, 
age, and metallicity. 

For the selections of candidate members of the different stellar populations,
we used the kinematical criteria defined
in \cite{bensby2003}, i.e. assuming that they
have Gaussian velocity distributions, different rotation 
velocities around the Galactic centre, and occupy certain fractions of the 
stellar content of the Solar neighbourhood. 
A shortcoming of this kinematical approach is the assumption that the 
distributions follow normal distributions. As noted in 
\cite{ruchti2010} these are first order approximations and the real 
functions may be more complex, which can also be seen in the GCS 
\citep{nordstrom2004}, where the velocity distributions are clearly not
Gaussian. A better understanding of the distribution 
functions may lead to a better decomposition of the stellar disk into 
sub-components \citep{binney2010}. However, for our purposes, these
kinematical criteria are, together with the metallicities [M/H] from the GCS,
a sufficient starting point to probe the thin and thick disks to their extremes.

\begin{figure*}
\resizebox{\hsize}{!}{
\includegraphics{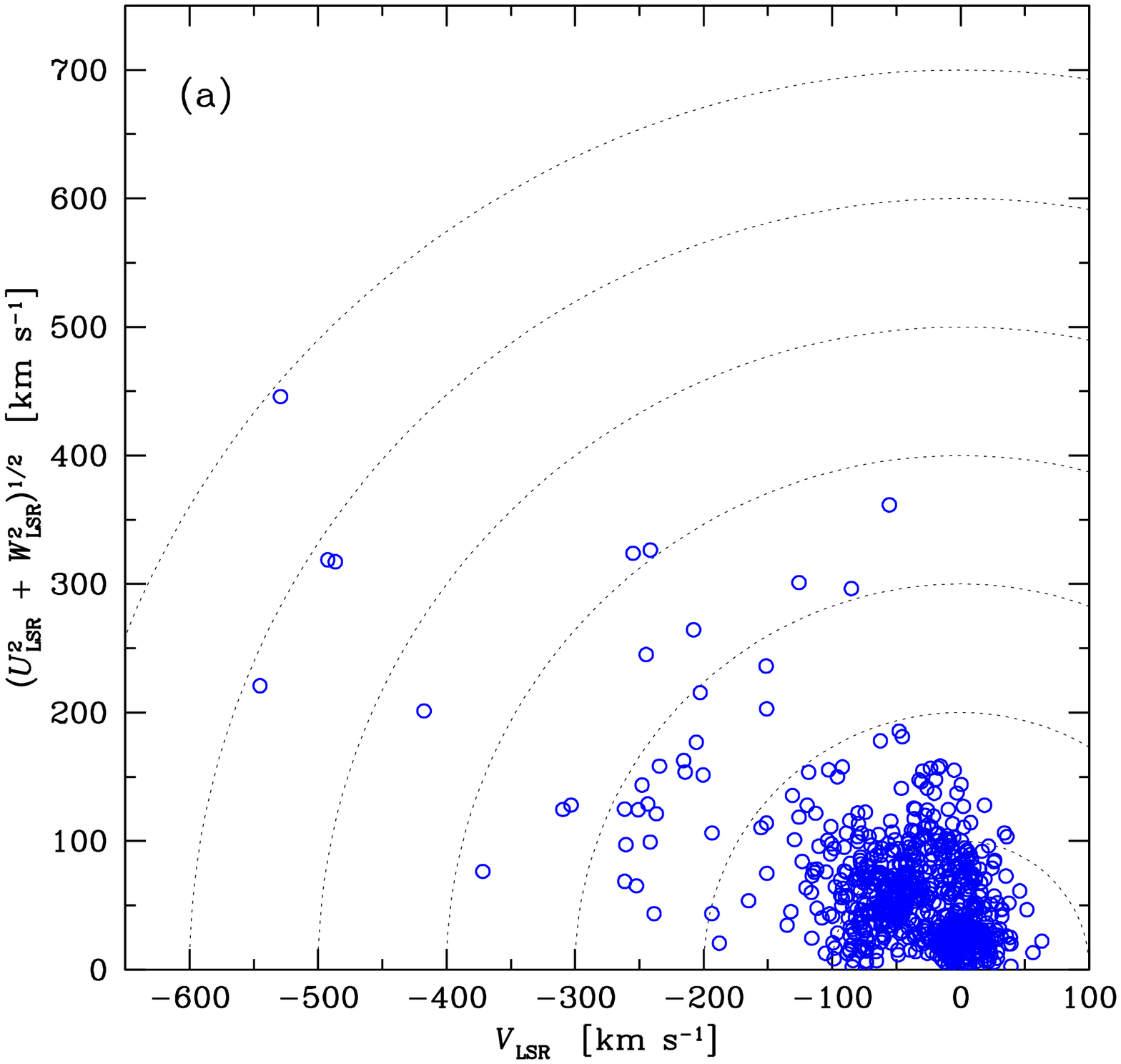}
\includegraphics{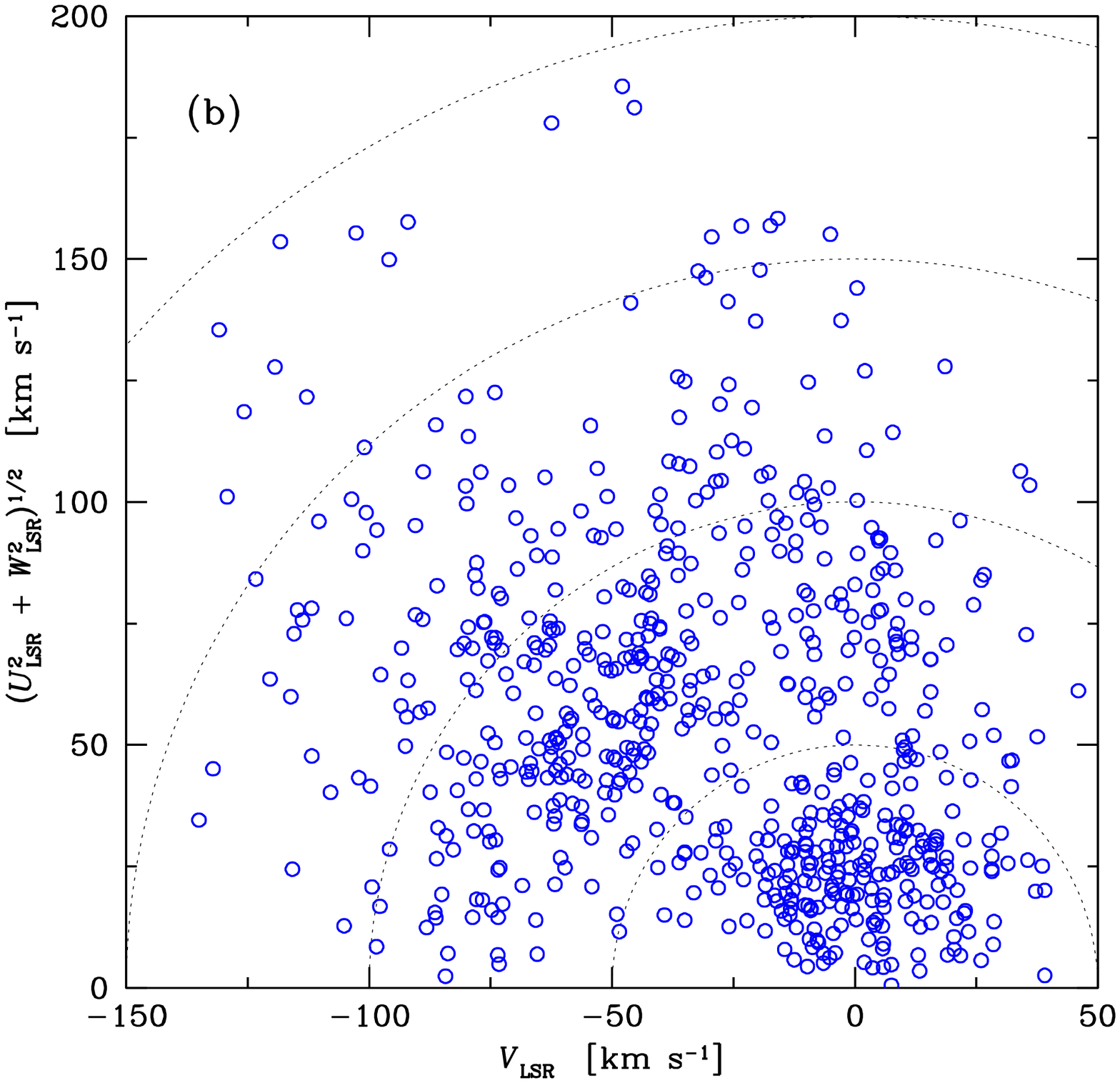}}
\caption{Toomre diagram of our program stars.  
(a) shows the full range of velocities while (b) zooms in
on the region where a majority of the sample is located.
Dotted lines show 
constant values of the
total space velocity, $v_{tot}=(\ulsr^2 + \vlsr^2 + \wlsr^2)^{1/2}$,
in steps of 100\,$\kms$ and $50\,\kms$, respectively, in the two plots. 
\label{fig:toomre}
}
\end{figure*}

Figure~\ref{fig:selection} shows the thick-to-thin disk
probability ratios\footnote{
The method to calculate the
probability ratios, e.g., how much more likely it is that a given 
star is a thick disk star than a thin disk star, is outlined in
Appendix~\ref{sec:kincriteria}.}
($TD/D$) versus the photometric metallicity, [M/H], from 
\cite{casagrande2011} for the $\sim$14\,000 stars in the GCS.
For a star to be a candidate thick disk star, we require it to have a 
probability at least two times that of being a thin disk star ($TD/D>2$), 
and vice versa for a candidate thin disk star $TD/D<0.5$. These probability 
ratios are marked by the two horizontal lines in Fig.~\ref{fig:selection}.
This plot is typical for how the candidate thin and thick disk stellar samples
were selected. The $TD/D$ distribution
of our sample of 714 stars is shown in
Fig.~\ref{fig:tddhist}a, and according to these kinematical criteria we have
387 stars with thin disk kinematics ($TD/D<0.5$), 203 stars with
thick disk kinematics ($TD/D>2$), 36 stars with halo kinematics ($TD/H<1$),
and 89 stars with kinematics in between those of the two disks.
Note that the probability ratios presented here are based on the
thin and thick disk normalisations and velocity dispersions given 
in Table~\ref{tab:dispersions}. As these numbers change, the $TD/D$
probability ratios will also change. For instance, the recent models
by \cite{binney2012} show that the thick disk might be kinematically
hotter vertically than radially, which is opposite to expectation from the numbers given
in Table~\ref{tab:dispersions}. The numbers given  here merely reflect
the way our sample was selected. 
The metallicity distributions of the three $TD/D$ samples are shown
in Fig.~\ref{fig:tddhist}b--d. There is a large
overlap in metallicity between them. The full sample of 714 stars
is further shown in Fig.~\ref{fig:uvwfeh} where the $\ulsr$, $\vlsr$, 
and $\wlsr$ velocities are plotted versus metallicity,
with all the GCS stars as grey dots in the background.
From these plots it is evident that our sample probes the whole GCS, and 
that we sample the extreme kinematics/metallicities;
the sample contains many stars with
hot kinematics at high metallicities and many stars with cold kinematics
at low metallicities.
Please note that the very highest metallicities in the plots in 
Fig.~\ref{fig:uvwfeh} in fact may
not correspond to high iron abundance but may result from a limitation
of photometric metallicity calibrations.

Another way of displaying the sample is
by a Toomre diagram, which is a representation of the combined vertical 
and  radial kinetic energies versus the rotational energy. This is shown 
for the 714 stars in Fig.~\ref{fig:toomre}. Low-velocity stars, within 
a total velocity $v_{\rm tot}\equiv(\ulsr^{2}+\vlsr^{2}+\wlsr^{2})^{1/2}$ 
of $50\,\kms$ are, to a first approximation, mainly thin disk stars, and 
stars with $70\lesssim v_{\rm tot}\lesssim180\,\kms$ are likely to be thick disk stars 
\citep[e.g.,][]{nissen2004}. Stars with $v_{\rm tot}>200\,\kms$ are likely halo 
stars. The slight excess of stars in Fig.~\ref{fig:toomre}
with $\vlsr\approx-50\,\kms$ and 
$\uw\approx50-70\,\kms$ is present because we have deliberately targeted
stars that can be associated with the Hercules 
stream \citep[e.g.,][]{famaey2005,bensby2007letter}

Please note that first results based on the current sample of 714 stars 
were published in \cite{bensby2007letter} about the Hercules stream 
(60 stars) and in \cite{bensby2007letter2} about the metal-rich limit of 
the thick disk (169 stars). The 102 stars from \cite{bensby2003,bensby2005}
are also present in the current sample. No tables or results for individual stars
have been published in the two letters in 2007, that only referred to the 
upcoming full publication (which is this paper). Hence we 
consider the data for all stars (except the 102 stars from 
\citealt{bensby2003,bensby2005}, but those have been re-analysed here) as new.

\section{Space velocities and galactic orbits}
\label{sect:orb}

Space velocities, $\ulsr$, $\vlsr$, and $\wlsr$\footnote{
$\ulsr$ is directed radially inwards towards the Galactic centre, 
$\vlsr$ along the direction of Galactic rotation, and $\wlsr$ 
vertically upwards towards the Galactic North pole.},
were calculated using positions from the Hipparcos catalogue
\citep{esa1997}, parallaxes from the new reduction of the Hipparcos
data by \cite{vanleeuwen2007}, proper motions from the Tycho-2 
catalogue \citep{hoeg2000}, and radial velocities from 
\cite{nordstrom2004} and, if not available in the GCS, from 
\cite{barbierbrossat1994} or \cite{barbierbrossat2000}.
To relate the space velocities to the Local Standard of Rest (LSR),
the Sun's velocity components relative to the LSR 
$(U_{\sun},\,V_{\sun},\,W_{\sun}) = (11.10,\,12.24,\,7.25)\,\kms$ 
from \cite{schonrich2010} were added. 

\begin{figure}
\resizebox{\hsize}{!}{
  \includegraphics[bb=18 420 592 718,clip]{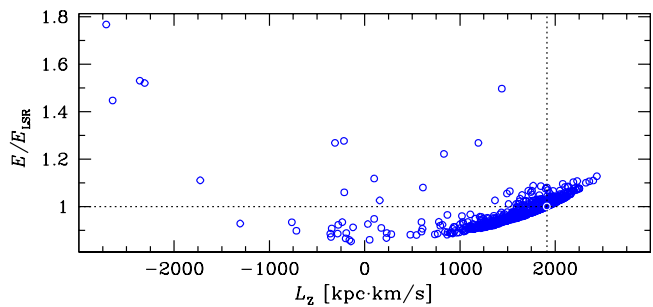}}
\caption{
\label{fig:energy}
Energy versus angular momentum $L_Z$. The energy has been normalised
to the local standard of rest (LSR), which is marked by the white
circle and the dotted lines.
}
\end{figure}

Galactic orbits were then calculated with the {\sc grinton} integrator
\citep{carraro2002,bedin2006} which uses the Milky Way potential 
model by \cite{allen1991b}. The model is time-independent, axisymmetric,
fully analytic, and consists of a spherical central bulge, a disk, 
a massive spherical halo, and has a total mass of $9\times10^{11}$ 
solar masses. 
When calculating $X,\,Y,\,{\rm and}\, Z$ for the stars, 8.5\,kpc was adopted as 
the Sun's distance to the Galactic centre, and 20\,pc for the Sun's 
distance above the Galactic plane \citep{humphreys1995,joshi2007}.
Output parameters from {\sc grinton} are: the minimum and maximum
distances from the Galactic centre $R_{\rm min}$ and $R_{\rm max}$ 
(i.e., the peri- and apocentric values); the maximum distance from the 
Galactic plane $Z_{\rm max}$; the eccentricity, 
$e=(R_{\rm max} - R_{\rm min})/(R_{\rm max} + R_{\rm min})$;
the total energy $E_{\rm tot}$; and the angular momentum $L_{z}$.
Figure~\ref{fig:energy} shows the total energy - angular momentum
plot (also commonly referred to as a Lindblad plot) for the sample.
As the value of $E_{\rm tot}$ is dependent on 
how the Galactic potential is normalised, and may be difficult to compare
between studies, we chose to normalise $E_{\rm tot}$ to the LSR, giving
$E/E_{\rm LSR}$. The parameters are listed for all stars 
in Table~\ref{tab:parameters} in the Appendix.

\begin{table*}
\centering
\setlength{\tabcolsep}{4mm}
 \caption{Observing runs.$^{\dagger}$ 
   \label{tab:observations}
        }
\begin{tabular}{ccrclrl}
\hline\hline
\noalign{\smallskip}
Telescope &
Instrument       &
\multicolumn{1}{c}{$R$}          &
Obs. mode         &
Date         &
\# of stars           &
Sun          \\
\noalign{\smallskip}
\hline
\noalign{\smallskip}
ESO\,1.5-m & FEROS &  48\,000 & visitor & 2000 Sep  & 31\phantom{00} &  -    \\
\multicolumn{1}{c}{"} & \multicolumn{1}{c}{"} & 48\,000 & visitor & 2001 Aug. & 31\phantom{00} & Sky   \\
\noalign{\smallskip}
ESO\,2.2-m & FEROS      & 48\,000 & service & 2005/2006      &  29\phantom{00} &  -    \\
\noalign{\smallskip}
ESO\,3.6-m & HARPS      & 120\,000 & visitor        &  2009 March          &  5\phantom{00}  &       \\
\noalign{\smallskip}
VLT & UVES  & 110\,000 & visitor & 2002 Jul & 4\phantom{00} &  -    \\
\multicolumn{1}{c}{"} & \multicolumn{1}{c}{"} & 110\,000 & service & 2003/2004 & 23\phantom{00} &  -   \\
\multicolumn{1}{c}{"} & \multicolumn{1}{c}{"} & 80\,000 & archive & UVES POP     & 31\phantom{00} & Sky$^{\ddagger}$   \\
\noalign{\smallskip}
NOT & SOFIN &  80\,000 & visitor & 2002 Aug & 11\phantom{00} &  -    \\
\multicolumn{1}{c}{"} & \multicolumn{1}{c}{"} & 80\,000 & visitor & 2002 Nov.   & 16\phantom{00} & Moon  \\
\multicolumn{1}{c}{"} & \multicolumn{1}{c}{"} & 80\,000 & service & 2003 June   &  9\phantom{00} &  -    \\
\multicolumn{1}{c}{"} & \multicolumn{1}{c}{"} & 80\,000 & service & 2004 Feb.   &  5\phantom{00} &  -    \\
\multicolumn{1}{c}{"} & \multicolumn{1}{c}{"} & 80\,000 & service & 2006 March   & 11\phantom{00} &  -    \\
\noalign{\smallskip}
NOT & FIES  & 67\,000  &  visitor       &  2008 July, Nov.      & 6\phantom{00}  &       \\
\noalign{\smallskip}
Magellan & MIKE  &  65\,000 & visitor & 2005 Aug.  & 61\phantom{00} &  -       \\
\multicolumn{1}{c}{"} & \multicolumn{1}{c}{"} & 65\,000 & visitor & 2006 Jan.   &  74\phantom{00} & Vesta    \\
\multicolumn{1}{c}{"} & \multicolumn{1}{c}{"} & 65\,000 & visitor & 2006 April   &  81\phantom{00} & Ganymede \\
\multicolumn{1}{c}{"} & \multicolumn{1}{c}{"} & 65\,000 & visitor & 2006 Aug.   & 158\phantom{00} & Ceres    \\
\multicolumn{1}{c}{"} & \multicolumn{1}{c}{"} & 42\,000 & visitor & 2007 April   &  49\phantom{00} & Ganymede \\
\multicolumn{1}{c}{"} & \multicolumn{1}{c}{"} & 55\,000 & visitor & 2007 May   &   6\phantom{00} &  -       \\
\multicolumn{1}{c}{"} & \multicolumn{1}{c}{"} & 55\,000 & visitor & 2007 July   &  13\phantom{00} &  -       \\
\multicolumn{1}{c}{"} & \multicolumn{1}{c}{"} & 55\,000 & visitor & 2007 Nov.   &  60\phantom{00} &  -       \\
\hline
\end{tabular}
\flushleft
$^{\dagger}${\scriptsize
The columns indicate the telescope and instrument with which the spectra were obtained;  
the spectral resolution ($R$);  observing mode; date the observations 
were carried out;  the number of stars observed ; and the sources for the solar
reference spectra that were obtained.}\\
$^{\ddagger}${\scriptsize
The UVES solar spectrum we use is the one publicly available
on ESO's web pages at 
\url{http://www.eso.org/observing/dfo/quality/UVES/pipeline/solar_spectrum.html}.}
\end{table*}

\section{Observations} \label{sec:observations}

High-resolution
and high signal-to-noise spectra of 848 nearby F and G dwarf and subgiant 
stars in the Solar neighbourhood were obtained during several observing 
runs between 2000 and 2009. Another 48 spectra were gathered 
from the ESO UVES POP database, giving a sample of, in total, 885 stars.  
A significant number (198) of the stars turned out to be 
spectroscopic binaries and/or had too wide spectral lines 
due to too high projected rotational velocities ($v \sin i$), making 
them unsuitable for detailed elemental abundance analysis based on
equivalent width measurements. The final stellar sample we analyse,
and for which stellar parameters, elemental abundances and 
stellar ages were determined, consists of 714 F and G dwarf stars.
Table~\ref{tab:rejected} in the Appendix lists the rejected stars 
and the reasons for which they were rejected.

For the first observing runs, between 2000 and 2002 with FEROS and SOFIN, 
the stars were selected from the catalogue by \cite{feltzing2001}.
Those 102 stars were published in \cite{bensby2003,bensby2005}.
The stars from observing runs in 2003 and onwards were selected from
the GCS. Except for 16 stars (from the
first FEROS and SOFIN runs), all stars in the current sample of 714 stars 
are present in the GCS. 

Table~\ref{tab:observations} lists the different observing
runs and additional details regarding the spectrographs and
data reductions are given below:

\paragraph{\sl FEROS:}

The Fibre-fed Extended Range Optical Spectrograph (FEROS,
\citealt{kaufer1999}) was used in visitor mode during two nights in
2000 and 2001  on the ESO\,1.52-m telescope on La Silla, and in service
mode 2005-2006 when the 
spectrograph had been moved to the ESO\,2.2-m telescope, also on La Silla.  
The data were reduced with the FEROS pipeline available at the time 
(based on MIDAS\footnote{ESO-MIDAS is the
acronym for the European Southern Observatory Munich Image Data
Analysis System, which is developed and maintained by the European
Southern Observatory.} routines).  For a detailed description of the data
reduction procedure we direct the reader to the FEROS-DRS  
manual\footnote{Available at \url{
http://www.eso.org/sci/facilities/lasilla/
instruments/feros/tools/DRS.html}}
and for a short outline to \cite{bensby2003}.  The final products are
complete  optical spectra (3800-9200\,{\AA}) with a resolving power of
$R\approx48\,000$.  The signal-to-noise ratios vary from about 150
in the 2000/2001 data to about 250 in the 2005/2006 data. 

\paragraph{\sl SOFIN:}

Several observing runs were carried out with the SOFIN spectrograph
\citep{ilyin2000} on the Nordic Optical Telescope (NOT) on La
Palma from 2002 to 2006.
The same two settings were used for all these runs, giving
high-resolution spectra with a resolving power of $R \approx 80\,000$
and a spectral coverage of the region between
4500--8800\,{\AA}, with small gaps between orders  
(see Table~2 in \citealt{bensby2005} for exact wavelength coverage). 
Signal-to-noise ratios are generally
around 250. Full details regarding data reductions can be 
found in \cite{ilyin2000} and  a brief outline in \cite{bensby2005}. 

\paragraph{\sl FIES:}

The FIES (Fiber feed Echelle Spectrograph) spectrograph on the NOT 
telescope on La Palma has a fixed wavelength coverage between 365.0 and 730.0\,nm. 
The resolution of FIES is $R = 67 000$. FIES is structurally isolated from the 
telescope dome and thermally isolated from the outside world. This means 
that the instrument is extremely stable which allows for precise 
and non-complicated measurements. On each night, calibration frames 
were taken before and after the observations. These calibrations consist 
of bias and flat-field images and Thorium-Argon (ThAr) spectra. We also 
observed a number of fast rotating B-stars during each night. These 
spectra were used to identify telluric lines in the spectra.
The spectra were reduced using FIEStool\footnote{Further information on FIEStool can be found at
{\tt www.not.iac.es/instruments/fies/fiestool/FIEStool.html}
where a downloadable version is available.} which is built on top of existing 
tasks from the echelle package in IRAF\footnote{IRAF is distributed by the National Optical Astronomy Observatory, which is operated by the Association of Universities for Research in Astronomy (AURA) under cooperative agreement with the National Science Foundation.} 
and provides a simple GUI to organise the data.
Signal-to-noise ratios are around $S/N\approx 400$.

\paragraph{\sl HARPS:}

The HARPS (High Accuracy Radial velocity Planet Searcher) spectrograph \citep{mayor2003} on 
the ESO\,3.6-m telescope on La Silla has a fixed wavelength coverage between 
378.0 and 691.0\,nm. It has two CCDs and there is therefore a gap between 
530.4 and 534.3\,nm. The resolution  is  $R \approx 120\,000$. As for FIES, the 
HARPS spectrograph is structurally isolated from the telescope dome and 
thermally isolated from the outside world. A number of fast rotating B-stars 
was observed each night. 

The HARPS data were reduced using the dedicated pipeline at the telescope 
during the observations. These reductions should be good enough for 
data-analysis, however, we found a persistent, semi-regular pattern in the 
extracted spectra. On first inspection it was thought that the pattern 
might be very regular, but attempts to remove it using fast fourier 
transforms (FFT) (by cutting out high frequency features and transfer 
the spectrum back to the wavelength) did not work even when higher 
order features were removed. The division with a B-star spectrum, that was
obtained each night, did remove these features completely.
Signal-to-noise ratios are  $S/N\approx 300-400$.

\paragraph{\sl MIKE:}

Observations were carried out with the Magellan Inamori Kyocera
Echelle (MIKE) spectrograph \citep{bernstein2003} during eight
observing runs in 2005-2007. A complete optical spectrum is captured on
two CCD:s (Blue CCD 3600-4800\,{\AA} and red CCD 4500-9300\,{\AA}).
Different slit widths of 0.35", 0.5", and 0.7" were used during the different runs, 
giving resolving powers of $R=65\,000$, 55\,000, and 42\,000 on the red CCD, 
and $R=80\,000$, 70\,000, and 53\,000 on the blue CCD, respectively.
All data
were reduced with the MIKE IDL pipeline\footnote{
Available at \url{http://web.mit.edu/~burles/www/MIKE/}} 
by Burles, Prochaska, and Bernstein. During each observing night
with MIKE, we always obtained spectra of rapidly rotating
B stars. These were used in the last stages of the data reduction
to divide out telluric lines and residuals from the fringing
pattern in the near infrared parts of the spectrum.
Signal-to-noise ratios are around or above 250.

\paragraph{\sl UVES:}

Two observing runs were carried out with the Ultraviolet-Visual
Echelle Spectrograph (UVES, \citealt{dekker2000}) on the ESO
Very Large Telescope (VLT) UT2 at the Paranal observatory.

First, four stars were observed as back-up targets during  an observing
run in 2002. Using image slicer \#3, and a rather red setting we got a
resolution of $R\approx 110\,000$, and a  wavelength coverage between
$5500-7500$\,{\AA} with a 100\,{\AA} gap around
6000\,{\AA}. These data were reduced with the
UVES pipeline available at the time (based on  MIDAS routines).
Second, 23 stars were observed in service mode in 2003/2004, using the
same setup as for the 2002 run. These data were reduced with the 
{\sc reduce} package \citep{piskunov2002}. 

Finally we obtained reduced spectra for 31 stars from the UVES
Paranal Observatory Project\footnote{Raw data as well as
reduced data can be downloaded from the UVES ESO archive using program
ID 266.D-5655(A), or from \url{http://www.sc.eso.org/santiago/uvespop/}}, \citealt{bagnulo} (UVES POP). The UVES
POP stars were observed with two instrument modes in order to cover  
almost completely the  wavelength
interval from 300 to 1000\,nm.  The spectral resolution is about
$R\approx 80\,000$, and for most  of  the spectra, the typical $S/N$ ratio
is 300 to 500  in the $V$ band.

\section{Abundance analysis}
\label{sec:analysis}

\begin{figure*}
\centering
\resizebox{\hsize}{!}{
\includegraphics[bb=80 145 565 718,clip]{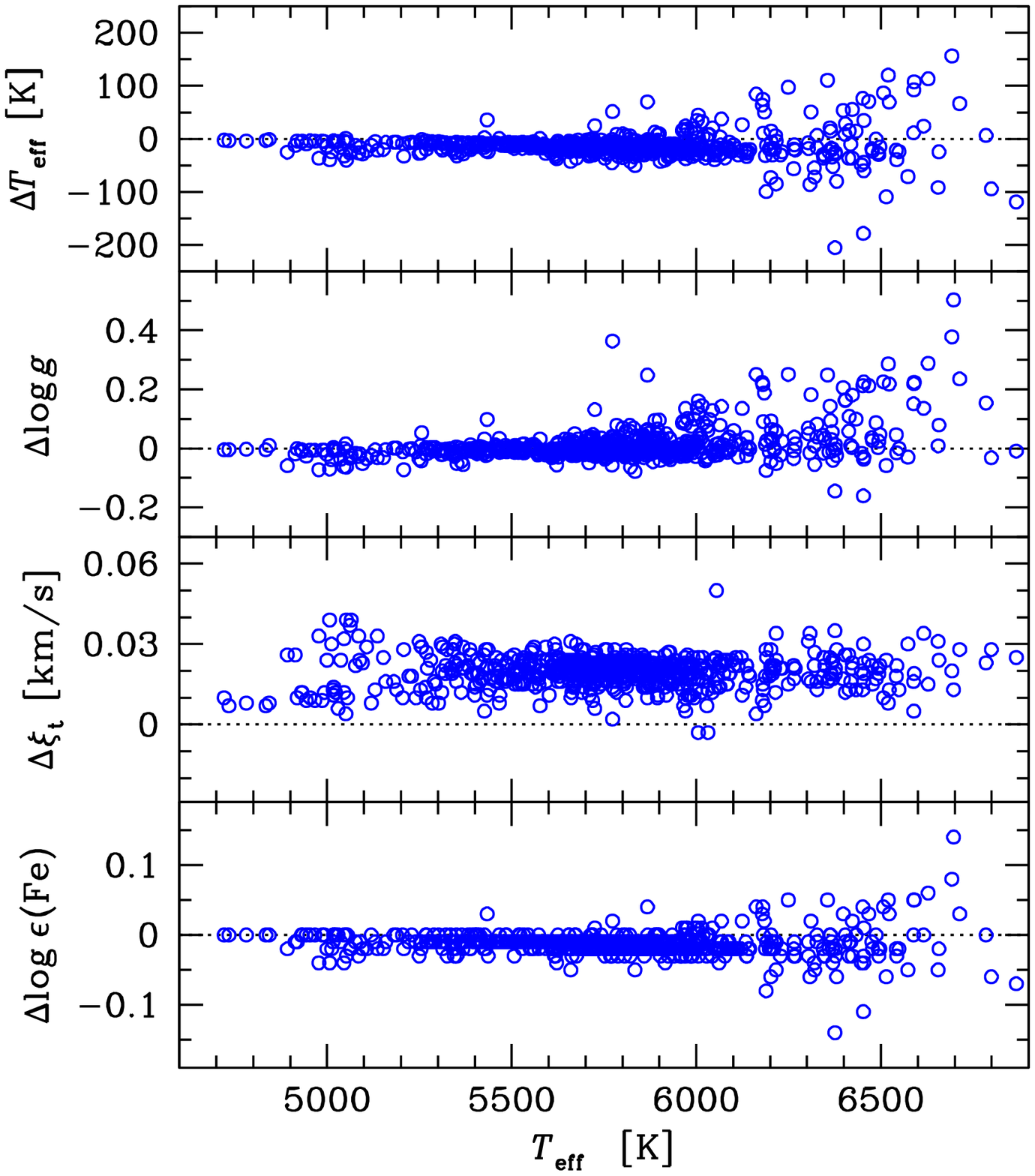}
\includegraphics[bb=170 145 565 718,clip]{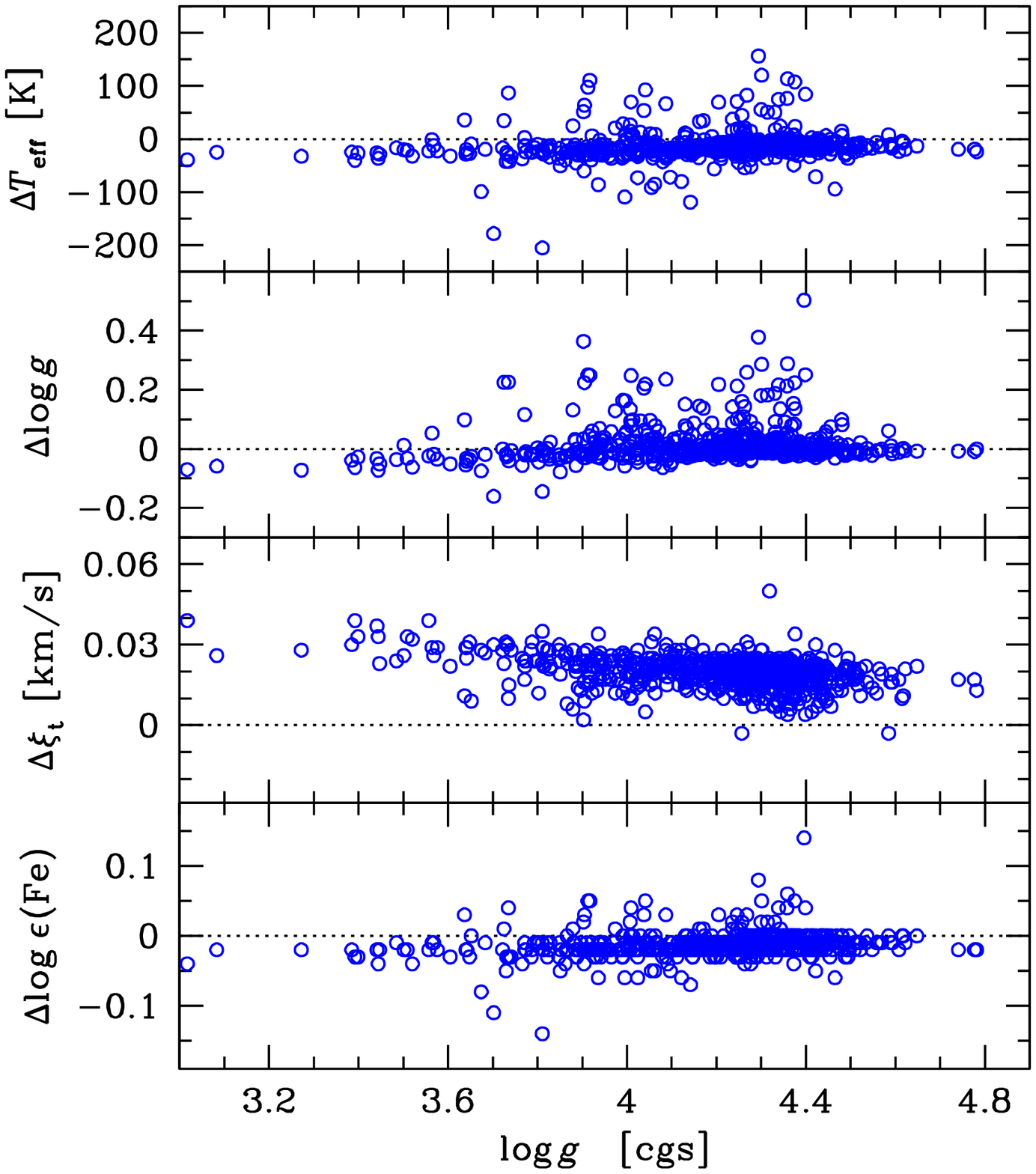}
\includegraphics[bb=170 145 590 718,clip]{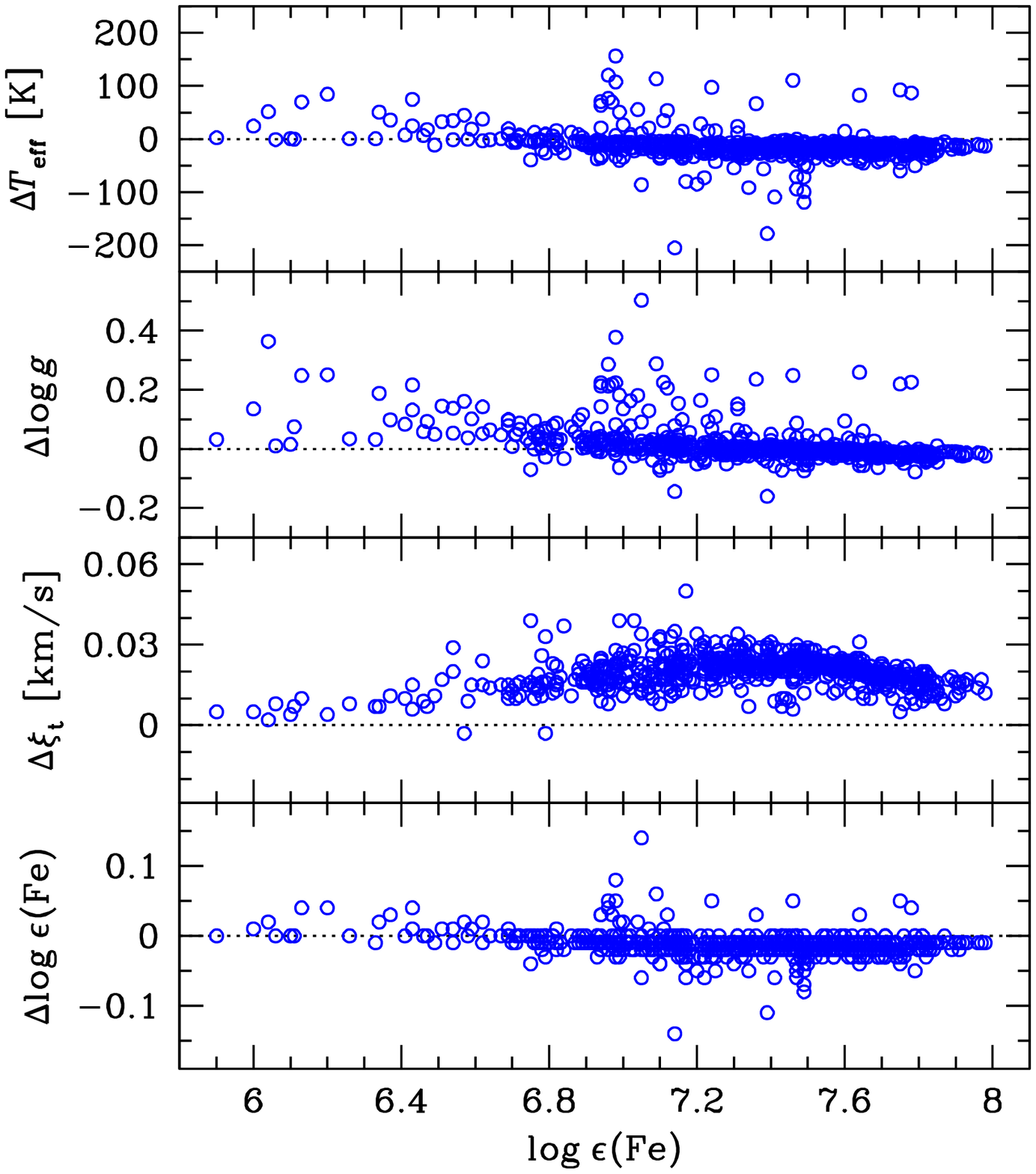}}
\caption{
The effects on the stellar parameters when including \ion{Fe}{i}
NLTE corrections from \cite{lind2012} in the analysis. 
The differences are given as NLTE values minus LTE values.
\label{fig:nlte}
        }
\end{figure*}
\begin{figure*}
\centering
\resizebox{\hsize}{!}{
\includegraphics[bb=80 390 565 718,clip]{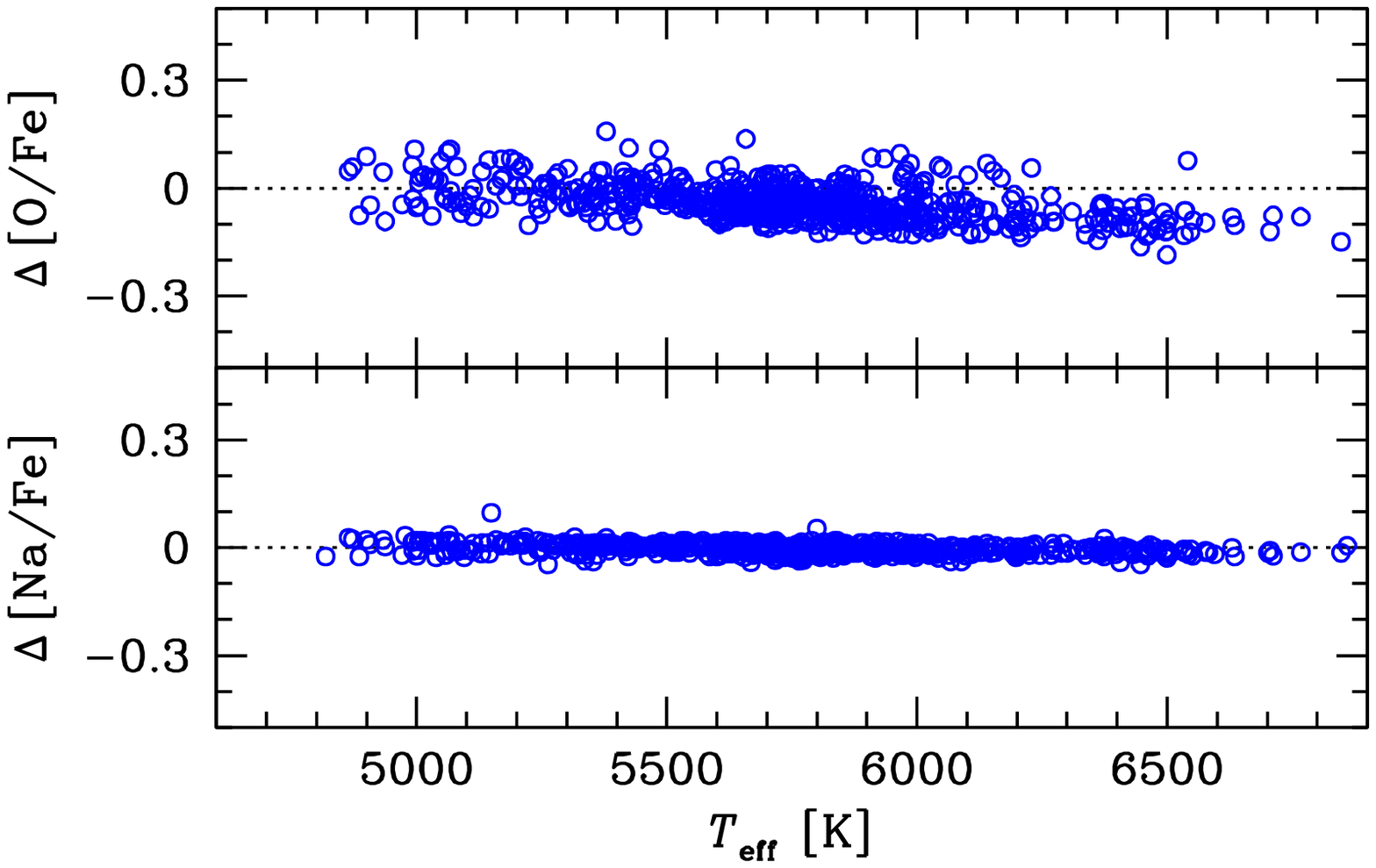}
\includegraphics[bb=170 390 565 718,clip]{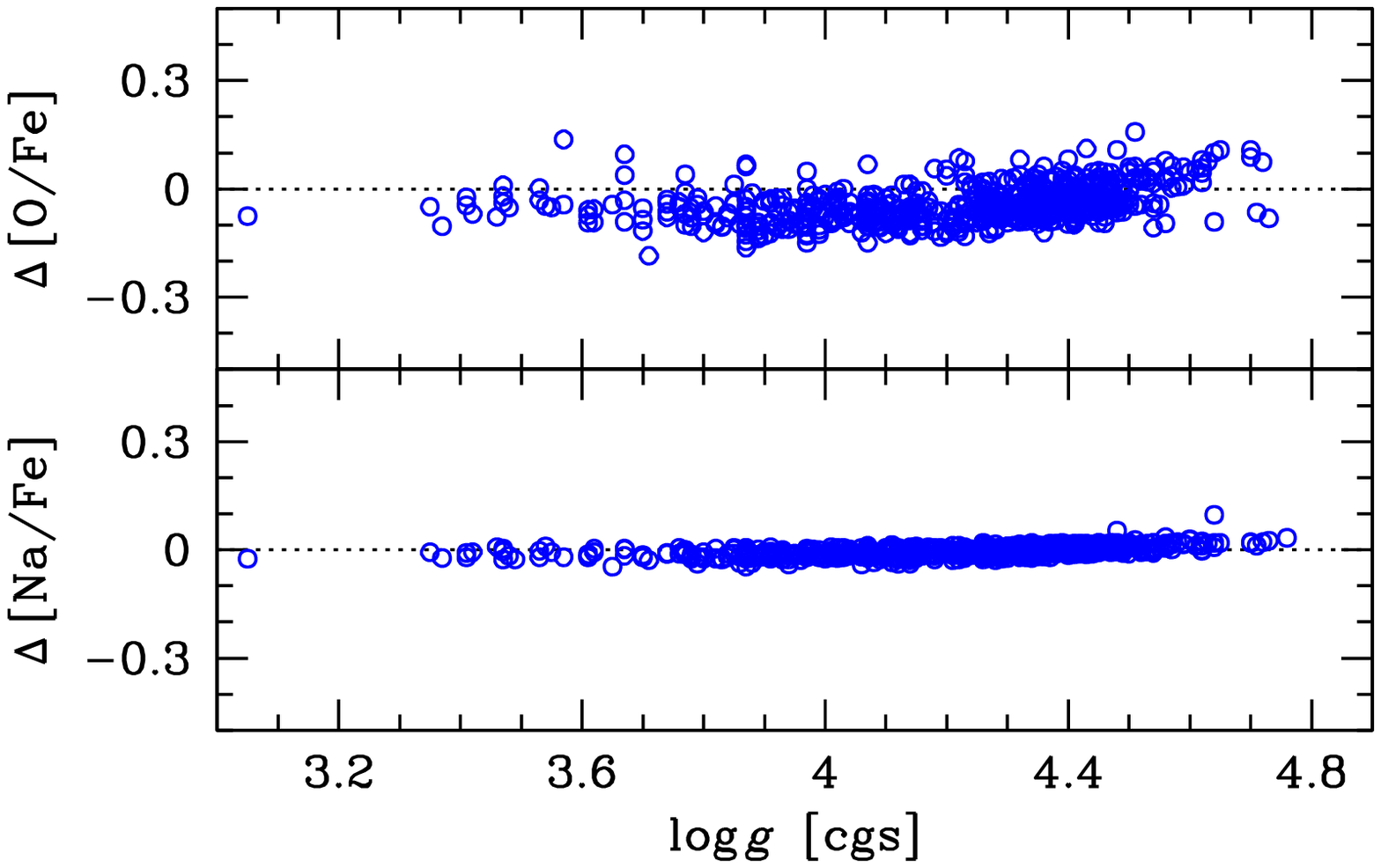}
\includegraphics[bb=170 390 590 718,clip]{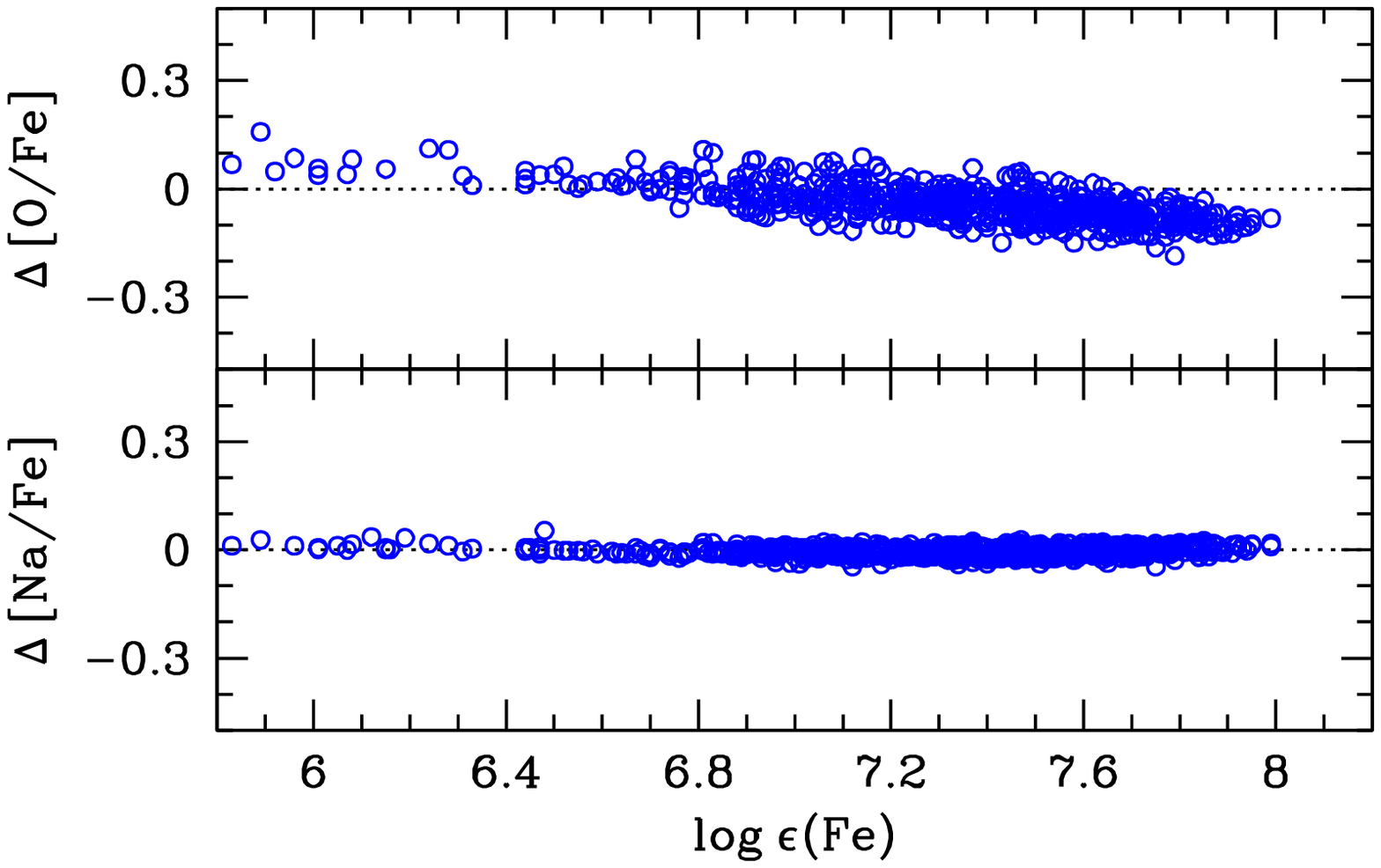}}
\caption{
The effects of the O and Na NLTE corrections. 
The differences are given as NLTE values minus LTE values.
\label{fig:onanlte}
        }
\end{figure*}

\subsection{Methodology}

The methodology to determine stellar parameters and elemental
abundances is essentially the same as in \cite{bensby2003,bensby2005}.
Briefly, it is based on equivalent width measurements and 
one-dimensional, plane-parallel, local thermodynamical equilibrium 
(LTE) model stellar atmospheres calculated with the Uppsala
MARCS code \citep{gustafsson1975,edvardsson1993,asplund1997}.  
For F and G dwarf stars, these models are satisfactory, and show 
little deviation from other models such as those calculated with the 
ATLAS code by  R.\,Kurucz and collaborators or the new version of the 
MARCS code  \citep{gustafsson2008}. A common way to determine 
stellar parameters is by requiring excitation balance of 
abundances from \ion{Fe}{i} lines to get the effective temperature
$(\teff$), and by requiring that abundances from \ion{Fe}{i}
lines are independent of reduced line strength to get the microturbulence
parameter ($\xi_{\rm t}$). The surface gravity ($\log g$) can be determined
from ionisation balance between abundances from \ion{Fe}{i} and 
\ion{Fe}{ii} lines, in which case the analysis is strictly
spectroscopic, or, if the stars have accurate distances, through the
formula that relates effective temperature and bolometric flux.
As the stars in the sample have parallaxes from the Hipparcos 
satellite \citep[as determined by][]{vanleeuwen2007},
we will in this paper investigate both methods to determine $\log g$,
and in Sect.~\ref{sec:ionbalance} we show that the two ways both have their 
strengths and weaknesses.

In total, for the sample of 714 stars, more than 300\,000 equivalent 
widths were measured by (the first author's right) hand using the IRAF 
task SPLOT by fitting 
Gaussian profiles to the observed line profiles. For some elements that 
often have quite strong lines (e.g., Mg, Ca, Na, and Ba), and if a 
Gaussian profile did not satisfactorily match the observed line profile,  
a Voigt profile was fitted to ensure that the wide wings and narrower
cores of those lines were properly accounted for. The continuum was set 
locally for each line.  To avoid saturation effects and non-linearities,
only \ion{Fe}{i} and \ion{Fe}{ii} lines with measured equivalent 
widths less than 90\,m{\AA} were used in the determination of the 
stellar parameters. The same effects mentioned can of course
also affect other abundances. But since the absolute Fe abundances
were used in the determination of stellar parameters it is extra important
not to include too strong Fe lines. For other elements, such as e.g.,
Mg and Ba that often have strong lines, we have no other option than to 
use the few available lines, and they often happen to be quite strong.
The effects might not be so severe in the end for these elements as 
the final abundances are normalised to the Sun, on a line-by-line basis.

Compared to our analysis in \cite{bensby2003,bensby2005} the current 
analysis contains the following changes and improvements:
\begin{itemize}
\item
The chemical compositions of the model atmospheres used in 
\cite{bensby2003,bensby2005} were scaled with metallicity relative 
to the standard solar abundances as given in 
\cite{asplundgrevessesauval2005}. To better reflect the 
actual compositions of the stars, the models are now
enhanced in the $\alpha$-elements (e.g., O, Mg, Si, Ca, Ti)
at sub-solar metallicities: $\rm [\alpha/Fe]=+0.4$\,dex for 
$\rm [Fe/H]\leq-1.0$; $\rm [\alpha/Fe]$ linearly decreasing
from $+0.4$ to 0 in the interval $\rm -1.0<[Fe/H]<0.0$.
\item 
Corrections for non-LTE effects for the \ion{Fe}{i} lines,
based on the calculations by \cite{lind2012}, are included on a 
line-by-line basis in each iterative step of the analysis. 
The effects on the stellar parameters is investigated in 
Sect.~\ref{sec:fe1_nlte}.
\item
The atomic line list used in \cite{bensby2003,bensby2005} has been
expanded with another $\sim$50 \ion{Fe}{i} lines from \cite{nave1994}.
These lines were selected on the basis that the derived
abundances for the Sun should agree within 0.05\,dex of the
average abundance from the $\sim150$ original \ion{Fe}{i} lines. 
The atomic data for the additional lines were sourced from the VALD 
database \citep{vald_1,vald_2,vald_3}. 
\item
We use several solar spectra, obtained from
different observing runs and different spectral resolutions 
(see Table~\ref{tab:observations}). 
This led us to revise some of the astrophysical $\log gf$ values given in 
\cite{bensby2003} so that the abundance from each line matches the
solar abundances given by \cite{asplund2009}. Exceptions
are \ion{Fe}{i}, \ion{Fe}{ii}, \ion{Ti}{i}, \ion{Ti}{ii},
and \ion{O}{i}, for which laboratory $\log gf$ values are used.
(More details regarding the choice of $\log gf$ values
can be found in \citealt{bensby2003}.)
The full line list totalling 498 lines for 13 elements
with updated atomic data is given in Table~\ref{tab:atomdata}, 
together with the measured solar equivalent widths.
\item
The atomic collisional broadening constants by 
\cite{barklem2001,barklem2005} have been included in the analysis. 
\end{itemize}

Furthermore, the analysis is strictly differential relative to the Sun. For this we used 
solar spectra that were obtained, reduced, and analysed using exactly
the same instruments and methods as were used for the stars in the 
sample. The spectra were obtained through observations of scattered 
solar light from the afternoon sky, the Moon, Jupiter's moon Ganymede,
and the asteroids Vesta and Ceres (see Table~\ref{tab:observations}).

The equivalent widths  measured in these different solar spectra agree 
well, with differences below 1-2\,\%. Using the equivalent widths 
from each of the different solar spectra, the atmospheric parameters for 
the Sun were determined and we find very good agreement.
$\teff$ varies between 5750\,K to 5798\,K, $\log g$ between 4.42 and 
4.45, and the Fe abundance between 7.56 and 7.59.
As the different solar spectra have 
been obtained during a period of six years, during which they also 
were measured, this indicates that the the way we have measured the 
equivalent widths has been consistent throughout the years.

Given the good agreement of the measured equivalent widths and 
of the stellar parameters from the different solar spectra, we find 
it unnecessary to use different solar spectra to normalise the 
different sets of observations. Instead we will use the average 
values of the measured equivalent widths from all seven solar 
spectra. Stellar parameters for the Sun based on the average 
equivalent widths are: $\teff = 5773$\,K, $\log g =4.42$, 
$\xi_{\rm t}=0.88\,\kms$, and $\log\epsilon(\rm Fe)=7.58$.

The final abundances are normalised relative to our solar values
on a line-by-line basis. 
In \cite{bensby2003,bensby2005} we 
used the mean abundance from all spectral lines to represent  
the abundance for a given element. Now we have chosen to use 
the median instead. The median is less sensitive to outliers, and 
especially for elements for which only a few lines were measured, 
the influence of one erroneously measured (or blended) line  
will be smaller. For Ti and Cr, abundances from both neutral 
and ionised lines were used in the calculation of the median value.

The final abundance ratios are given in Table~\ref{tab:parameters}
which also gives the standard deviation from the 
median value (line-to-line scatter) and the number of lines used 
when computing the median value.
To avoid systematic errors due to the updates and changes listed above,
the 102 stars in \cite{bensby2003,bensby2005} have been re-analysed.

\begin{figure}
\centering
\resizebox{\hsize}{!}{
\includegraphics{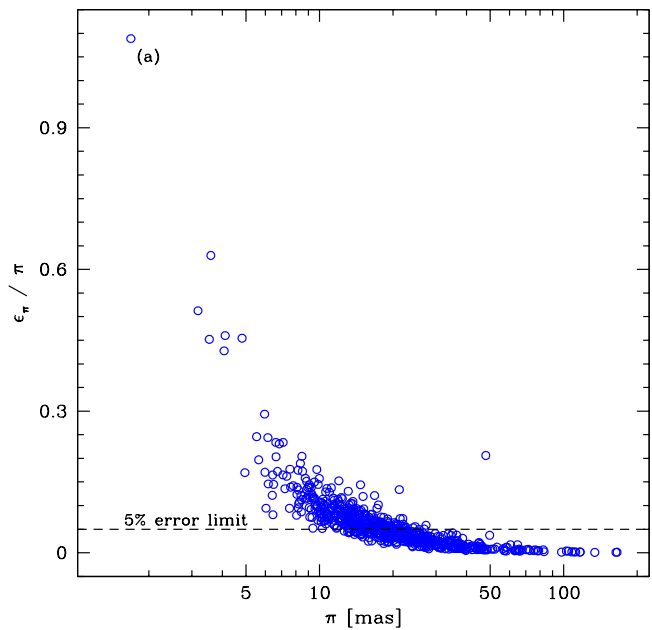}}
\caption{
The fractional uncertainty in parallax versus the parallax
from \cite{vanleeuwen2007} for the 714 stars in the sample.
339 stars have errors larger than 5\,\% and 95 stars have errors
larger than 10\,\%.
\label{fig:parallax}
        }
\end{figure}

\subsection{NLTE corrections}

\begin{figure*}
\resizebox{\hsize}{!}{
\includegraphics[bb=18 144 570 718,clip]{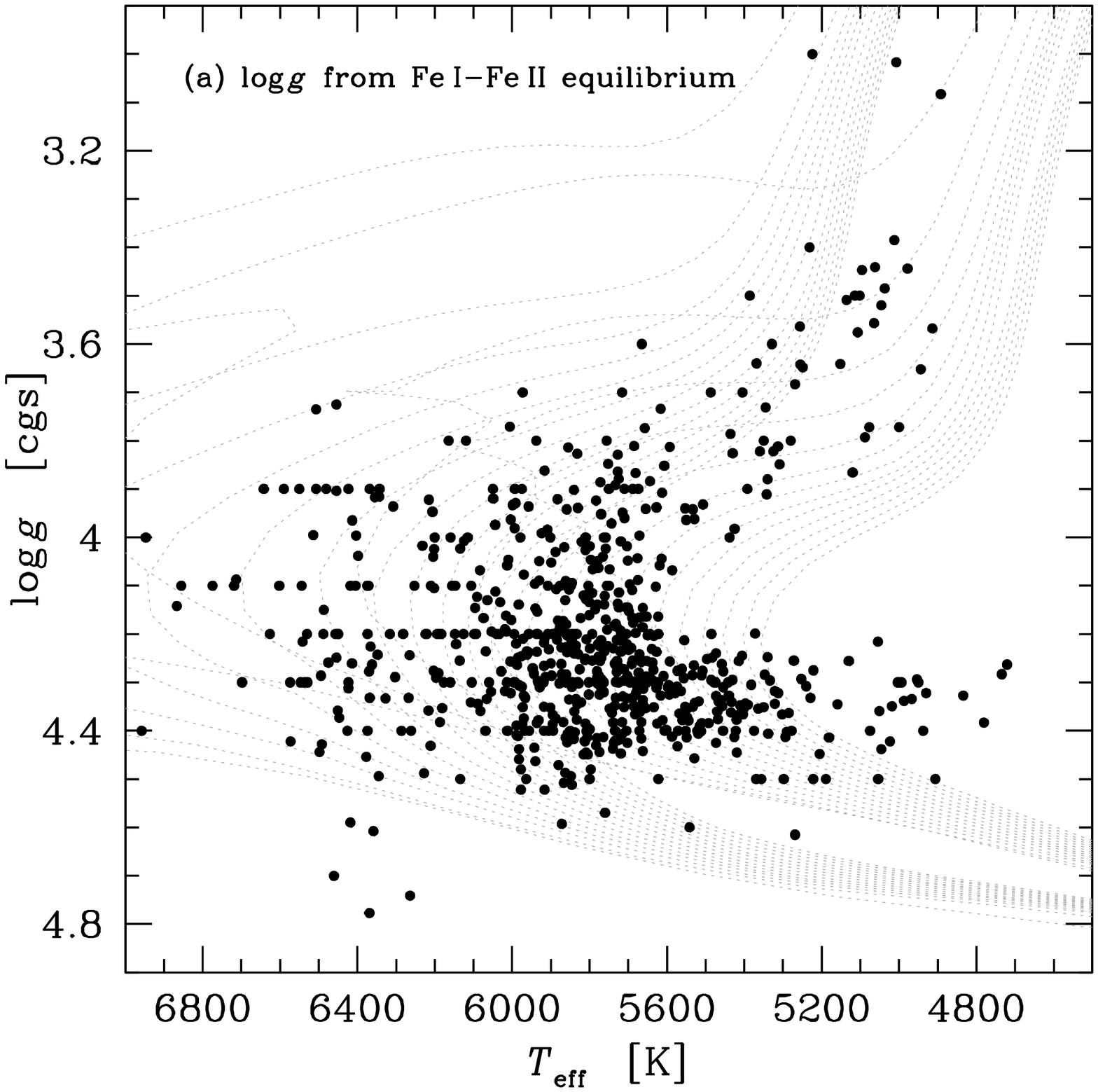}
\includegraphics[bb=75 144 570 718,clip]{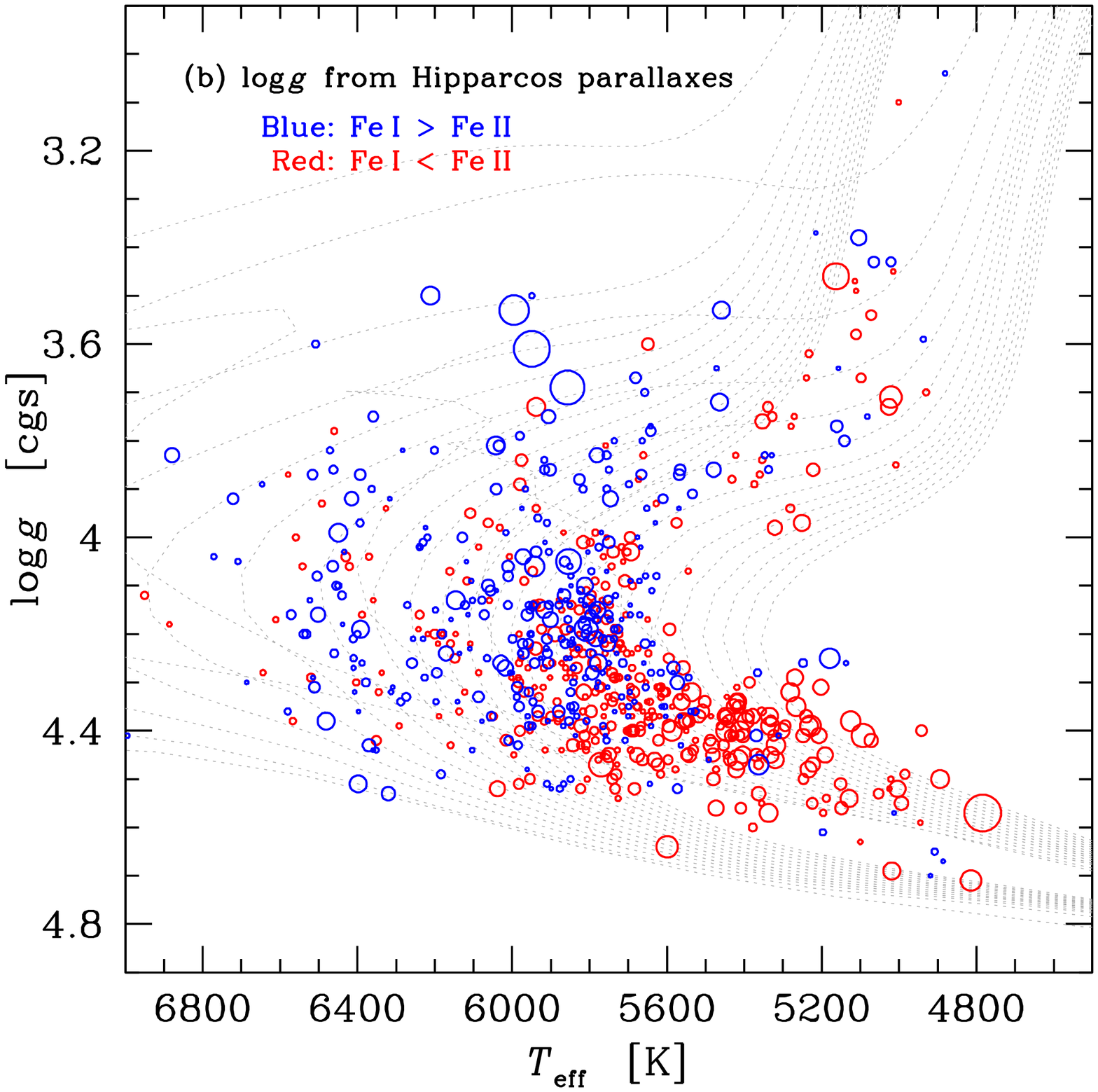}
\includegraphics[bb=75 144 592 718,clip]{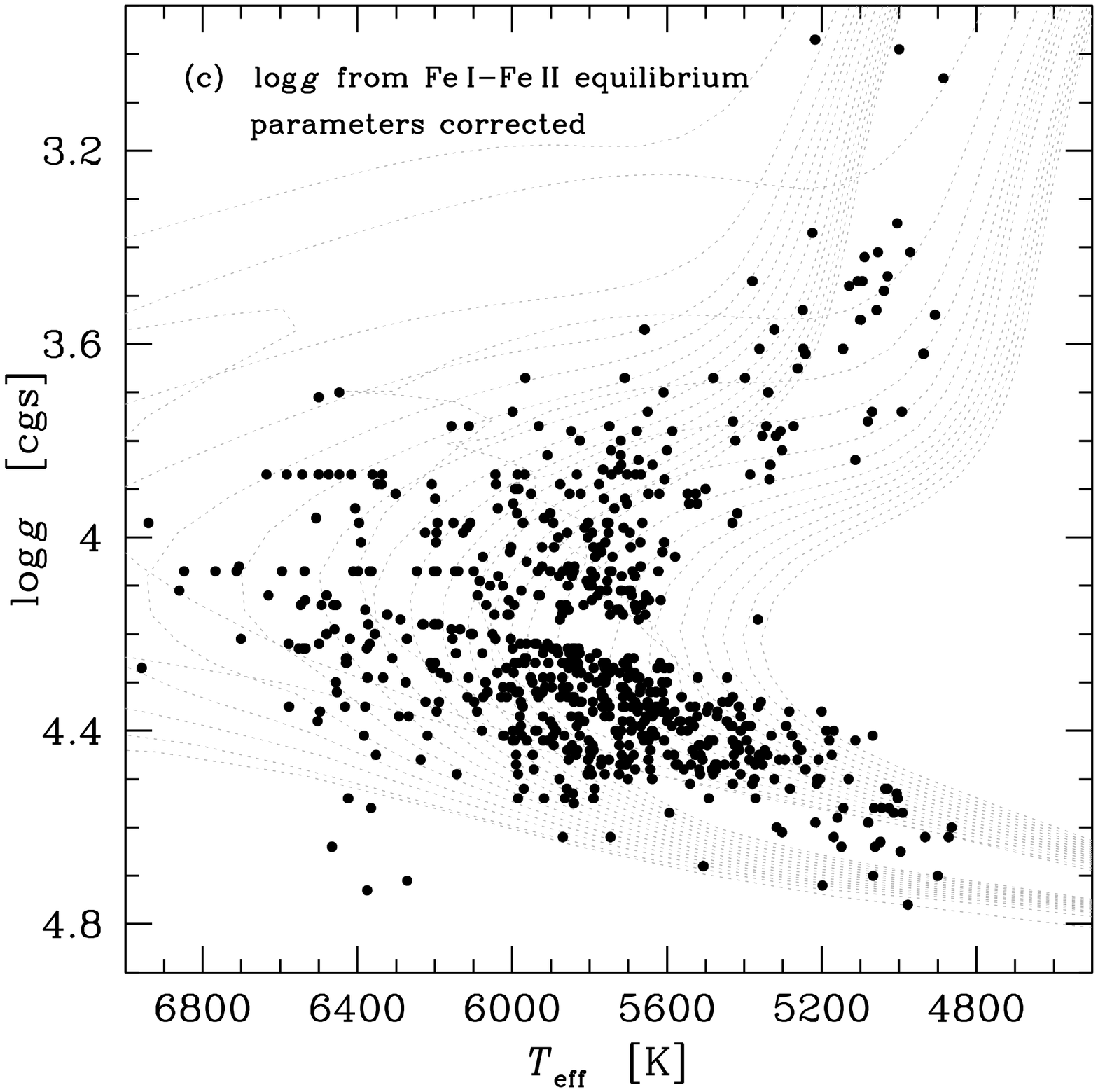}
}
\caption{HR diagram for the sample when (a) $\log g$ is
based on \ion{Fe}{i}-\ion{Fe}{ii} ionisation
equilibrium, and (b) when $\log g$ is based on
Hipparcos parallaxes. In (b) the sizes of the 
circles are scaled with the difference between \ion{Fe}{i} and \ion{Fe}{ii}
abundances. Red circles mark those stars where the \ion{Fe}{i} abundances
are lower than the \ion{Fe}{ii} abundances,
and vice versa for the blue circles. The
$\alpha$-enhanced Yonsei-Yale (Y2) 
isochrones by \cite{demarque2004}
have metallicities of $\rm [Fe/H]=-1$ and $+0.3$\,dex, respectively, and are 
shown from 1 to 15 Gyr in steps of 1\,Gyr.
\label{fig:hr}
}
\end{figure*}

\paragraph{\bf \ion{Fe}{i}:}
\label{sec:fe1_nlte}

Abundances based on \ion{Fe}{i} lines are sensitive to departures
from the assumption of LTE, while abundances from \ion{Fe}{ii} lines
generally are not \citep[e.g.,][]{thevenin1999,melendez2009,lind2012}.
As \ion{Fe}{i} lines play a key role in our analysis and the
determination of stellar parameters, it is important to investigate 
this and if possible, to make corrections accounting for the effects. 
We have done that by using the NLTE calculations
for \ion{Fe}{i} lines by \cite{lind2012}.
Using an IDL script kindly provided by K.~Lind, the corrections were 
applied in real-time on a line-by-line basis in the process of 
determining the stellar parameters.

Stellar parameters were also determined without applying the \ion{Fe}{i}
NLTE corrections and Fig.~\ref{fig:nlte} shows how the
stellar parameters change. The differences are usually very small,
but we do see a larger scatter in $\teff$, log\,$g$, and Fe abundance
for stars with effective temperatures above approximately 6100\,K. 
There might also
be slight systematic trends with surface gravity, however, too small
to be statistically significant.

The average effects on the stellar parameters are (NLTE values minus 
LTE values, and excluding stars with $\teff>6100$\,K in parentheses):
$\Delta \teff=-12\,(-14) \pm 28\,(12)$\,K,
$\Delta \log g=+0.012\,(+0.002) \pm 0.059\,(0.035)$,
$\Delta \log\,({\rm Fe})=-0.013\,(-0.013) \pm 0.016\,(0.008)$, and
$\Delta \xi_{\rm t} = +0.019\,(+0.020) \pm 0.006\,(0.006)$.
For the Sun, the effect on the Fe abundance when applying the NLTE
corrections is $-0.01$\,dex. This means that for the whole sample of 714
stars the average metallicity becomes $-0.003$\,dex lower after including
the \ion{Fe}{i} NLTE corrections in the analysis. While this is
a truly minuscule effect, the effects on temperatures and surface 
gravities could have some impact on stellar ages, and possibly also
when determining abundances for elements like Li, which is very 
temperature-sensitive. The stars for which we see significant effects are
those that are warmer than about 6100\,K.

\paragraph{\bf Oxygen and sodium:}

The oxygen abundances have been determined from the infrared triplet
lines at 777\,nm\footnote{
The forbidden oxygen line at 
630\,nm line was not analysed here since the analysis in this
paper is purely based on equivalent width measurements. Furthermore, the 
spectral range of the UVES 2002/2004 as well as the FIES and HARPS
spectra (in total 38 stars) does not cover the 777\,nm triplet lines
and hence the number of stars with oxygen abundances is lower than 714.}. 
These lines are known to be strongly affected by deviations
from LTE \citep[e.g.,][]{kiselman1993,asplund2009}. To correct our
oxygen abundances for NLTE effects, we apply the empirical
formula from \cite{bensby2004}, who analysed
the forbidden oxygen line at 630\,nm, which is a very robust
indicator of the oxygen abundance, unaffected by departures from
LTE \citep[e.g.,][]{kiselman1993,asplund2009}.

For sodium we applied the NLTE corrections from \cite{lind2011},
using an IDL script that was kindly provided by Karin Lind.

How the NLTE corrections affect the [O/Fe] and [Na/Fe] abundance
ratios is shown in Fig.~\ref{fig:onanlte}.

\subsection{Surface gravity}
\label{sec:ionbalance}

Two widely used methods to determine
the surface gravity are derived from ionisation balance between
\ion{Fe}{i} and \ion{Fe}{ii}, and from basic principles through 
the relationship between bolometric flux, temperature, and gravity
(see, e.g., Eq.~4 in \citealt{bensby2003}). 
The latter requires that the distance to the star is known, and in 
our case all stars have distances based on Hipparcos parallaxes 
from the new reduction by \cite{vanleeuwen2007}.

There are some indications that by using parallaxes to determine 
$\log g$ from basic principles, one introduces an external source 
of uncertainty, independent of the spectra. For instance, studies 
of solar analogs have shown that a 
purely spectroscopic approach (i.e. $\teff$ from excitation balance 
of abundances from \ion{Fe}{i} lines and $\log g$ from ionisation 
balance of abundances from \ion{Fe}{i} and \ion{Fe}{ii} lines) has
better precision than when using $\log g$ based on parallaxes 
\citep[e.g.,][]{ramirez2009}. Another advantage of using a purely 
spectroscopic approach in our case is that the uncertainties will 
be essentially distance-independent. This is so because the sample contains 
relatively bright stars ($V<9$), and as a majority have been observed 
with large, 6-8\,m class telescopes, the exposure times are short and
the spectra have high signal-to-noise independent of the magnitude (or distance) 
of the star. If the
parallax method is used, the uncertainties increase with distance,
as is seen in Fig.~\ref{fig:parallax}, which shows the 
fractional parallax errors versus the parallaxes for our stars:
there is a clear increase in the parallax error with distance. 
The sample contains 329 stars that have  fractional errors in the 
parallax larger than 5\,\% and 89 stars larger the 10\,\%.
Furthermore, for stars with large parallax uncertainties
the Lutz-Kelker bias can be severe
and is impossible to correct for on an individual basis.

Therefore, we start by analysing our sample using ionisation balance
to get the surface gravity. Figure~\ref{fig:hr}a shows the resulting 
HR diagram, and at a first glance, it appears peculiar in the sense 
that the lower main sequence is horizontal rather than declining. 
As there are many stars that fall in regions unoccupied by 
isochrones, and as the whole appearance is somewhat ``uncomfortable'',
we redetermine the stellar parameters, but this time using the 
Hipparcos parallaxes to get the surface gravity.  
The resulting HR diagram, in Fig.~\ref{fig:hr}b, having
gravities based on Hipparcos parallaxes,
shows a declining main sequence (as expected). 
It should be noted that the inclusion of the \ion{Fe}{i} NLTE 
corrections are far too small to have an effect on the gravities of 
the magnitude to produce the flat lower main sequence.

\begin{figure}
\resizebox{\hsize}{!}{
\includegraphics[bb=18 200 570 718,clip]{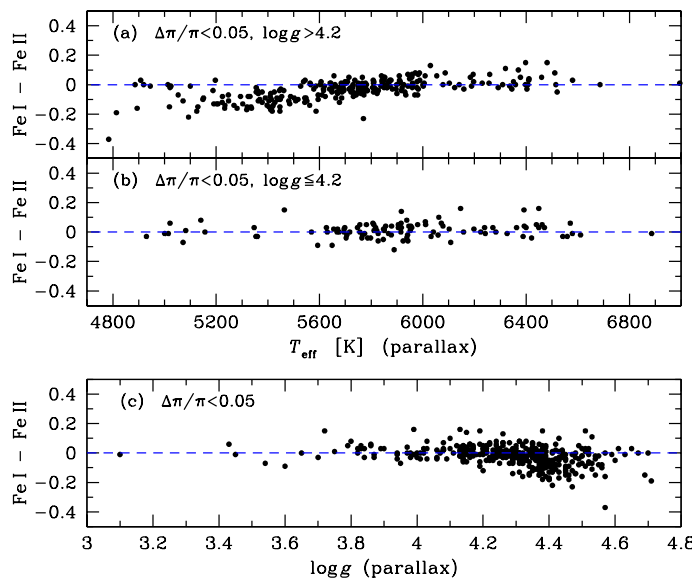}}
\caption{The difference in abundances from \ion{Fe}{i} and \ion{Fe}{ii}
lines versus effective temperature (a and b,),
and versus surface gravity (c). 
Only stars with relative errors in their parallaxes
smaller than 5\,\% are included, and the parameters are the ones
when $\log g$ is determined from the Hipparcos parallax.
\label{fig:decide1}
}
\end{figure}
\begin{figure}
\resizebox{\hsize}{!}{
\includegraphics[bb=18 400 570 718,clip]{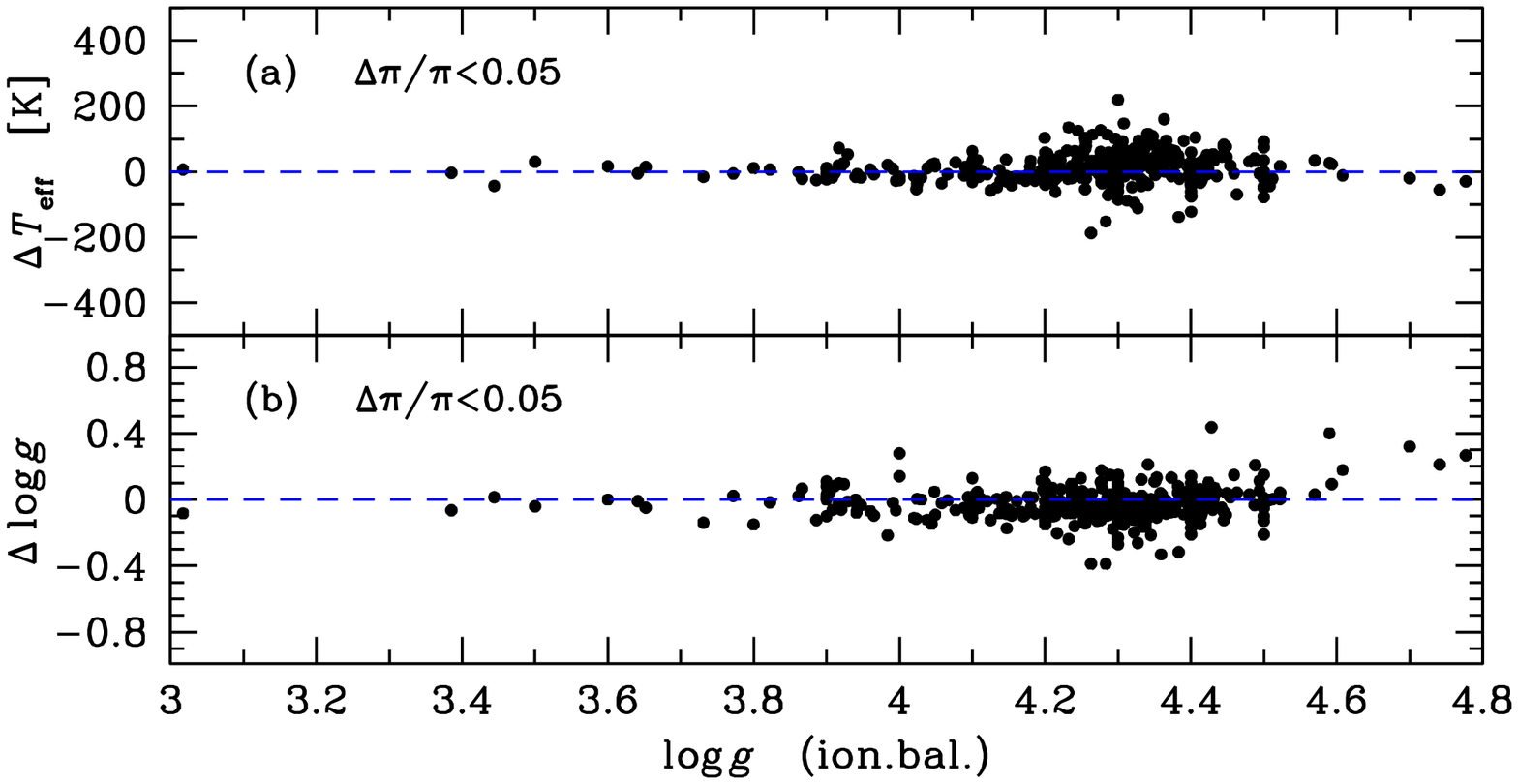}}
\resizebox{\hsize}{!}{
\includegraphics[bb=18 400 570 710,clip]{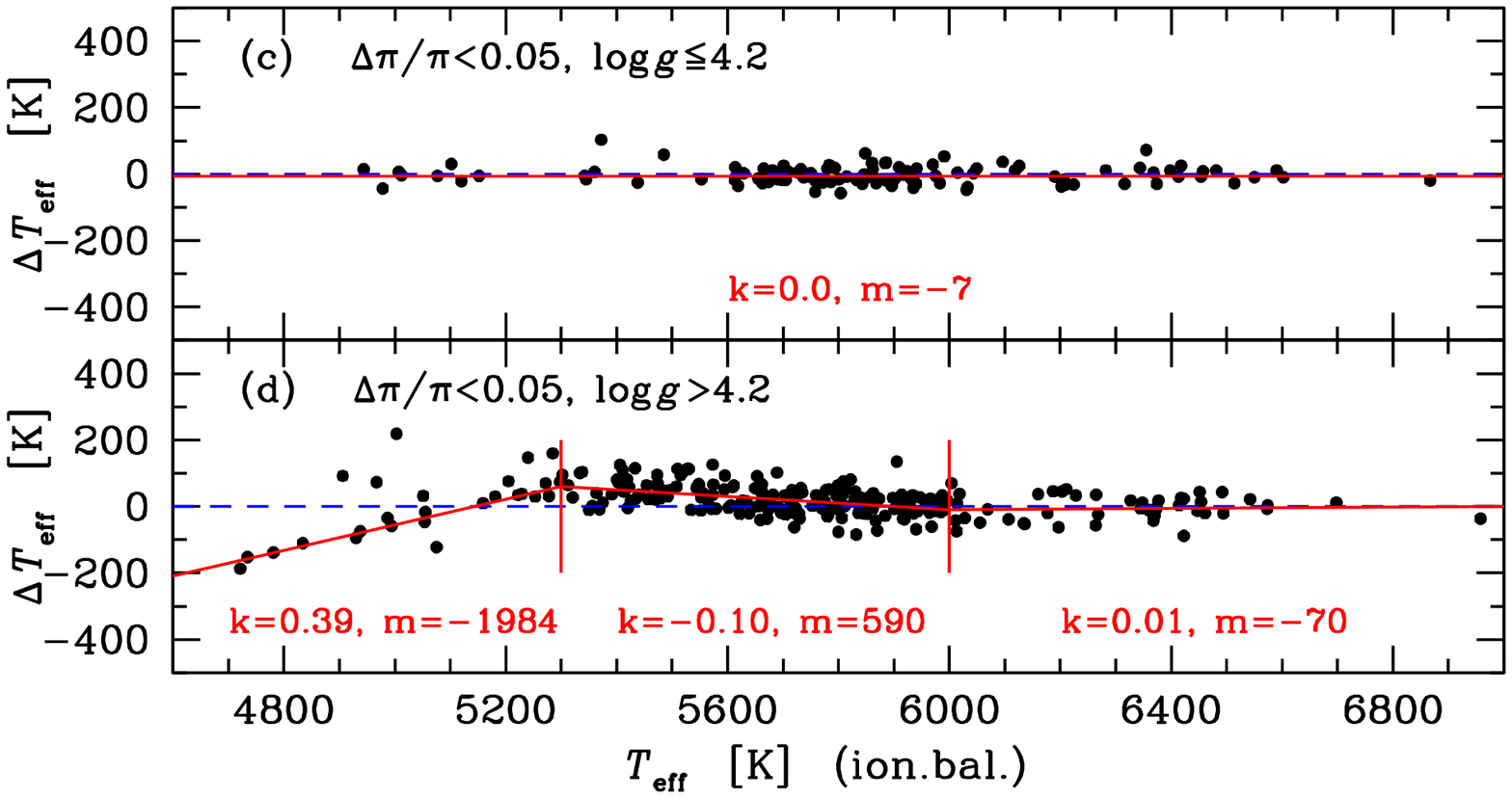}}
\resizebox{\hsize}{!}{
\includegraphics[bb=18 400 570 710,clip]{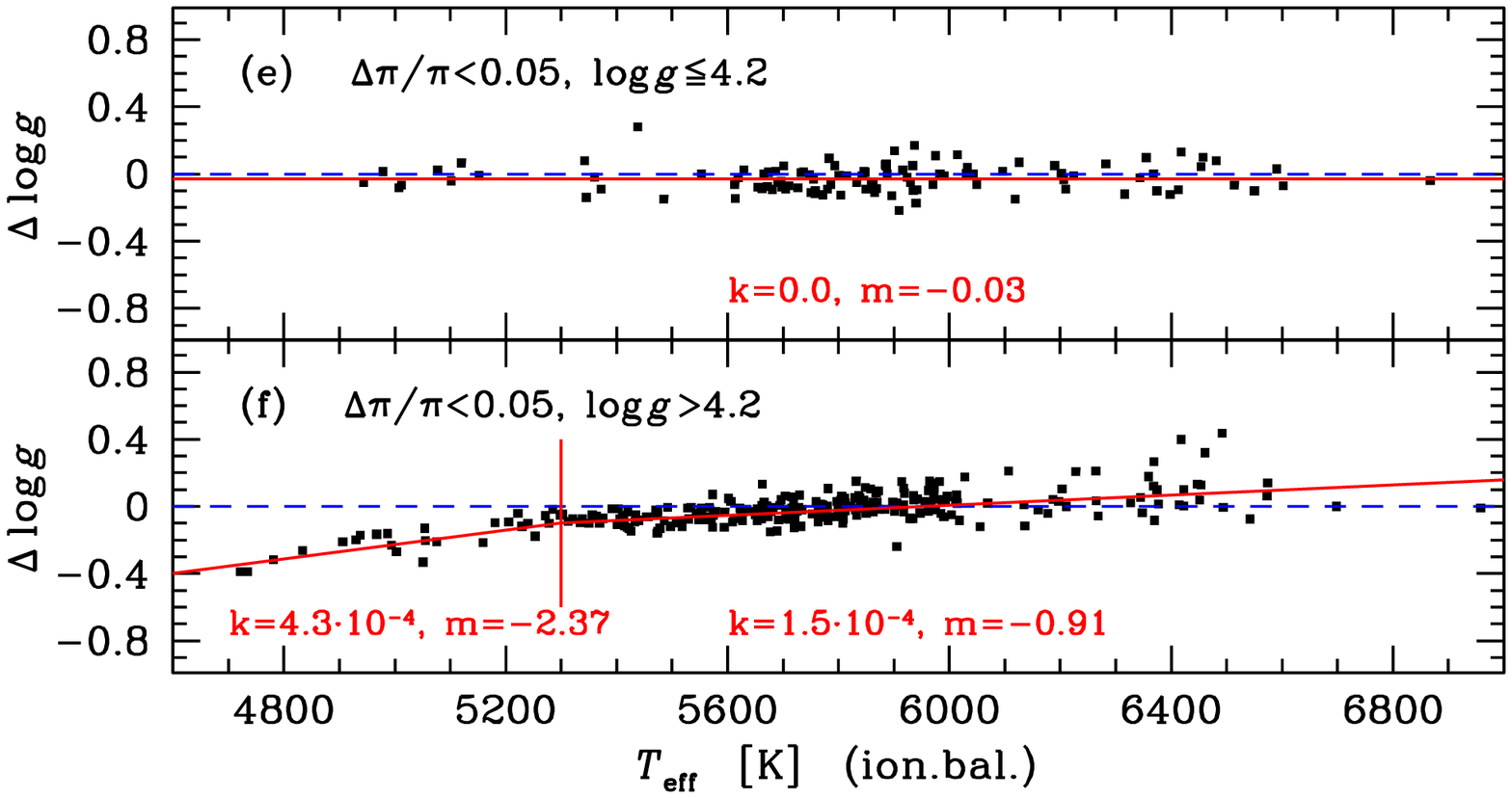}}
\caption{The difference in effective temperature (upper panels)
and surface gravity (lower panels) from the ionisation balance and
Hipparcos parallax methods for stars with a relative parallax uncertainty 
less than 5\,\%. 
\label{fig:decide2}
}
\end{figure}

\subsection{Investigating the flat lower main sequence}

To further investigate the difference in the two methods for determining
the surface gravity, the stars in Fig.~\ref{fig:hr}b have been 
encoded in red if the resulting 
\ion{Fe}{i} abundances are lower than the \ion{Fe}{ii} abundances, 
and blue if the opposite is true. The sizes of the circles are 
scaled with the magnitude of the difference between \ion{Fe}{i} 
and \ion{Fe}{ii} abundances. What we see is that on the lower 
main sequence essentially all stars appear to be red, i.e. the 
\ion{Fe}{i} abundances are lower than the \ion{Fe}{ii} abundances. 
In other parts of the HR diagram, there is a mixture of red and 
blue circles. This is further illustrated
in Fig.~\ref{fig:decide1}, where we plot the difference between 
\ion{Fe}{i} and \ion{Fe}{ii} abundances versus $\teff$ and $\log g$
for all stars that have relative parallax uncertainties less than 5\,\%.
The stars above the turn-off ($\log g<4.2$) show perfectly flat
trends with both $\teff$ and $\log g$, while many stars
below the turn-off with $\teff\lesssim 5600$\,K show large discrepancies
between \ion{Fe}{i} and \ion{Fe}{ii}. There is also a declining trend
in \ion{Fe}{i}--\ion{Fe}{ii} with $\log g$ that also
increases in dispersion with $\log g$. In summary, it appears
that essentially all 
stars with $\log g>4.2$ and $\teff<5650$ do not show ionisation 
equilibrium between \ion{Fe}{i} and \ion{Fe}{ii} when determining 
the surface gravity from Hipparcos parallaxes.

\begin{figure*}
\resizebox{\hsize}{!}{
\includegraphics[bb=18 144 570 718,clip]{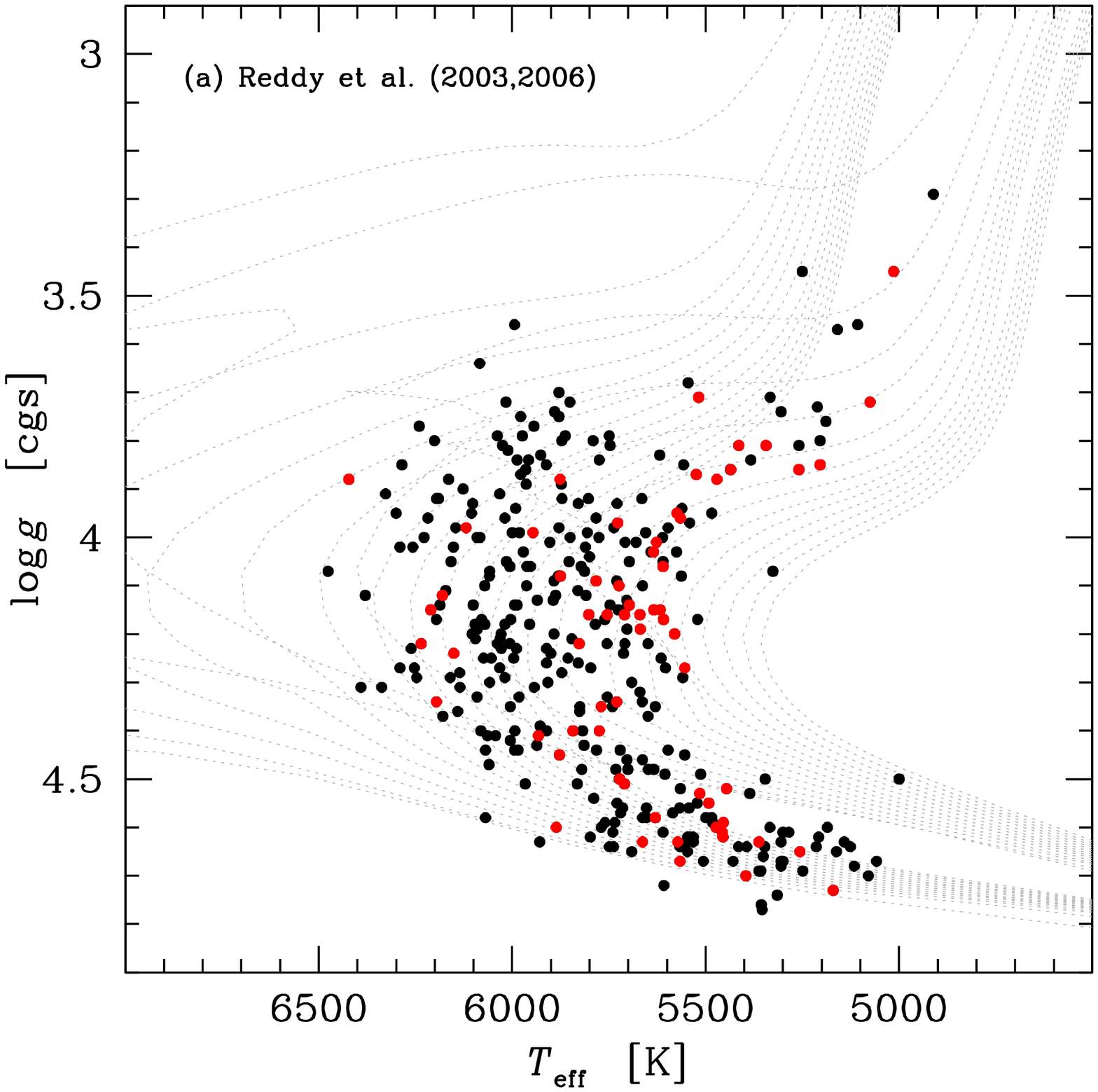}
\includegraphics[bb=75 144 570 718,clip]{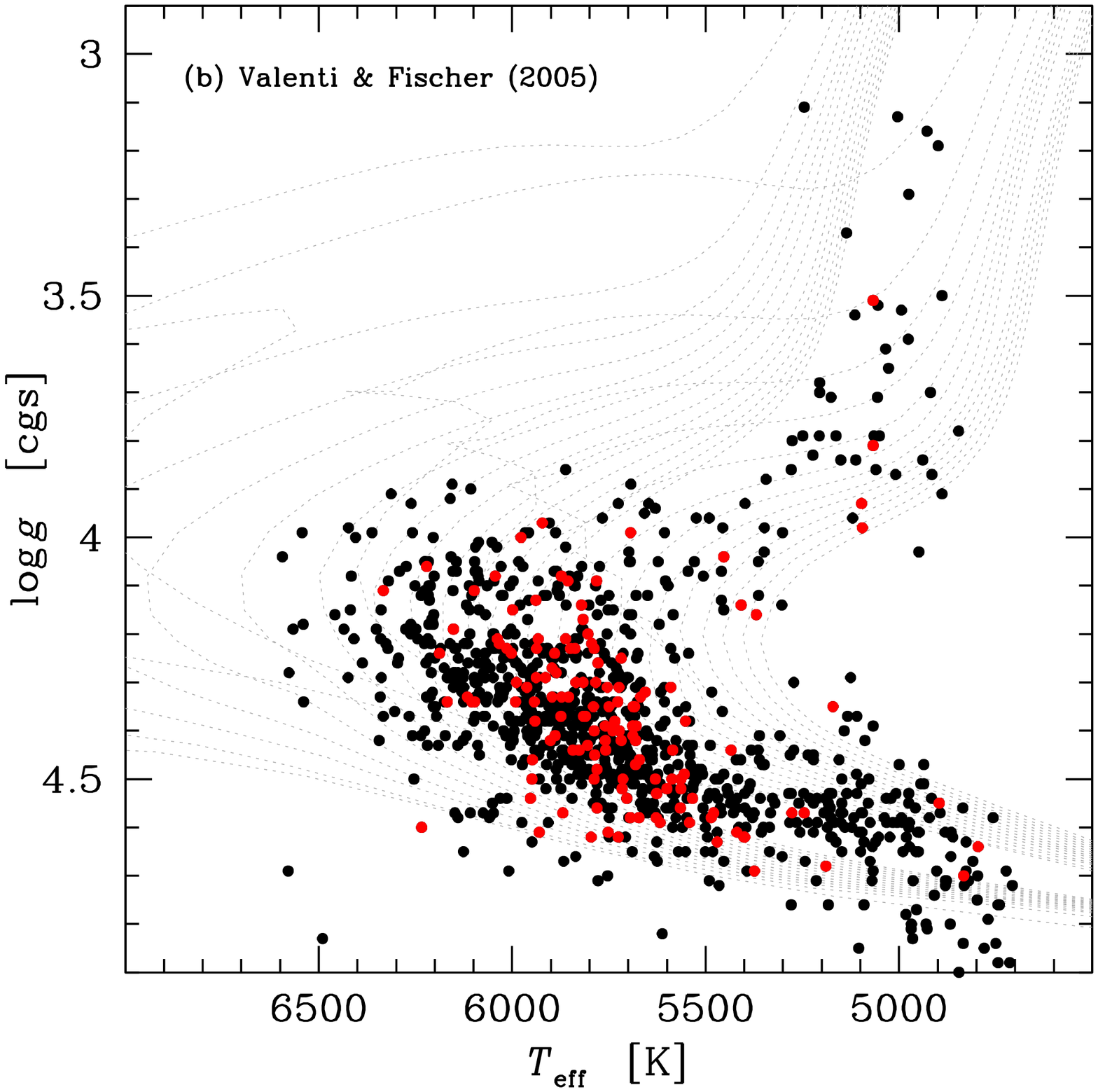}
\includegraphics[bb=75 144 592 718,clip]{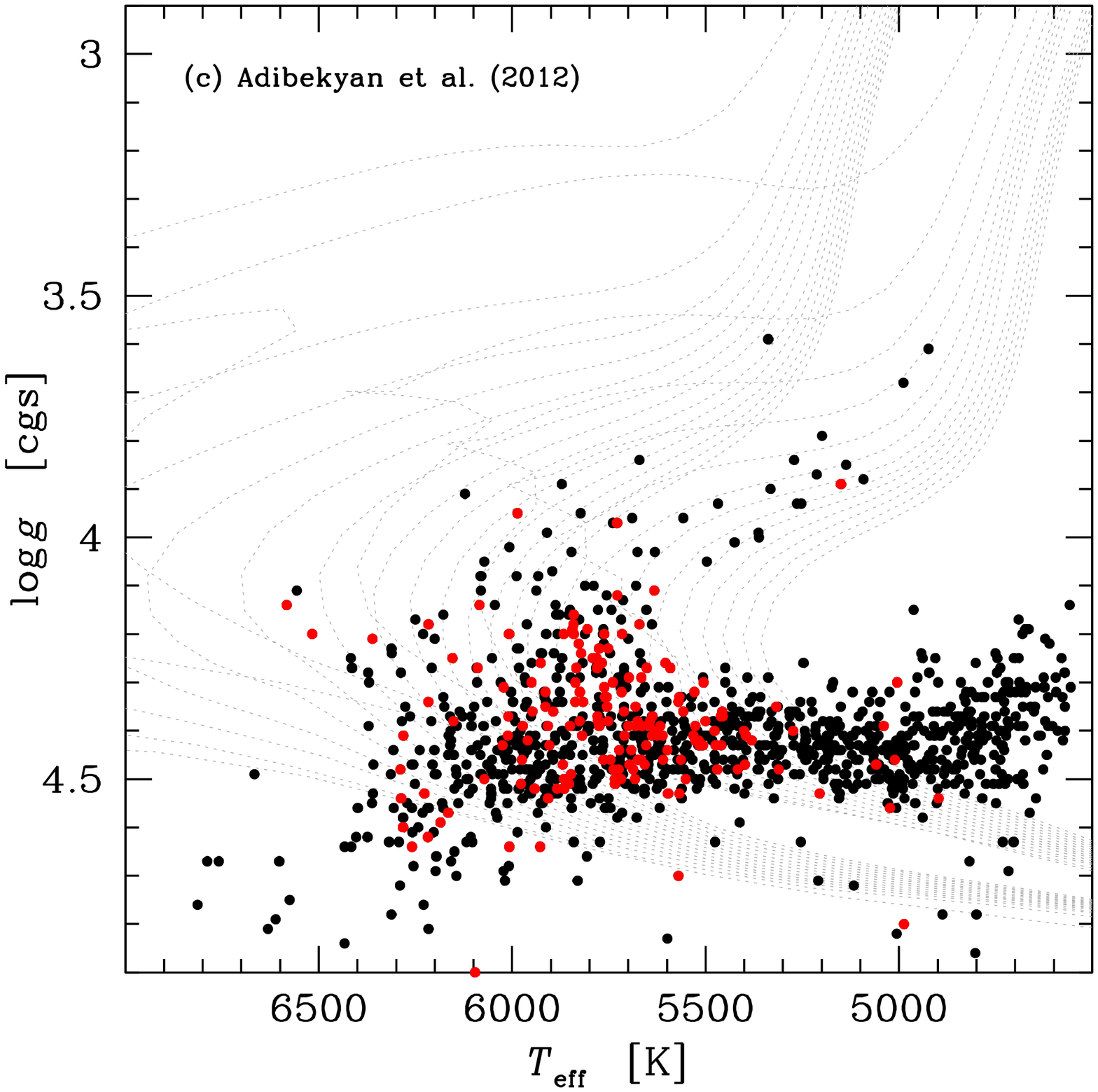}
}
\caption{HR diagrams for the \cite{reddy2003,reddy2006} sample, the 
\cite{valenti2005} sample, and the \cite{adibekyan2012} sample.
Overlapping stars from this study are marked by red solid circles.
The $\alpha$-enhanced Yonsei-Yale (Y2) isochrones by \cite{demarque2004} 
have metallicities of $\rm [Fe/H]=-1$ and $+0.3$\,dex, respectively, and are 
shown from 1 to 15 Gyr in steps of 1\,Gyr.
\label{fig:hr2}
}
\end{figure*}

The question is now which method to use to get a consistent analysis.
As a significant fraction of the stars in the sample have parallax
uncertainties larger than 5\,\% (see Fig.~\ref{fig:parallax}), the
best way would be to use ionisation balance. Ionisation balance also has
a large advantage over the parallax method as it is based on the 
stellar spectrum only, and can be utilised even when the distance
to the star is not known to high precision, such as in the case
of microlensed dwarf stars in the Galactic bulge. On the other hand, 
from the analysis of nearby stars with very good Hipparcos parallaxes, it is
evident that ionisation balance has its limitations, and mainly on
the lower main sequence for stars with $\log g \gtrsim 4.2$ and 
$\teff \lesssim 5600$\,K. Figures~\ref{fig:decide2}a and b shows the differences
between the effective temperatures and surface gravities that the 
two methods generate as a function of the surface gravity. For $\log g\lesssim4.2$
there appears to be a slight constant offset, while for higher $\log g$ there might
be a rising trend, although it is difficult to say as the dispersion also increases.
The differences as a function of effective temperature shown in 
Figs.~\ref{fig:decide2}c-f have a clearer appearance. The sample is split
at $\log g=4.2$, which is roughly in the turn-off region. For stars above the
turn-off ($\log g \lesssim 4.2$) there appears to be a constant small offset in both 
$\teff$ and $\log g$; $-7$\,K and $-0.03$\,dex, respectively. Below the turn-off
the situation appears more complicated and we make linear regressions to the regions
defined in Figs.~\ref{fig:decide2}d and f. The corrections $\Delta T$ and $\Delta G$
to be applied (subtracted) to the ionisation balance parameters are given in 
Table~\ref{tab:corrections}.

\begin{table}
\centering
\setlength{\tabcolsep}{1.8mm}
\caption{
\label{tab:corrections}
        Corrections to be applied to the parameters from
        ionisation balance. $\teff (corr) = \teff - \Delta T$ and $\log g (corr)=\log g - \Delta G$ where $\Delta T=k_T * \teff + m_T$ and $\Delta G = k_G *\teff + m_G$ with the parameters given in the table below. Relationships are illustrated in Figs.~\ref{fig:decide2}c-f.
        }
\small
\begin{tabular}{cccccc}
\hline \hline\noalign{\smallskip}
    $\teff$ interval    & 
    $\log g$ interval   &
    $k_T$                   &
    $m_T$   &
    $k_G$   &
    $m_G$   \\
\noalign{\smallskip}
\hline\noalign{\smallskip}
$4600 - 7000$ & $3.0 - 4.2$ &    0.00   & $-$7     &  0.00  &   -0.03  \\
$4600 - 5300$ & $4.2 - 4.8$ &    0.39   & $-$1984  &  $4.3\cdot10^{-4}$   &   $-$2.4  \\
$5300 - 6000$ & $4.2 - 4.8$ & $-$0.11   &   590    &  $1.5\cdot10^{-4}$   &   $-$0.91  \\
$6000 - 7000$ & $4.2 - 4.8$ &    0.025  & $-$70    &  $1.5\cdot10^{-4}$   &   $-$0.91  \\
\noalign{\smallskip}
\hline
\end{tabular}
\flushleft
\end{table}

The HR diagram based on the corrected ionisation balance parameters 
is shown in Fig.~\ref{fig:hr}c.
The gap that can be seen at $\log g \approx 4.2$ is an artefact due to that the
corrections are different for stars below and above the turn-off.

After having identified these ionisation balance issues
on the lower main sequence for our sample, it is interesting to 
see whether flat main sequences
are present in other similar high-resolution spectroscopic studies
of the Galactic disk. For that, we choose three studies: first, the sample
of 355 dwarf stars from \cite{reddy2003,reddy2006} where stellar parameters
are determined from the infrared IRFM flux method and Hipparcos parallaxes;
second, the sample of 1040 dwarf stars from \cite{valenti2005}
where stellar parameters are determined through $\chi^2$-minimisation 
between observed spectrum and synthesised spectrum in selected wavelength bands
using the SME software; and third,
the sample of 1111 dwarf stars from \cite{adibekyan2012} who, like us,
use ionisation and excitation balance to determine stellar parameters.
The HR diagrams for these studies are shown in Fig.~\ref{fig:hr2}. 
For the \cite{reddy2003,reddy2006} and \cite{valenti2005} studies,
which do not utilise ionisation balance, the HR diagrams appear normal,
with declining main sequences. The HR diagram for the
\cite{adibekyan2012} sample, on the other hand, shows an extremely
flat relation, where $\log g$ is even slightly rising with decreasing temperature.

What the causes are for the flat main sequence is not all clear.
It is possible that they arise due to limitations of the models
that cannot properly handle excitation balance and/or ionisation balance.
Or it could be that  NLTE effects and/or 3D effects play roles, or a 
combination of all of these. It is
beyond the scope of the current paper to further investigate this,
and we will for now settle with the empirical corrections in 
Table~\ref{tab:corrections}.
We will report stellar parameters for all three varieties
(ionisation balance, parallaxes, corrected ionisation balance), but
elemental abundances and stellar ages will only be reported for the 
corrected ionisation balance values, which is also what will be used 
in the remainder of the paper. All parameters are reported in Table~\ref{tab:parameters}.

\subsection{Systematic errors}

As the analysis is strictly differential relative to the Sun, systematic errors 
should largely cancel out and the internal precision should be good.
This is seen through the good agreement between equivalent width
measurements and stellar parameters that we derive for the Sun based 
on the spectra from the different spectrographs and observing runs.
Systematic shifts relative to other studies are more difficult,
as methods, model atmospheres, atomic data, and methods for normalisation 
to the Sun, might differ. To check and compare our 
results we have made a detailed comparison of our stellar 
parameters and elemental abundances to three recent and large studies
of the Galactic stellar disk. First we have chosen the studies
by \cite{reddy2003,reddy2006}, consisting of stars observed from the
Northern hemisphere at the MacDonald Observatory. In total this sample
consists of 355 kinematically selected F and G dwarf stars that nicely
would complement our sample, which mainly has been observed from
the Southern hemisphere. With \cite{reddy2003,reddy2006} we have 64
stars in common. Next, we have chosen the study by \cite{adibekyan2012}
who have done a detailed abundance analysis of 1111 stars observed
with the HARPS spectrograph on the ESO\,3.6-m telescope on La Silla. 
With \cite{adibekyan2012} we have 168 
stars in common. And finally, we have chosen the \cite{valenti2005}
study of 1040 F, G, and K dwarfs from the Keck, Lick, and AAT planet 
search programs, with which we have 140 stars in common.
The stars in common with each of these studies are marked in red 
in the HR diagrams in Fig.~\ref{fig:hr2}.

\begin{figure}
\resizebox{\hsize}{!}{
\includegraphics[bb=18 142 592 718,clip]{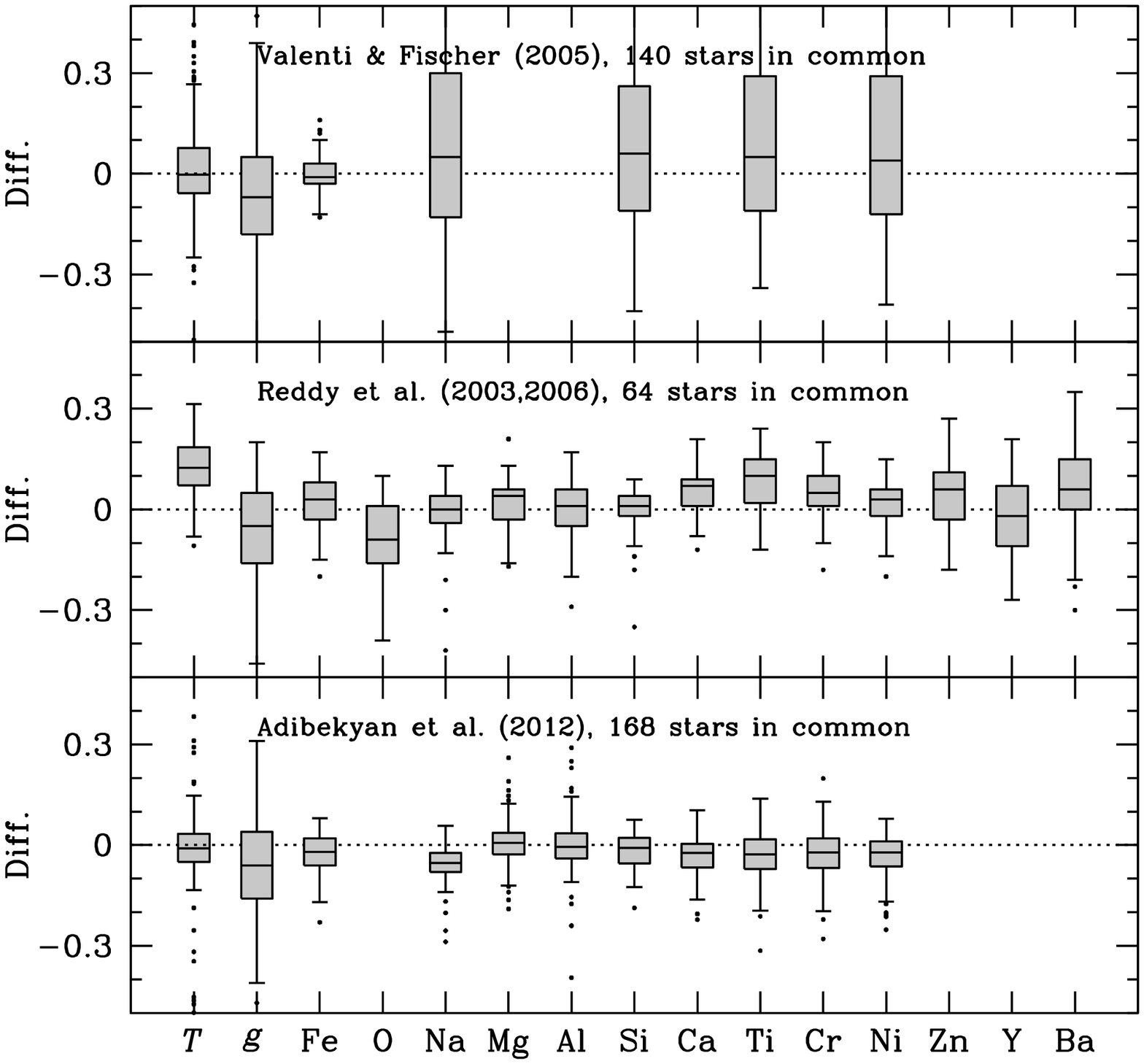}}
\caption{Comparison of abundances ($[X/{\rm H}]$) for stars in common
between this study and those of 
\cite{valenti2005}, \cite{reddy2003,reddy2006}, and \cite{adibekyan2012}.
The elements ($X$) are indicated
on the abscissa. The two left-most boxes in each panel show the $\teff$ (denoted
by $T$) and $\log g$ (denoted by $g$) comparisons. Please note  that 
the scale on the ordinate for the temperature should be multiplied by a 
factor 100. The differences are given  as our values minus their values, 
and the differences are also listed in Table~\ref{tab:systematics}.
In the boxplots the central horizontal line represents the median value. 
The lower and upper quartiles are represented by the outer edges of 
the boxes, i.e., the box encloses 50\,\% of the sample. 
The whiskers extend to the farthest data point that lies within 
1.5 times the inter-quartile distance. Those stars that do not 
fall within the reach of the whiskers are regarded as outliers 
and are marked by dots. 
\label{fig:systematics}
        }
\end{figure}
\begin{table}
\centering
\setlength{\tabcolsep}{4mm}
 \caption{Comparisons of stars in common with 
 \cite{reddy2003,reddy2006}, \cite{adibekyan2012}, and \cite{valenti2005}.
 The differences are given as values from this work minus the other studies.
 The values given are the median value as well as the 1-$\sigma$ dispersion
 around the median.
   \label{tab:systematics}
        }
\tiny
\begin{tabular}{lccc}
\hline\hline
\noalign{\smallskip}
            & R03/06 & A12 & VF05\\
\noalign{\smallskip}
\hline
\noalign{\smallskip}
 \# of stars         & 355                 &       1111            & 1040  \\
 overlap             &  64                 &       168             & 140   \\
 $\Delta\teff$       &  $  +124\pm57   $   &      $-10\pm42$        & $-2\pm 67$ \\
 $\Delta\log g$      &  $-0.05\pm0.10$   &      $-0.06\pm0.10$ & $-0.07\pm0.12$ \\
 $\rm \Delta [Fe/H]$ &  $+0.03\pm0.05$   &      $-0.02\pm0.04$ & $-0.01\pm0.03$ \\
 $\rm \Delta [O/H]$  &  $-0.09\pm0.08$   &                       & \\
 $\rm \Delta [Na/H]$ &  \phantom{0}$0.00\pm0.04$   &      $-0.05\pm0.03$ & $+0.05\pm0.21$ \\
 $\rm \Delta [Mg/H]$ &  $+0.04\pm0.04$   &      $+0.01\pm0.03$ & \\
 $\rm \Delta [Al/H]$ &  $+0.01\pm0.05$   &      $-0.01\pm0.04$ & \\
 $\rm \Delta [Si/H]$ &  $+0.01\pm0.03$   &      $-0.01\pm0.04$ & $+0.06\pm0.18$    \\
 $\rm \Delta [Ca/H]$ &  $+0.07\pm0.04$   &      $-0.02\pm0.03$ &     \\
 $\rm \Delta [Ti/H]$ &  $+0.10\pm0.06$   &      $-0.03\pm0.04$ & $+0.05\pm0.20$\\
 $\rm \Delta [Cr/H]$ &  $+0.05\pm0.04$   &      $-0.02\pm0.04$ & \\
 $\rm \Delta [Ni/H]$ &  $+0.03\pm0.04$   &      $-0.02\pm0.04$ & $+0.04\pm0.20$\\
 $\rm \Delta [Zn/H]$ &  $+0.06\pm0.07$   &                       & \\
 $\rm \Delta [Y/H]$  &  $-0.02\pm0.09$   &                       & \\
 $\rm \Delta [Ba/H]$ &  $+0.06\pm0.07$   &                       & \\
\noalign{\smallskip}
\hline
\end{tabular}
\end{table}

Figure~\ref{fig:systematics} and Table~\ref{tab:systematics} show the 
comparisons to the \cite{reddy2003,reddy2006}, \cite{adibekyan2012},
and \cite{valenti2005} studies. The comparisons are very favourable 
and we see that our results compare reasonably well. 
With a few exceptions, the median difference in the 
abundance ratios are well below 0.1\,dex. The main difference
lies in the comparison of
the Na, Si, Ti, and Ni abundances from \cite{valenti2005} where the
dispersion is much larger than in the comparisons to 
\cite{reddy2003,reddy2006} and \cite{adibekyan2012}.
Note that most stars in common with \cite{adibekyan2012}
are located in the turn-off region and not on the lower main sequence
(see Fig.~\ref{fig:hr2}), so systematics due to the flat main sequence issue 
should not be significant.

\subsection{Random errors}
\label{sec:uncertainties}

An error analysis, as outlined in 
\cite{epstein2010}, has been performed for all stars.
The method accounts for abundance spreads (line-to-line scatter) 
as well as how the abundances for each element react to changes in 
the stellar parameters. The details of the method are given in
Appendix~\ref{sec:uncertainties2}.

Figure~\ref{fig:uncert} shows the uncertainties for the stellar
parameters and the abundance ratios as a function of
temperature, gravity, and metallicity. The uncertainties are reasonably
small and it is only for low effective temperatures
(below about 5400\,K), higher gravities, and at the highest [Fe/H] where
they start to become substantial. Interestingly, contrary to the
other $\alpha$-elements, the uncertainty in
[Ti/Fe] stays low and flat for essentially all parameters.
In the upcoming sections (from Sect.~\ref{sec:criteria} and onwards)
where we investigate different properties
of the Galactic disk we will therefore mainly utilise
the Ti results.

Uncertainties in the stellar parameters and in the abundance ratios
([$X$/Fe] and [$X$/Ti]) are given in Table~\ref{tab:parameters}.

\begin{figure*}
\resizebox{\hsize}{!}{
\includegraphics[bb=18 440 405 730,clip]{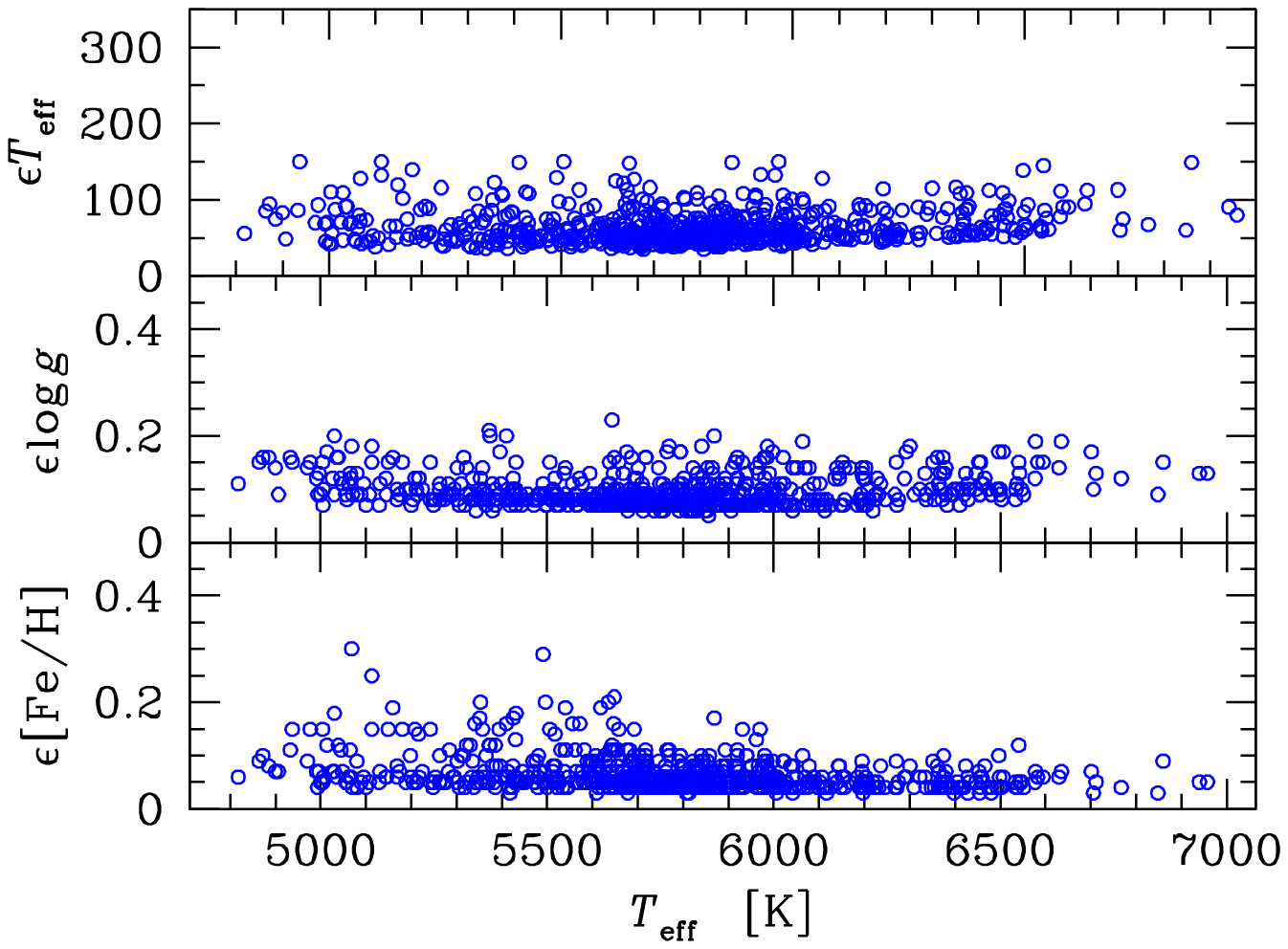}
\includegraphics[bb=80 440 405 730,clip]{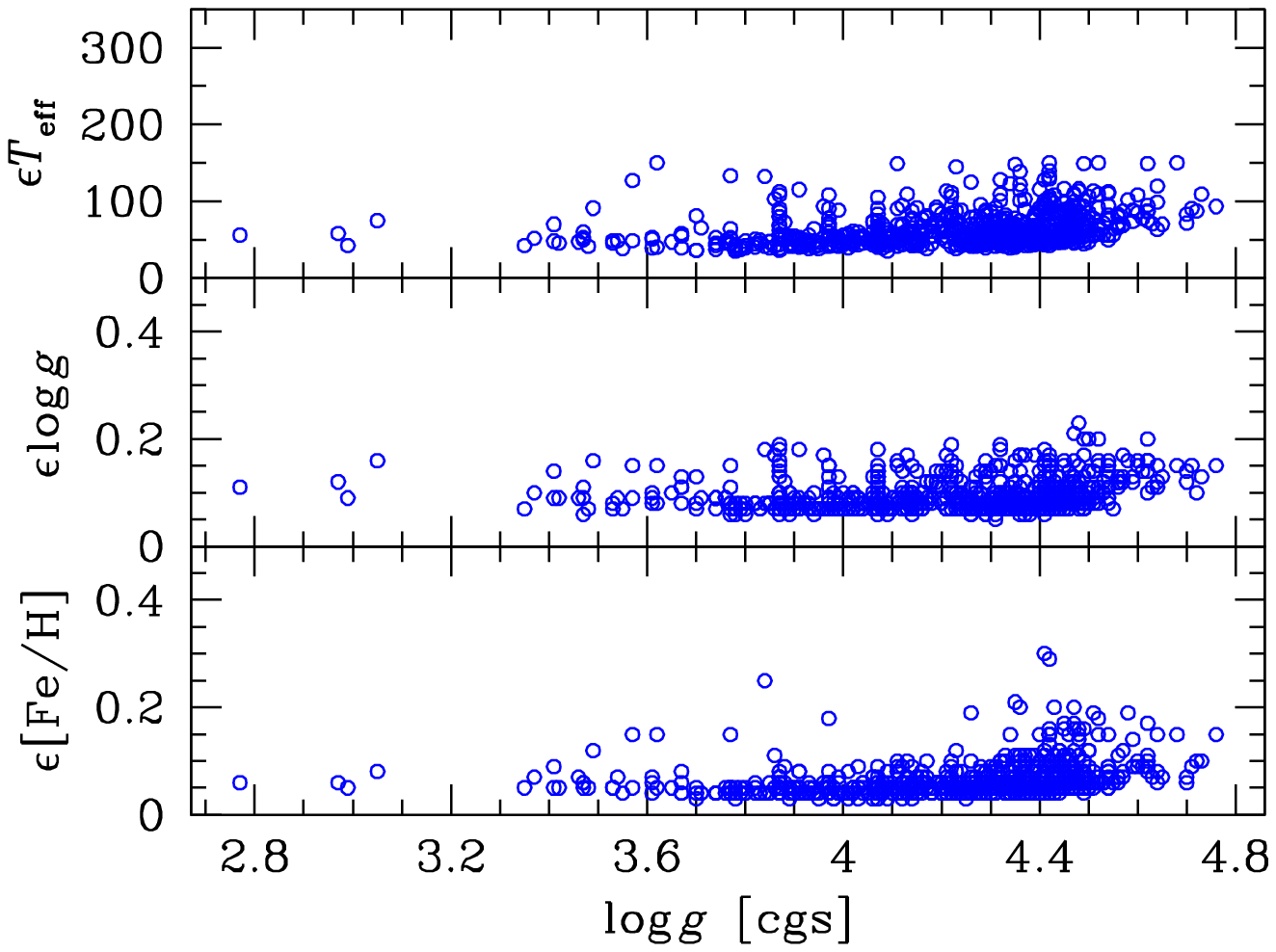}
\includegraphics[bb=80 440 425 730,clip]{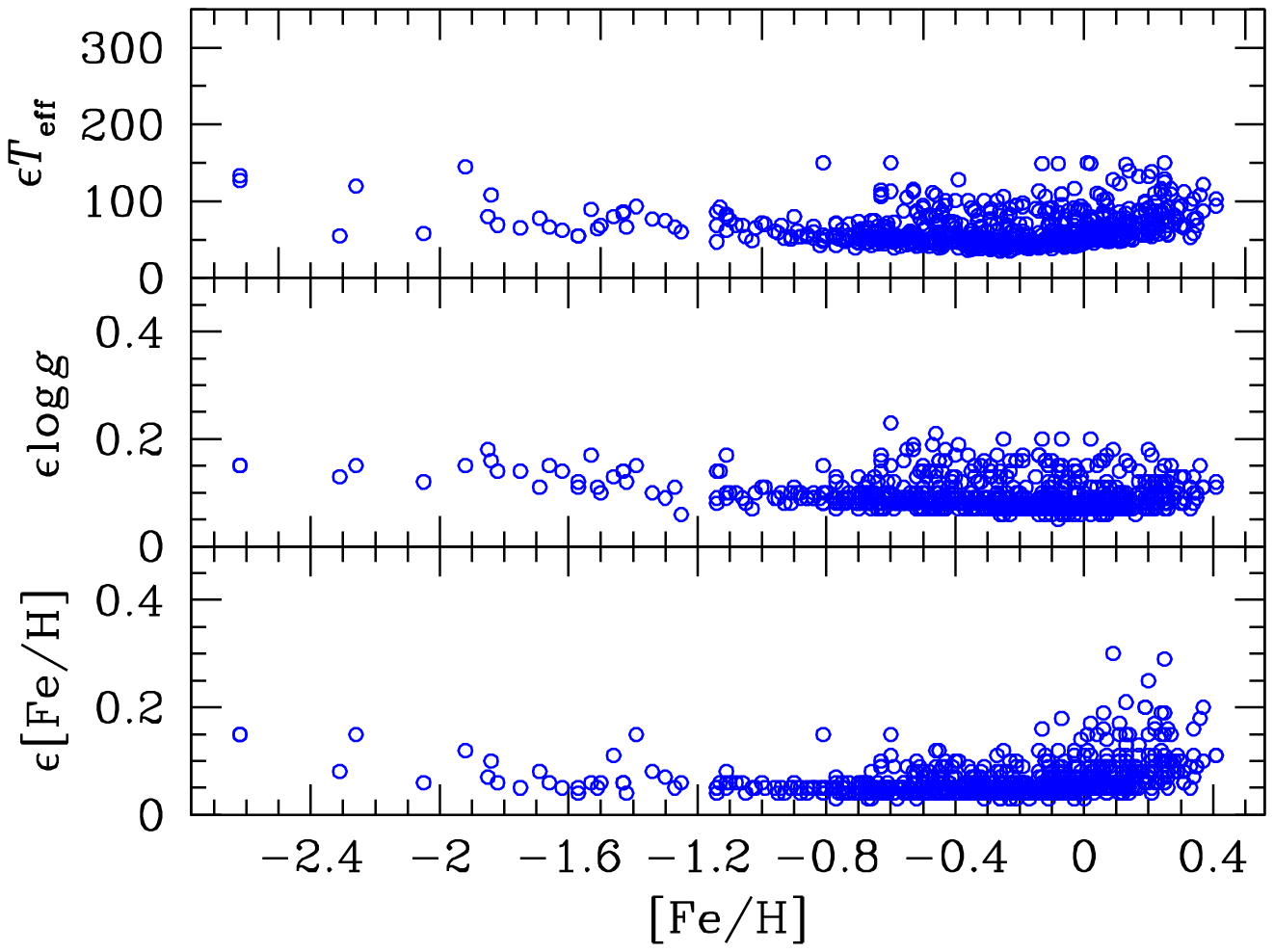}}
\resizebox{\hsize}{!}{
\includegraphics[bb=18 200 405 690,clip]{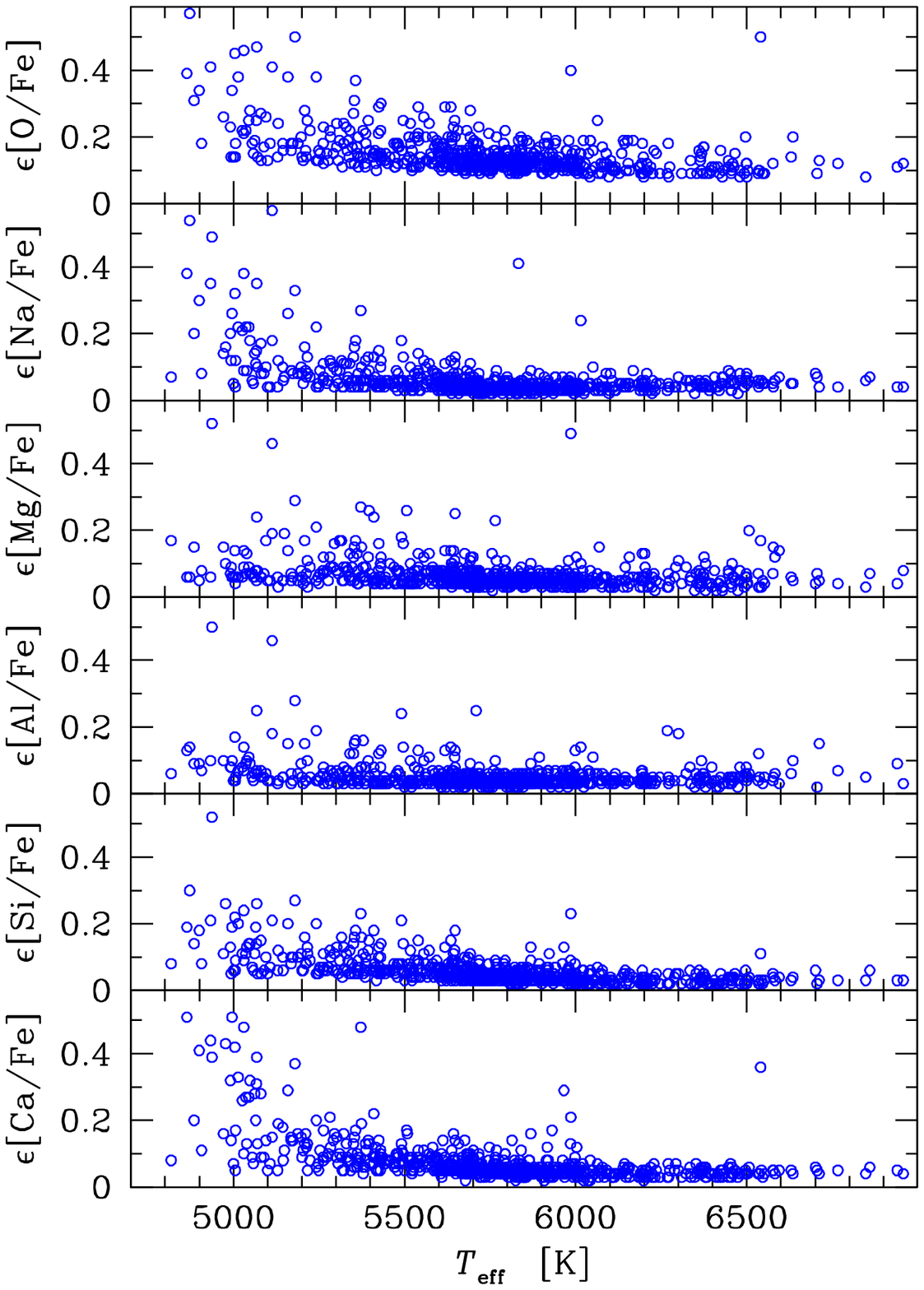}
\includegraphics[bb=80 200 405 690,clip]{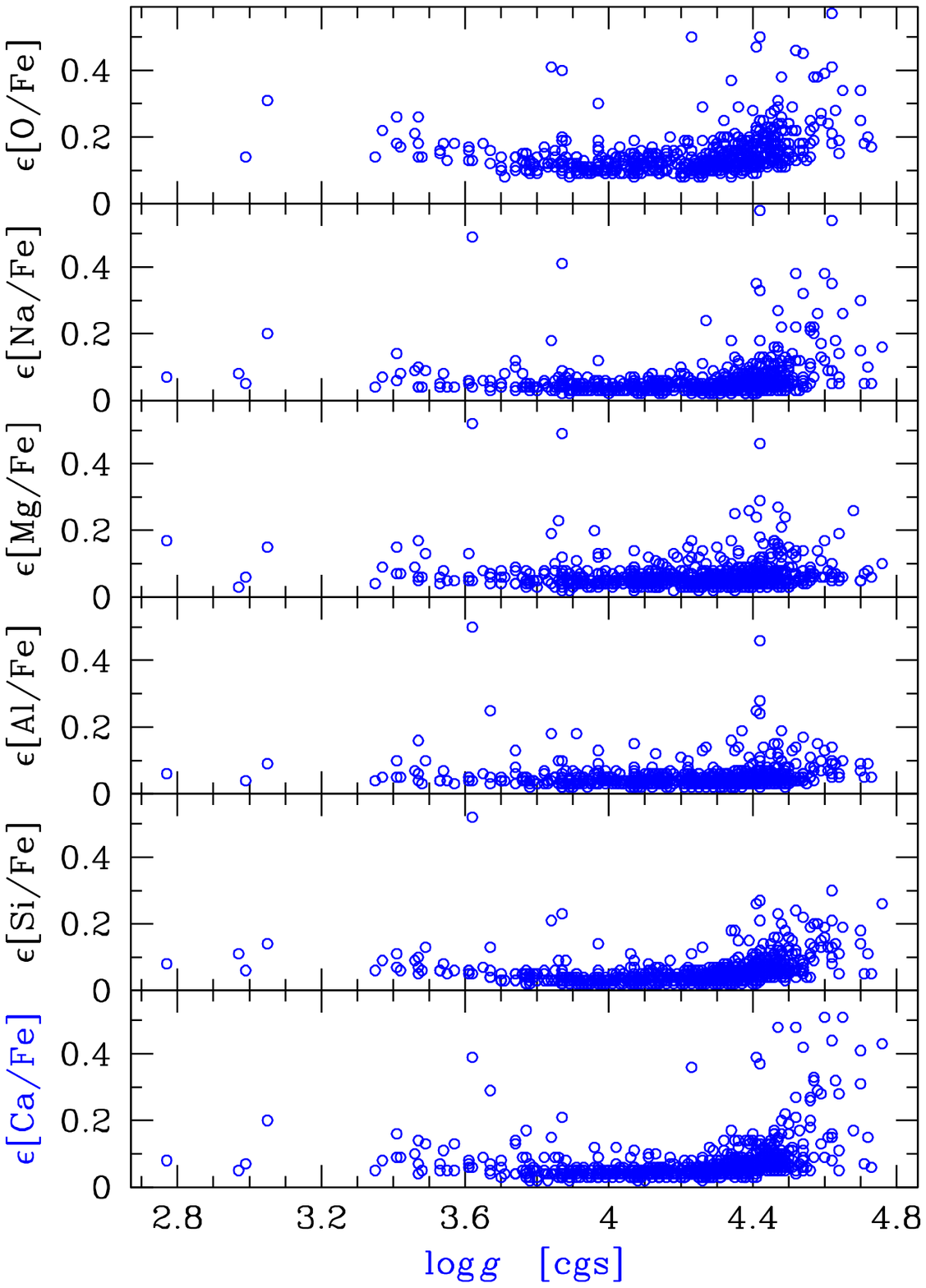}
\includegraphics[bb=80 200 425 690,clip]{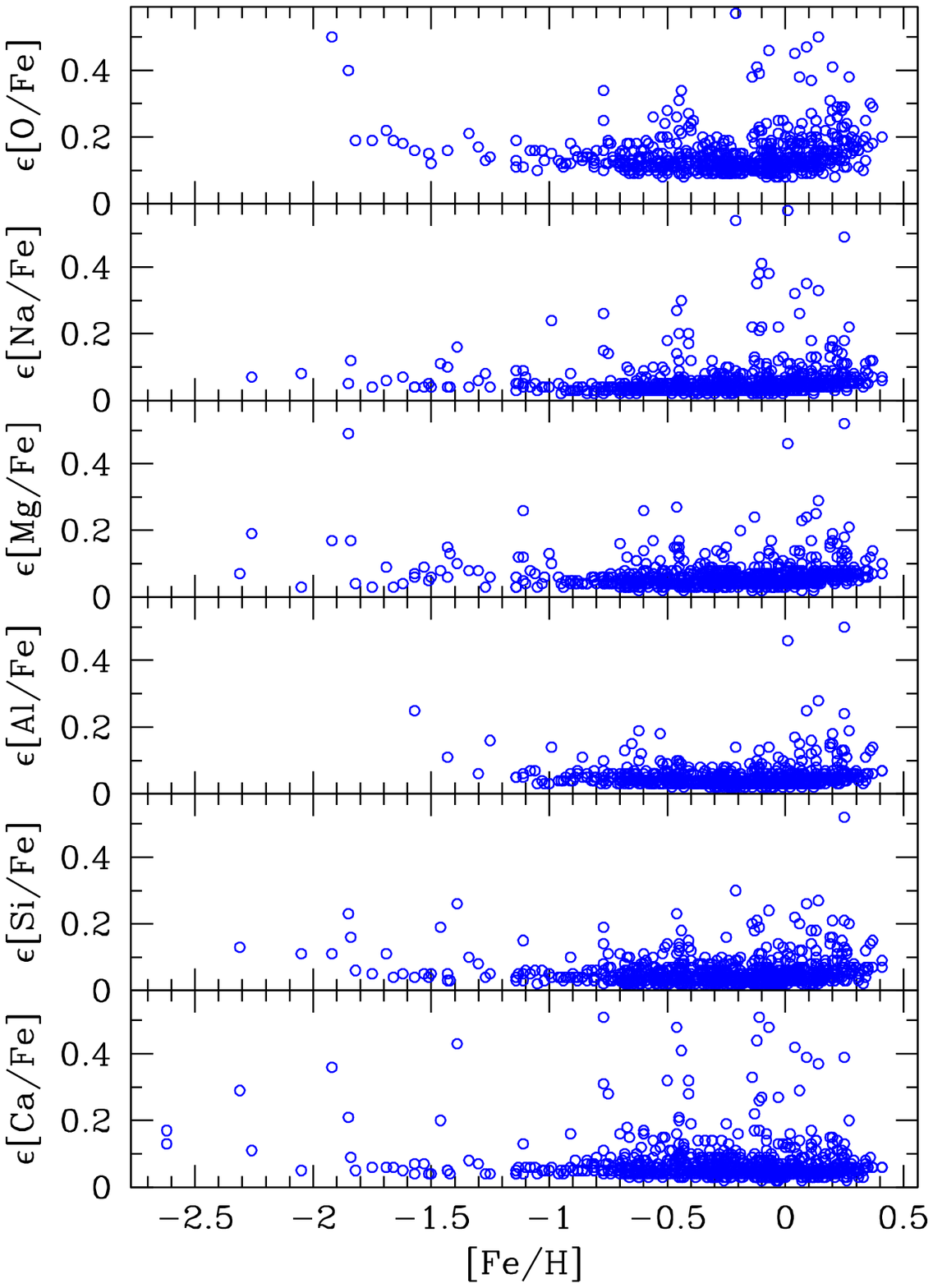}}
\resizebox{\hsize}{!}{
\includegraphics[bb=18 144 405 690,clip]{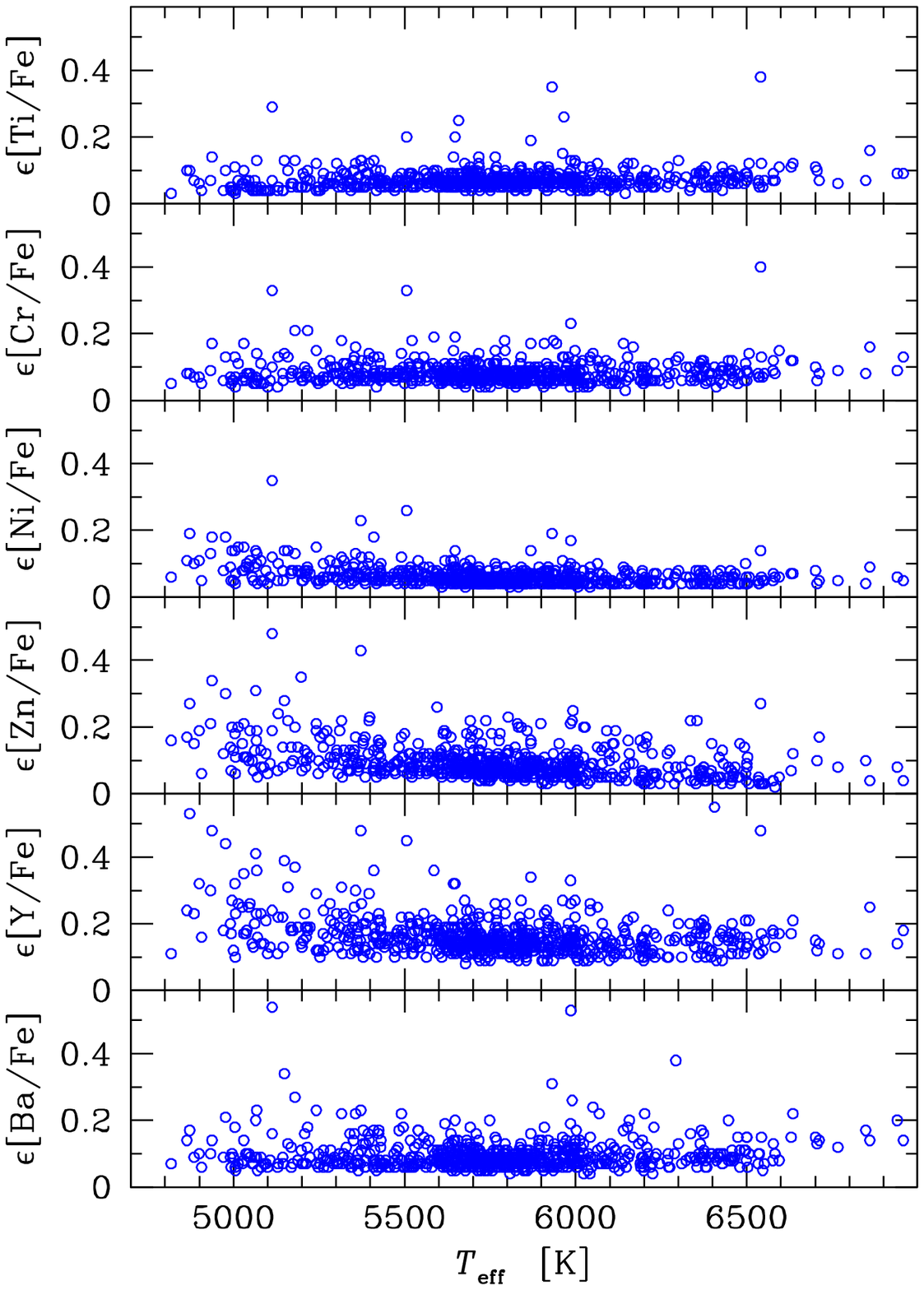}
\includegraphics[bb=80 144 405 690,clip]{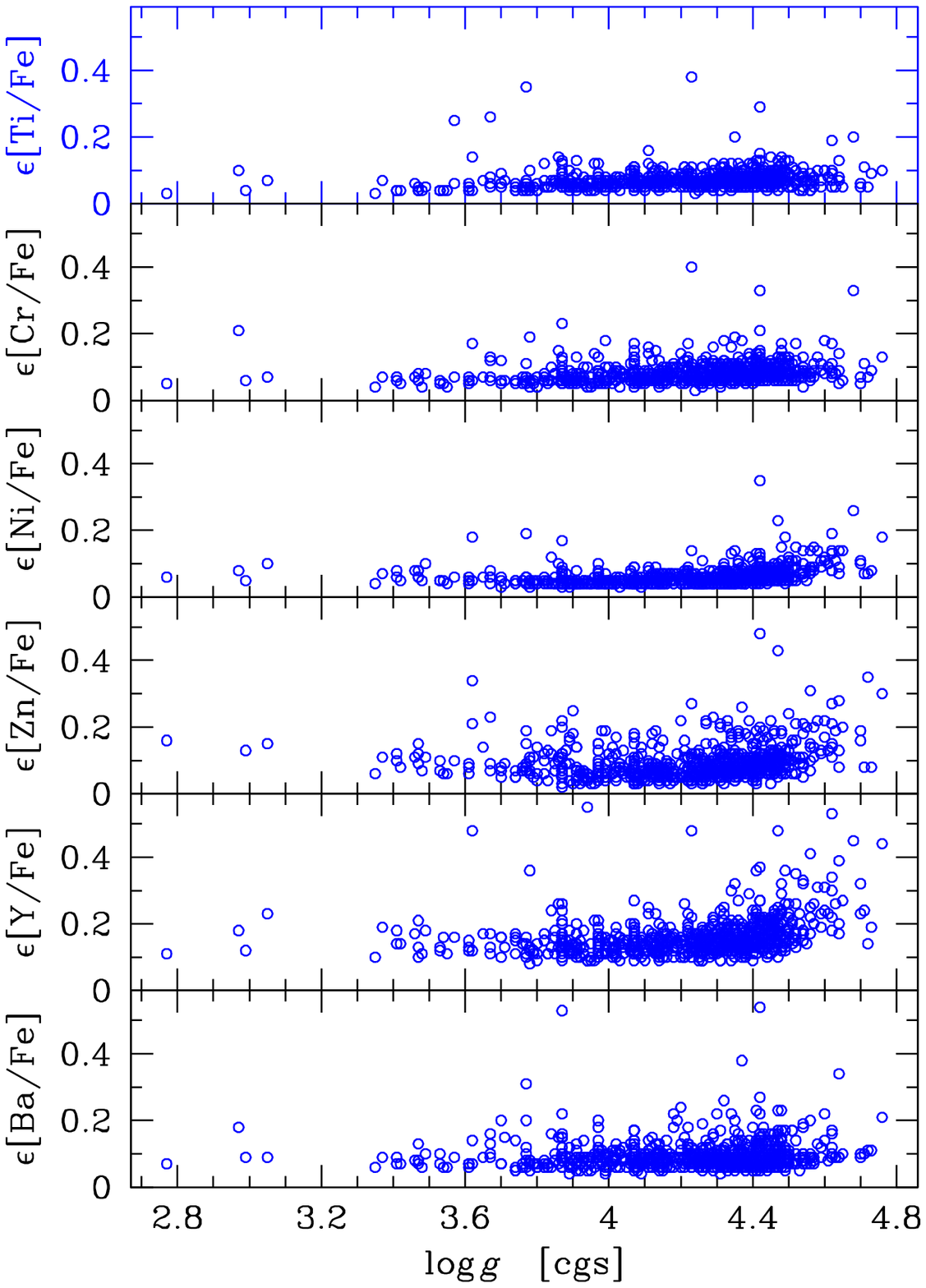}
\includegraphics[bb=80 144 425 690,clip]{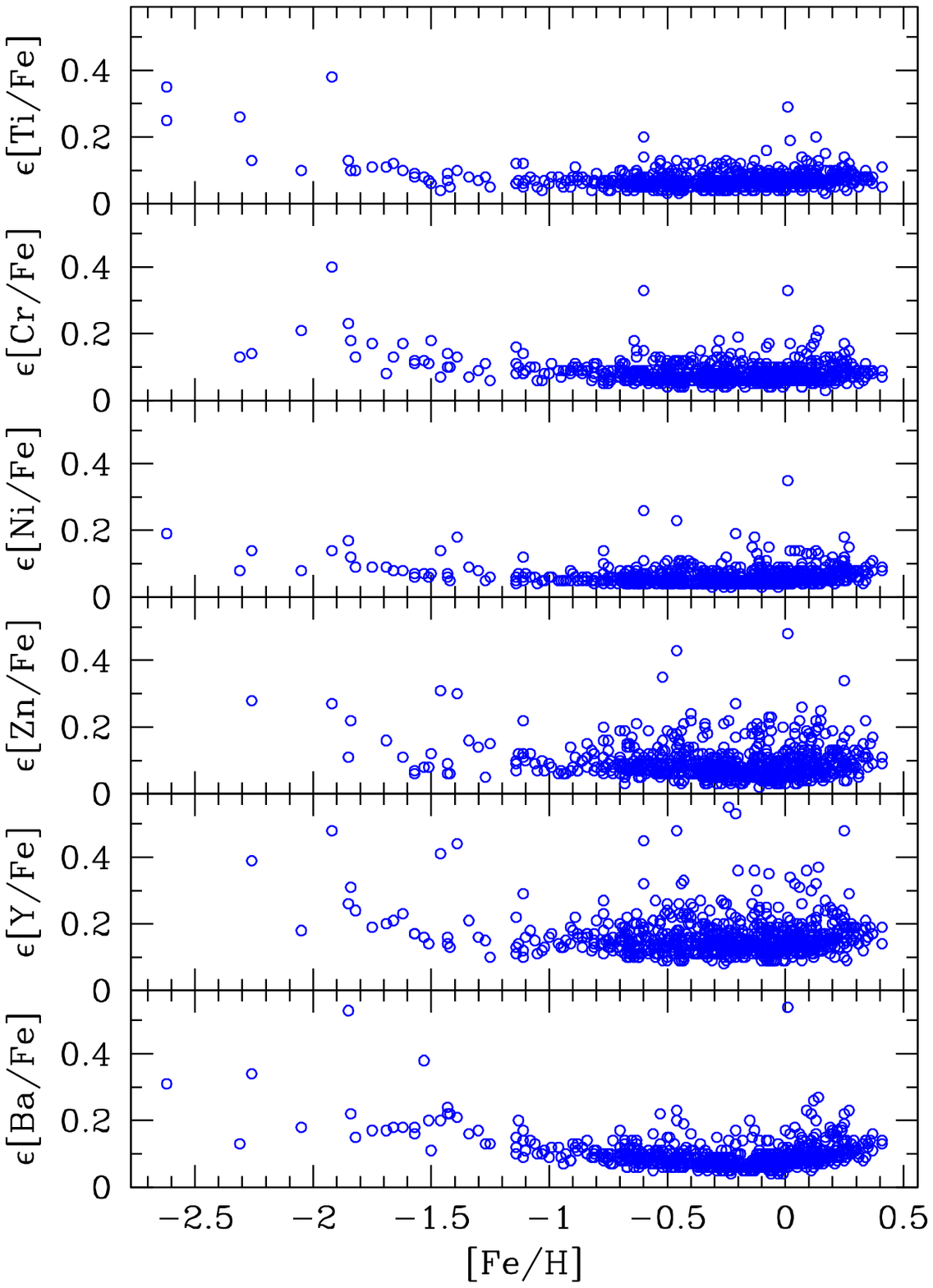}}
\caption{
Uncertainties in stellar parameters and abundance ratios
as a function of $\teff$, $\log g$, and [Fe/H].
\label{fig:uncert}
        }
\end{figure*}

\begin{figure*}
\resizebox{\hsize}{!}{
\includegraphics[bb=-90 200 700 385,clip]{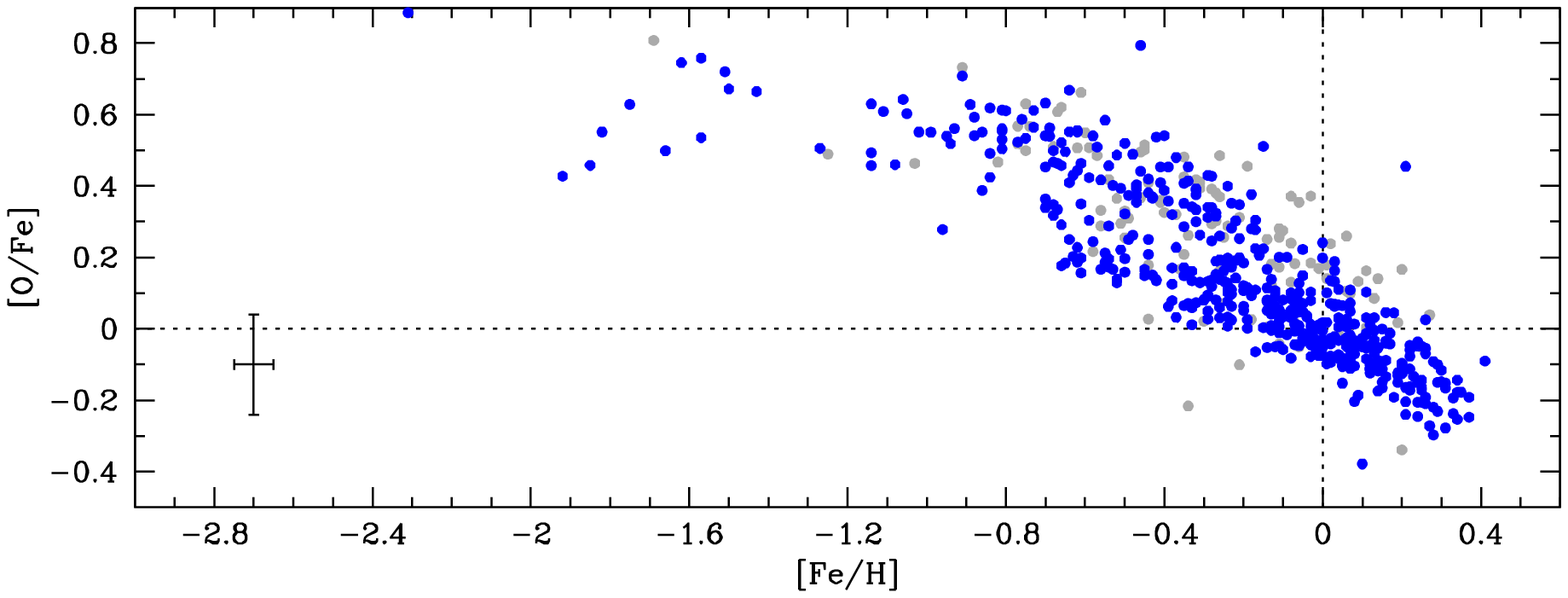}}
\resizebox{\hsize}{!}{
\includegraphics[bb=-90 200 700 350,clip]{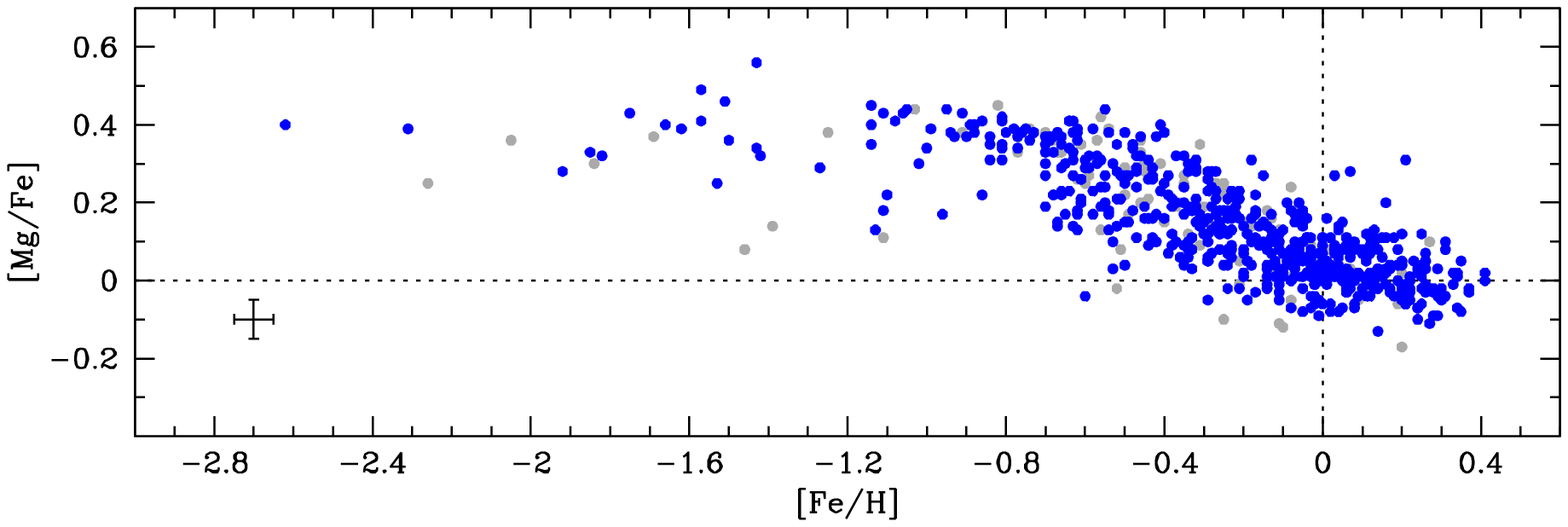}}
\resizebox{\hsize}{!}{
\includegraphics[bb=-90 200 700 350,clip]{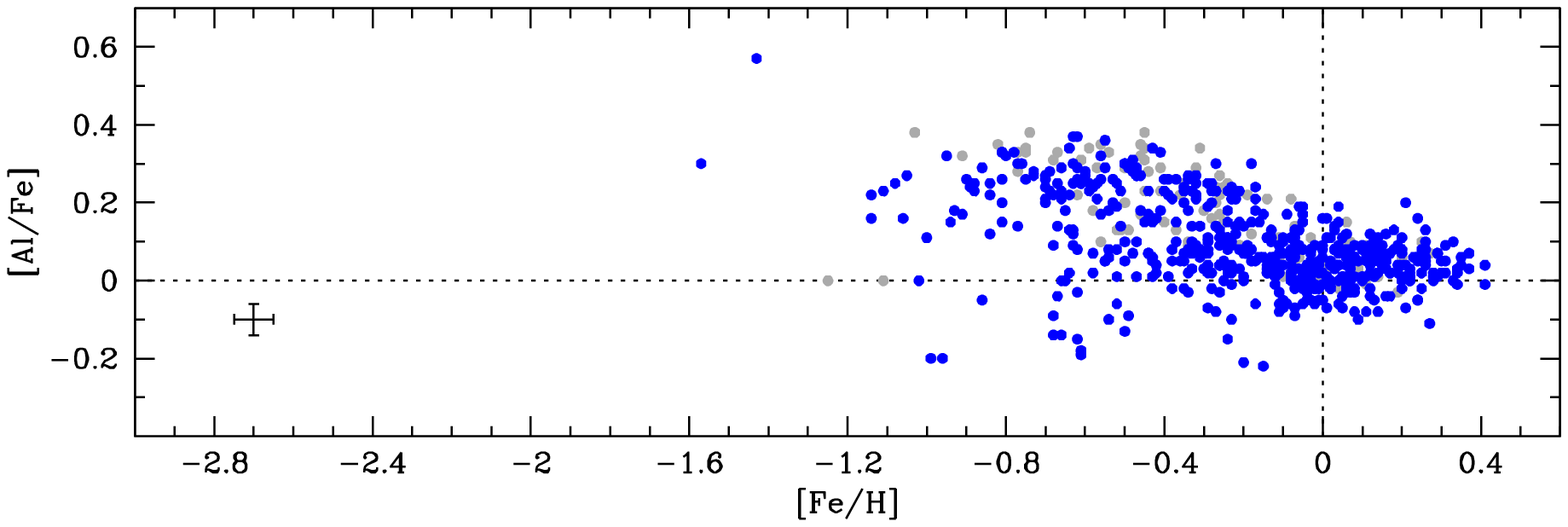}}
\resizebox{\hsize}{!}{
\includegraphics[bb=-90 200 700 350,clip]{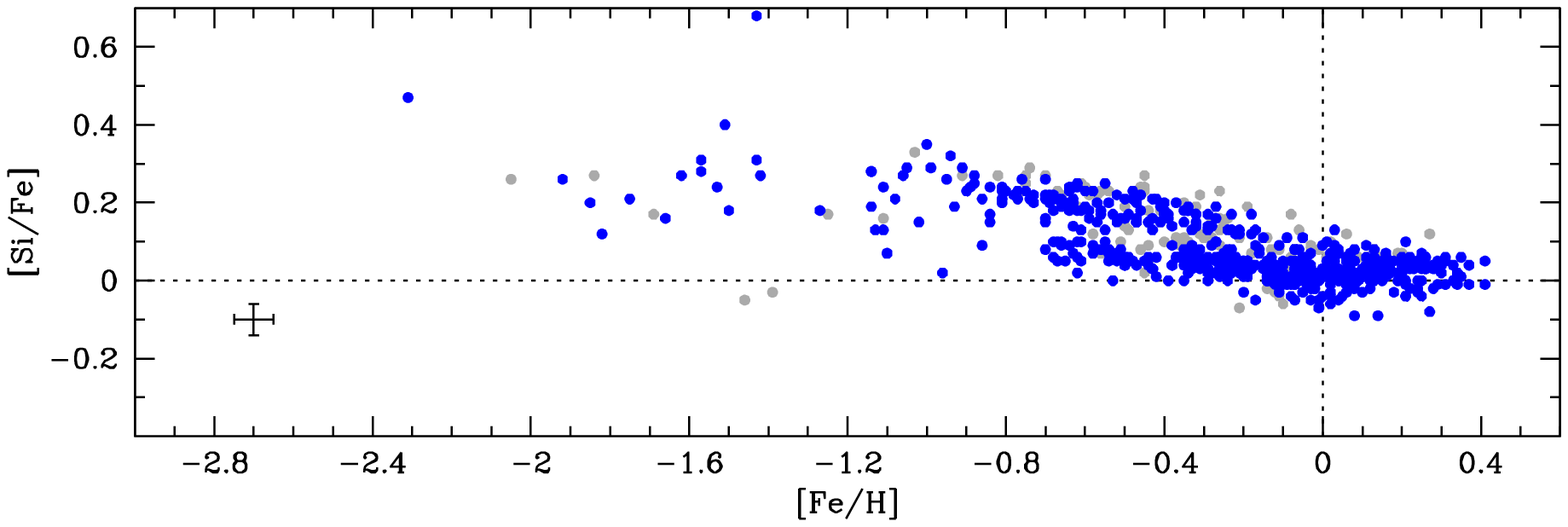}}
\resizebox{\hsize}{!}{
\includegraphics[bb=-90 200 700 350,clip]{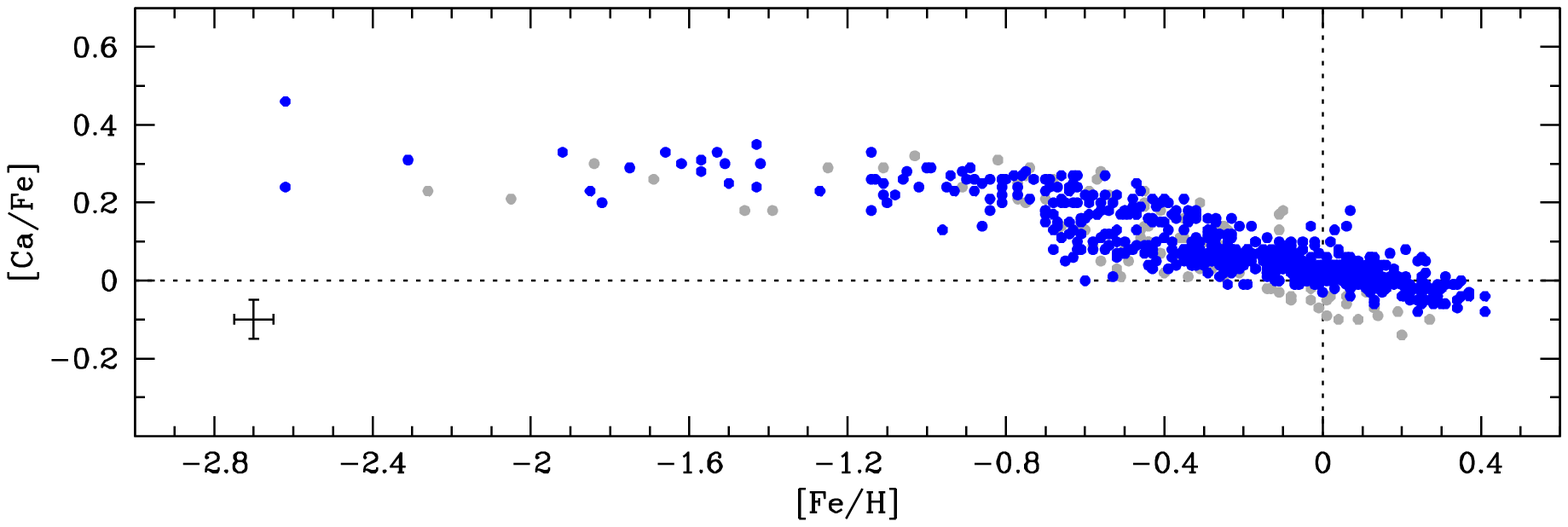}}
\resizebox{\hsize}{!}{
\includegraphics[bb=-90 165 700 350,clip]{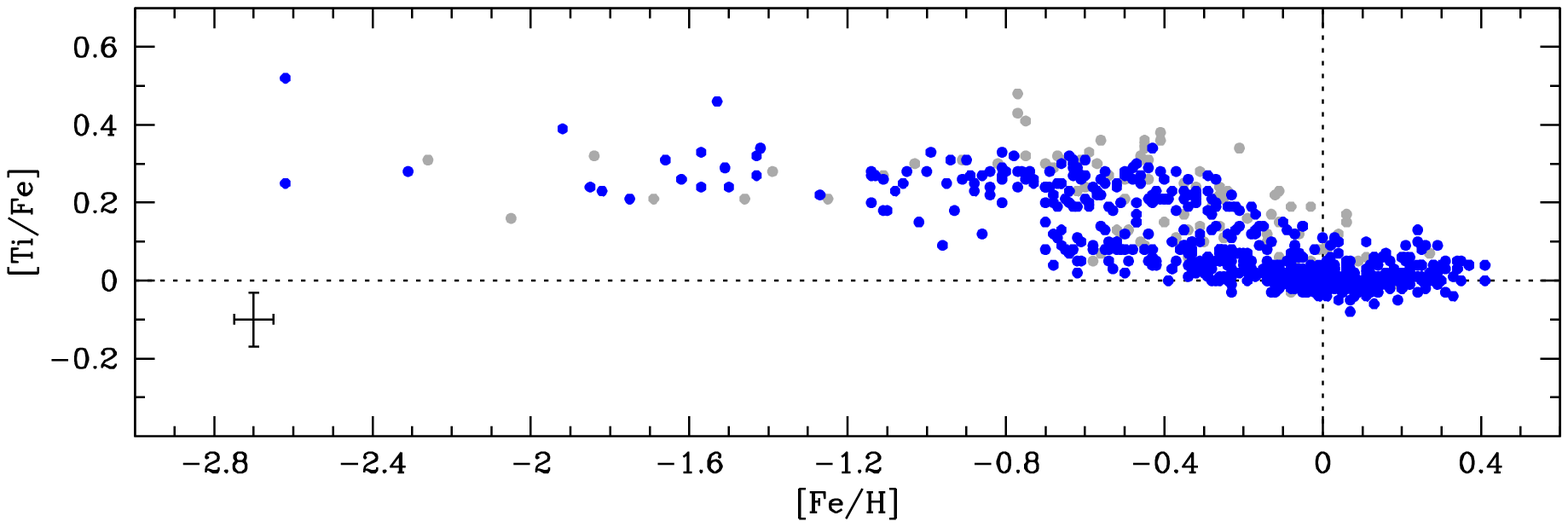}}
\caption{[$X$/Fe] versus [Fe/H] plots for the $\alpha$-elements 
(O, Mg, Si, Ca and Ti) and the light element Al.
The full sample of 714 stars is shown and black dots show the 604 stars 
with $\teff>5400$\,K and grey dots the stars with $\teff<5400$\,K.
A typical error bar is shown in each plot.
\label{fig:abundancetrends}
}
\end{figure*}
\begin{figure*}
\resizebox{\hsize}{!}{
\includegraphics[bb=-90 200 700 370,clip]{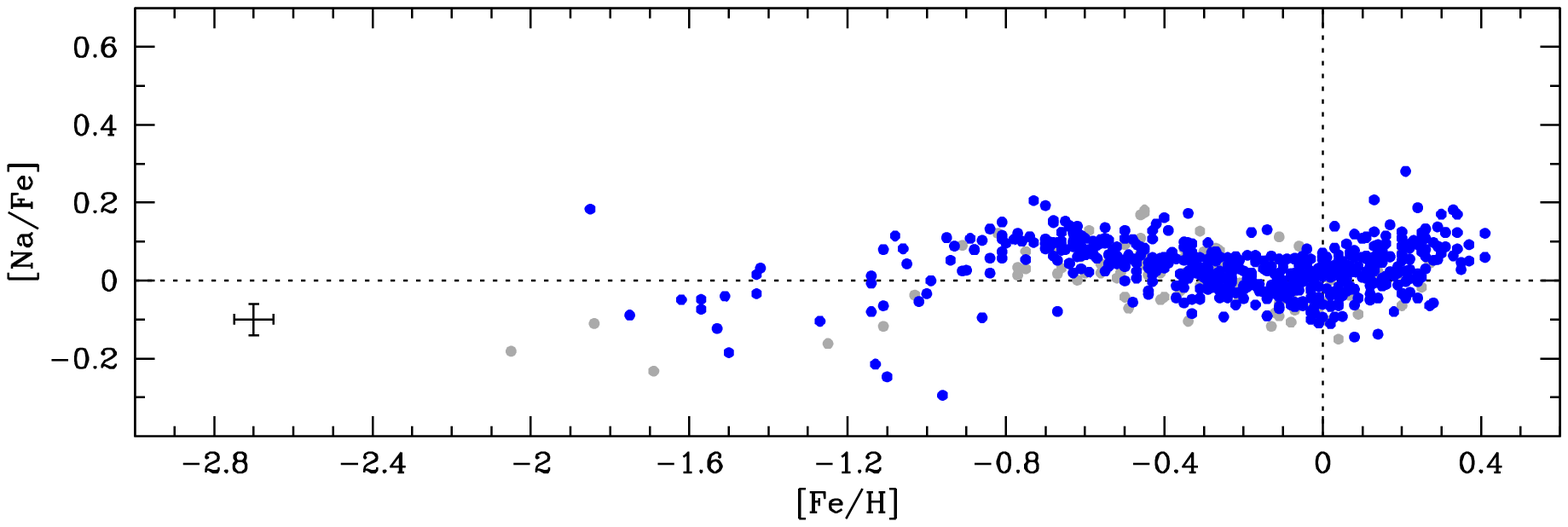}}
\resizebox{\hsize}{!}{
\includegraphics[bb=-90 200 700 350,clip]{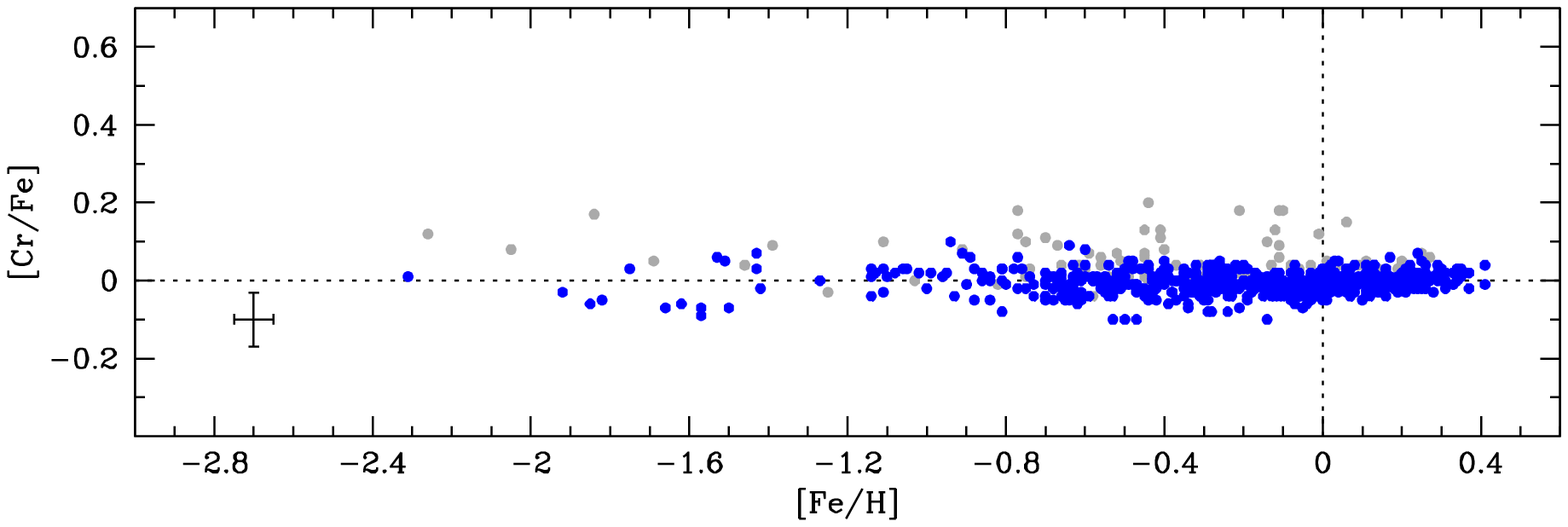}}
\resizebox{\hsize}{!}{
\includegraphics[bb=-90 200 700 350,clip]{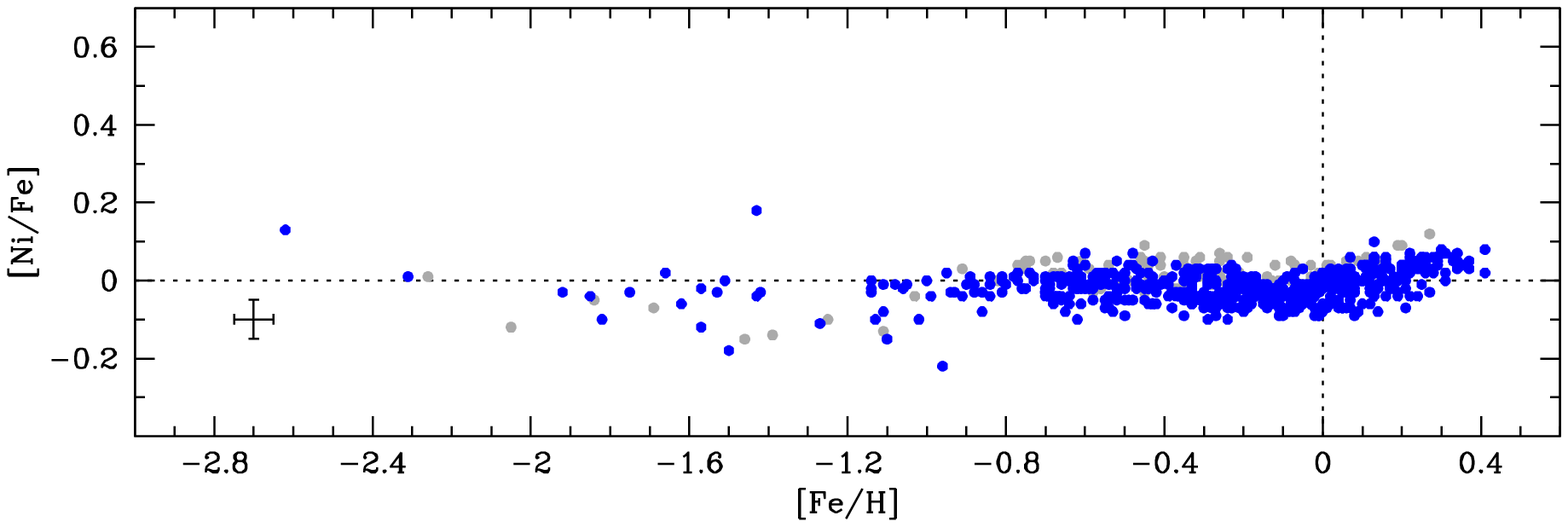}}
\resizebox{\hsize}{!}{
\includegraphics[bb=-90 200 700 350,clip]{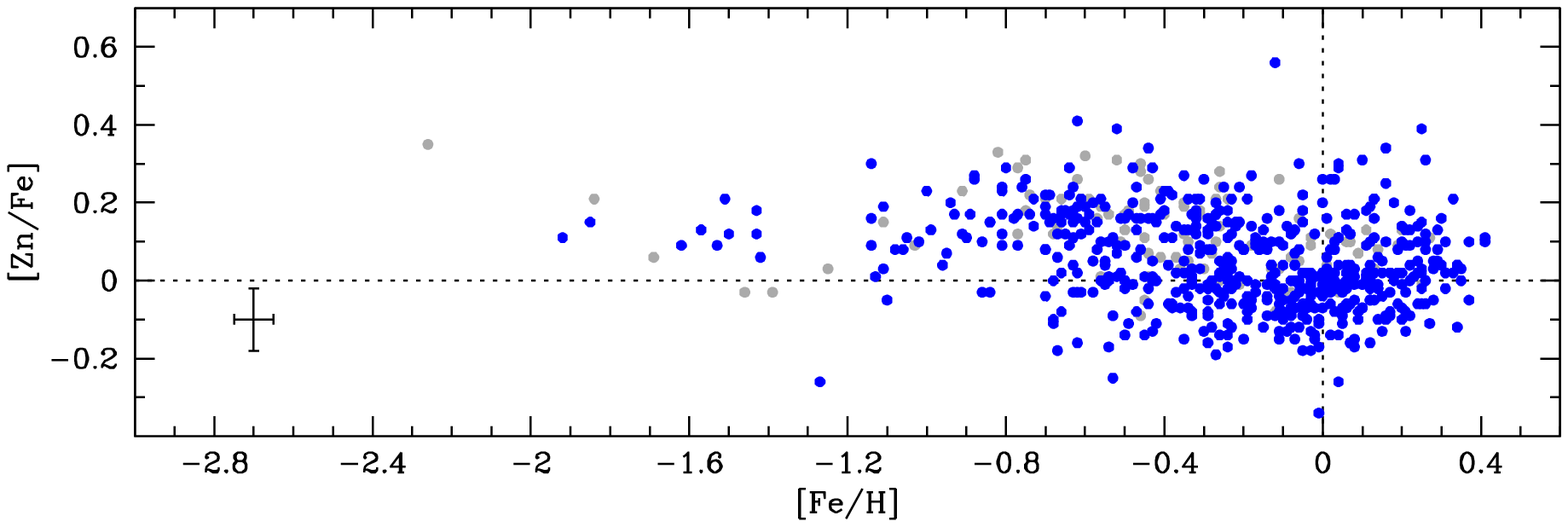}}
\resizebox{\hsize}{!}{
\includegraphics[bb=-90 200 700 375,clip]{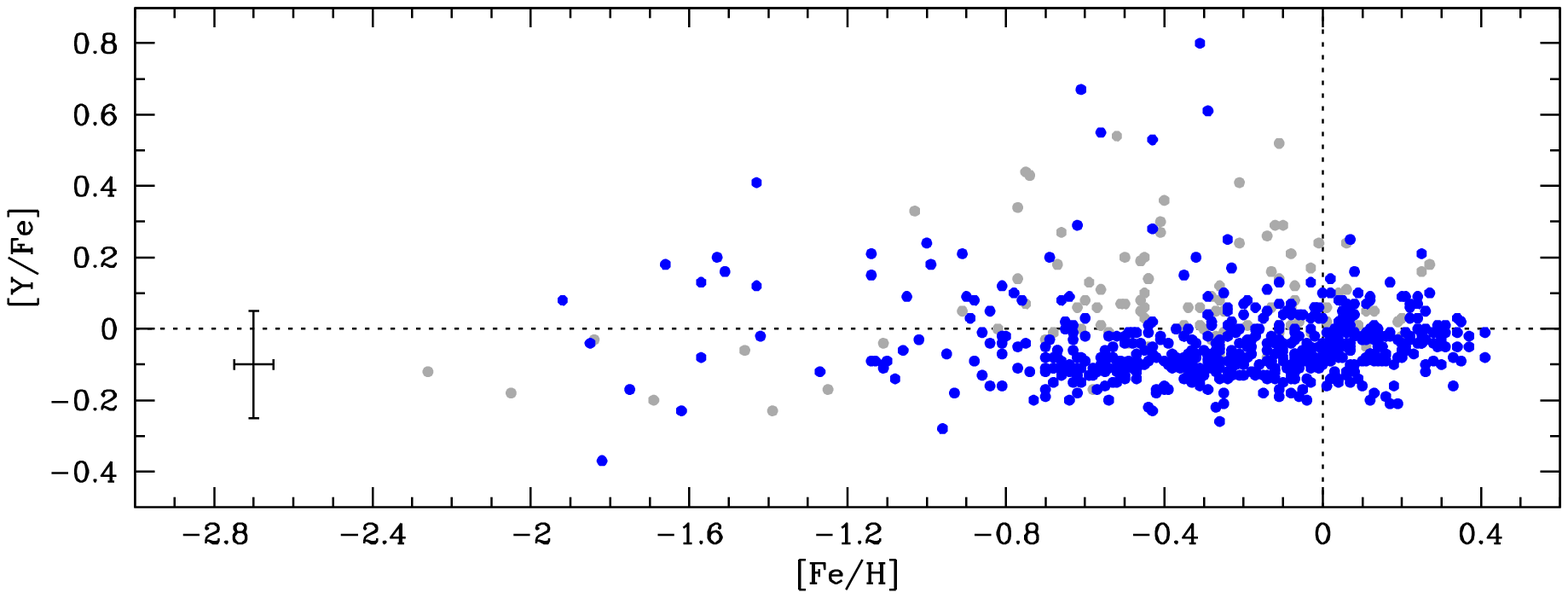}}
\resizebox{\hsize}{!}{
\includegraphics[bb=-90 165 700 375,clip]{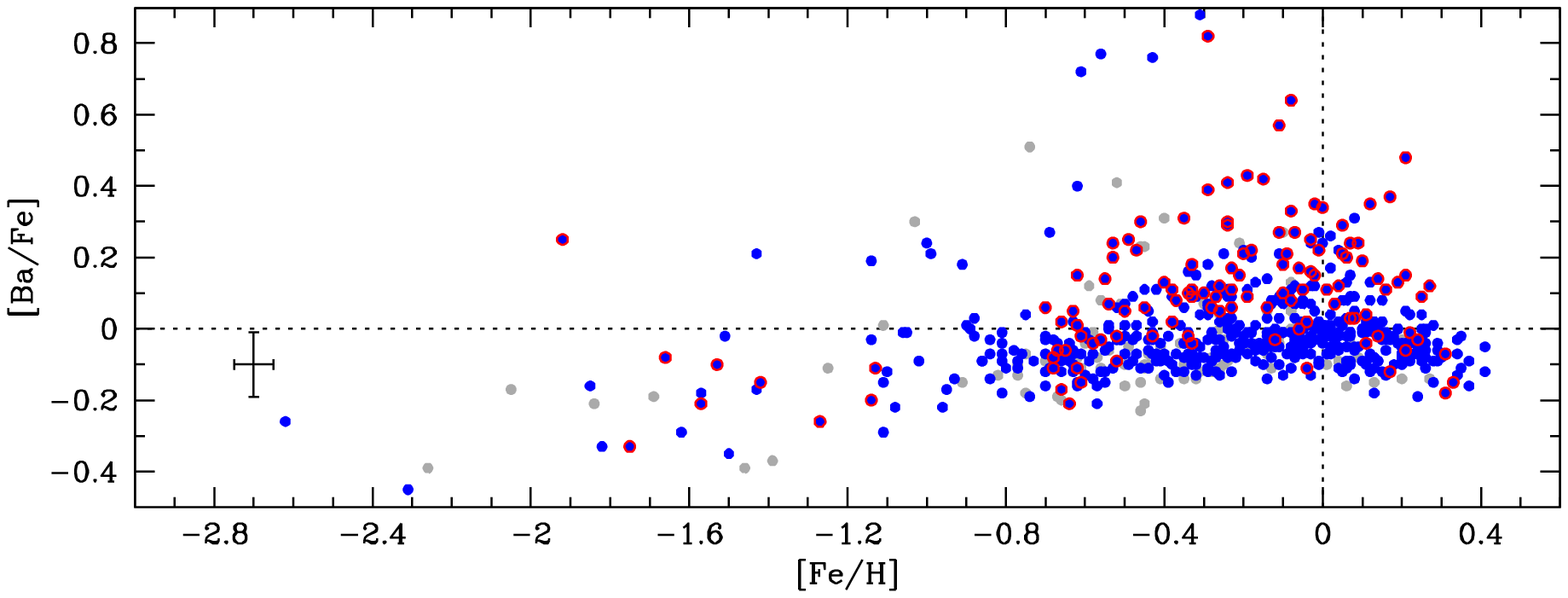}}
\caption{[$X$/Fe] versus [Fe/H] plots for the light element Na, the
iron-peak elements Cr, Ni, and Zn, and the s-process elements Y and Ba.
The full sample of 714 stars is shown and black dots show the 604 stars 
with $\teff>5400$\,K and grey dots the stars with $\teff<5400$\,K.
For Ba, stars with $\teff>6100$\,K have been
identified by red circles. A typical error bar is shown in each plot.
\label{fig:abundancetrends2}
}
\end{figure*}

\subsection{Age determination}
\label{sect:ages}

Stellar ages were determined from a fine grid of 
$\alpha$-enhanced Yonsei-Yale (Y2) 
isochrones by \cite{demarque2004}, adopting $\rm [\alpha/Fe] = 0$ for 
$\rm [Fe/H] > 0$, $\rm [\alpha/Fe] = -0.3 \times [Fe/H$] for 
$\rm -1 \leq [Fe/H] \leq 0$, and $\rm [\alpha/Fe] = +0.3$ for 
$\rm [Fe/H] < -1$. Taking the errors in effective temperature,  
surface gravity,
and metallicity into account, an age probability distribution (APD) was constructed
for each star. The most likely age, as well as lower and upper age estimates,
was estimated from this APD as described
in \cite{melendez2012} and a short outline in \cite{bensby2011}. 
In a similar manner, stellar masses
were determined as well. Ages, masses, and their
associated uncertainties  are given in Table~\ref{tab:parameters}.

\section{Elemental abundance results}
\label{sec:abundancetrends}

Studies of elemental abundances in nearby stars are important. 
\cite{gratton2000} showed that stars on cold
disk orbits have lower [$\alpha$/Fe] than stars that move
on halo and thick disk-like orbits.  
\cite{fuhrmann1998,fuhrmann2000unpubl,fuhrmann2004,fuhrmann2008,fuhrmann2011} showed
that stars very close to the Sun trace two distinct abundance trends. 
Several recent studies have obtained elemental abundances for stars that have typical
thin and thick disk kinematics; these also show distinct trends 
\citep[e.g.,][]{bensby2003,bensby2005,bensby2007letter2,reddy2003,reddy2006,adibekyan2012}.
However, recently there has been quite some debate about whether the Milky Way has a 
distinct thick disk or whether it forms a continuum with the thin disk 
\citep[see, e.g.,][]{bovy2012}.

Figures~\ref{fig:abundancetrends} and \ref{fig:abundancetrends2}
show the resulting abundance trends for the full sample of
714 stars. In Sect.~\ref{sec:uncertainties} we saw that the uncertainties
tend to increase for stars at lower temperatures and higher surface 
gravities. By restricting the sample to those stars with $\teff>5400$\,K
many of the stars with high uncertainties will be avoided.
604 of the 714 stars in the sample have $\teff>5400$\,K. 
In the abundance plots in Figs.~\ref{fig:abundancetrends} and 
\ref{fig:abundancetrends2} we have therefore marked the stars that have 
temperatures lower than 5400\,K by grey dots.

The abundance plots for oxygen, Mg, Si, Ca, and Ti show a flat plateau in 
[$X$/Fe] for stars more metal-poor than $\rm [Fe/H]\lesssim-0.5$. 
At higher [Fe/H] there is a general downward trend.
From $\rm [Fe/H]\approx -0.7$ and upwards, there appears to be two
abundance trends. 
At super-solar metallicities [O/Fe], and possibly also [Ca/Fe]
and [Mg/Fe] show downward trends with [Fe/H], while the 
[Si/Fe] and [Ti/Fe] trends are practically flat.

Na and Al are light odd-Z elements and we see that Al behaves like an 
$\alpha$-element, showing all the characteristics that the genuine 
$\alpha$-elements do, i.e., a flat plateau at lower [Fe/H] that at 
higher [Fe/H] starts to decline toward solar values. The
[Na/Fe] trend shows less dispersion than [Al/Fe] and there is no 
resemblance with Al or the $\alpha$-elements. Instead
[Na/Fe] is almost solar, with a slightly curved appearance, 
that rises at super-solar [Fe/H].

\begin{figure}
\resizebox{\hsize}{!}{
\includegraphics[bb=18 144 592 718,clip]{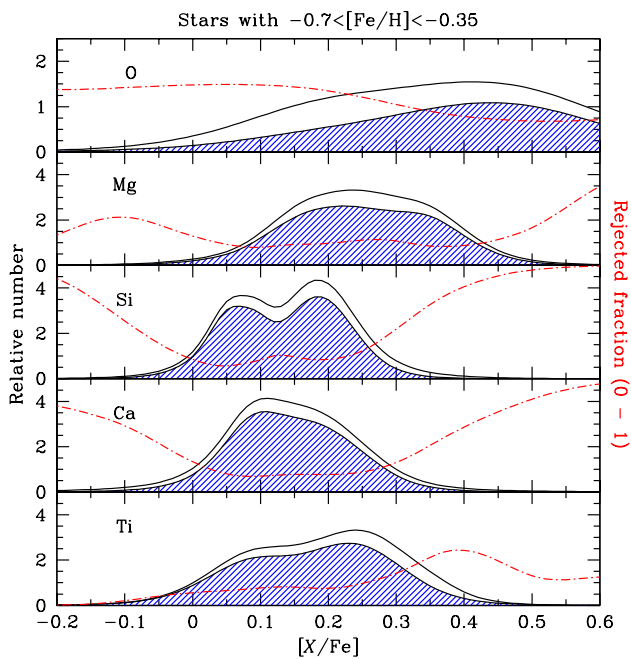}}
\caption{Generalised abundance ratio
histograms for the O, Mg, Si, Ca, and Ti for stars in the
interval $\rm -0.7<[Fe/H]<-0.35$. Shaded histograms show stars
with $\teff>5400$\,K.
The red dash-dotted lines show the fraction of rejected stars
when selecting stars with $\teff<5400$\,K.
}
\label{fig:gaphist}
\end{figure}

Both [Ni/Fe] and [Cr/Fe] show internally
extremely small dispersions and vary essentially in lock-step with [Fe/H].
The only discernible pattern is that the [Ni/Fe] ratio is slightly below
solar values at $\rm [Fe/H]<0$ and that it then shows a shallow increase
at $\rm [Fe/H]>0$. The latter feature turns out to be an important feature
when determining oxygen abundances from the forbidden [O{\sc i}] line
at 6300\.{\AA} which is blended with \ion{Ni}{i} lines
\citep[see][]{bensby2004}. We note that the few stars in the [Cr/Fe] plot
that lie slightly above the very flat trend of the bulk of stars 
are all stars that fall outside the selected temperature interval 
(grey coloured).

Zn is the second even-Z element beyond the iron-peak and albeit with a 
scatter, we find a somewhat declining [Zn/Fe] trend with metallicity, 
from being slightly elevated at $\rm [Fe/H]\lesssim -0.5$, to being solar
at $\rm [Fe/H]\gtrsim0$. There is also a slight resemblance with the
$\alpha$-elements. At lower metallicities $\rm [Fe/H]<-1$,
$\rm [Zn/Fe]$ is roughly constant at $\approx 0.1$, 
meaning that Zn could serve as a good
proxy for Fe in metal-poor damped Lyman alpha systems as it can be observed 
in damped Lyman alpha systems without being depleted by interstellar/galactic 
dust \citep[e.g.,][]{kobayashi2006}).

Both Y and Ba are $s$-process elements and we see that are slightly
under-abundant relative to Fe. 
We note that most of the stars in the [Ba/Fe] plot
that have high Ba abundances around solar [Fe/H] disappear when 
discarding stars 
with $\rm \teff<5400$ temperature range. The same is also true for
[Y/Fe]. Furthermore, Ba is known to suffer from NLTE effects
at higher $\teff$ \citep[e.g.,][]{korotin2011} and stars in the 
[Ba/Fe] plot with $\teff>6100$\,K have therefore been marked
with red circles. We see that essentially all stars with high 
[Ba/Fe] around solar metallicities also have high temperatures.

\begin{figure}
\resizebox{\hsize}{!}{
\includegraphics[bb=18 165 592 555,clip]{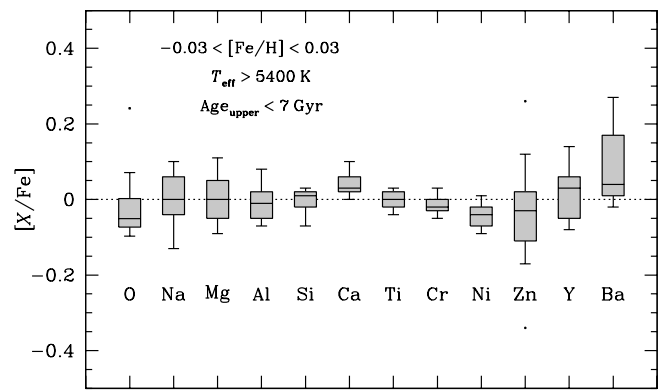}}
\caption{Boxplots showing the distribution of abundance ratios 
for 16 stars with $\teff>5400$\,K
and an upper age limit of 7\,Gyr, in a narrow
metallicity range around that of the Sun ($\pm 0.03$\,dex).  
Due to large NLTE effects for Ba at higher temperatures, the Ba box
has been further restricted to $\teff<6000$\,K.
In the boxplots the central horizontal line represents the median value. 
The lower and upper quartiles are represented by the outer edges of 
the boxes, i.e., the box encloses 50\,\% of the sample. 
The whiskers extend to the farthest data point that lies within 
1.5 times the inter-quartile distance. Those stars that do not 
fall within the reach of the whiskers are regarded as outliers 
and are marked by dots. 
}
\label{fig:normalsun}
\end{figure}

\subsection{The [$\alpha$/Fe] distribution at intermediate [Fe/H]}

The region where the potential gap, or bimodality, between the 
thin and thick disk 
abundance trends is largest is for metallicities in the interval
$-0.7\lesssim \rm [Fe/H]\lesssim-0.35$. Due to observational uncertainties
and the magnitude of the astrophysical signature,
this gap appears clear for some elements and less so for others.
From the abundance trend plots in Figs.~\ref{fig:abundancetrends}
and \ref{fig:abundancetrends2}, it is evident that the scatter decrease
and that the abundance trends become more well-defined when only including stars
with $\teff>5400$\,K. To further highlight the effects, Fig.~\ref{fig:gaphist} 
shows the generalised [$X$/Fe] histograms for O, Mg, Si, Ca, and Ti for stars in the
metallicity range $\rm -0.7<[Fe/H]<-0.35$. The empty and shaded histograms
show the distributions when including or discarding the stars with $\teff<5400$\,K,
respectively. The red dash-dotted lines show the fraction of the sample
as a function of $[X/{\rm Fe}]$ that gets rejected when selecting stars
with $\teff<5400$\,K . Especially for Si, and perhaps
Mg and Ti, one sees that a higher fraction
of the ``bad'' stars with $\teff<5400$\,K  are located
in the gap area and that the potential bimodality
becomes clearer when discarding the stars that are more prone to 
uncertainties. For all abundance trends there is a large fraction
of ``bad'' stars at the lower and upper limits of the abundance
ratios, i.e., these stars increase the dispersion in the plots.
This demonstrates that uncertainties
potentially can wash out differences between stellar populations
\citep[see also][for a quantitative analysis]{lindegren2013}.

\subsection{The abundance pattern of the Sun}

Several studies give opposing results regarding the Sun's
abundance pattern relative to what is seen for the
Galactic disk \citep[e.g.,][]{melendez2009sun,ramirez2010}. 
In Fig.~\ref{fig:normalsun}
we show the abundance ratios for young disk stars in a narrow 
metallicity range around that of the Sun ($\pm 0.05$\,dex) that
have upper age estimates below 7\,Gyr, and discarding stars
that are more susceptible to uncertainties
(i.e. only keeping stars with $\teff>5400$\,K, 
see \ref{sec:uncertainties}). For most
of the abundance ratios the Sun appears to be ``normal'', i.e.,
the lines showing the median values in the boxplots are close to zero. For the 
few abundance ratios where the central 50\,\% fall either above or
below a value of zero, the median line is still within 0.05\,dex of the Sun.
Based on this, the Sun appears not to be too different from
the bulk of young disk stars in the immediate Solar neighbourhood.

\subsection{Statistical definitions of stellar populations based on kinematics}
\label{sec:criteria}

Many recent studies of the stellar disk in the Milky Way have aimed to 
characterise the elemental abundances for stars that are thought to 
belong to the thick and thin disks. It thus became important 
to select stars that likely belong, respectively, to the two disks. 
An expedient way to 
do this is to use kinematical criteria such as the one from
\cite{bensby2003,bensby2005}, and which is outlined in 
Appendix~\ref{sec:kincriteria}.

Figure~\ref{fig:tifetdd} shows the [Ti/Fe]-[Fe/H] abundance trends 
for five different cuts in the thick disk-to-thin disk
probability ratios $(TD/D)$
that indicate how much more likely it is that
a star belongs to the thick disk than the thin disk.
The top panel shows the stars that are at least ten times more likely
to be thick disk stars, while the bottom panel shows stars that are at least
ten times more likely to be thin disk stars. The three panels 
in the middle show probabilities in between, with the middle one
containing stars that cannot easily be classified as either thin disk
or thick disk.
What we see is that even with these very extreme 
definitions of the samples there is a significant overlap in the sense
that there are stars with either classification that fall on the
abundance trend normally associated with the other population
\citep[see also, e.g.,][]{fuhrmann1998,bensby2003,reddy2006}. 
This is an obvious consequence of this kinematical 
classification, as stars from the low-velocity tail of the thick disk
will be classified as thin disk stars, while stars from the high-velocity tail
of the thin disk will be classified as thick disk stars
(assuming that there are two distinct trends for the elemental abundances).

\begin{figure}
\resizebox{\hsize}{!}{
\includegraphics[bb=25 160 425 710,clip]{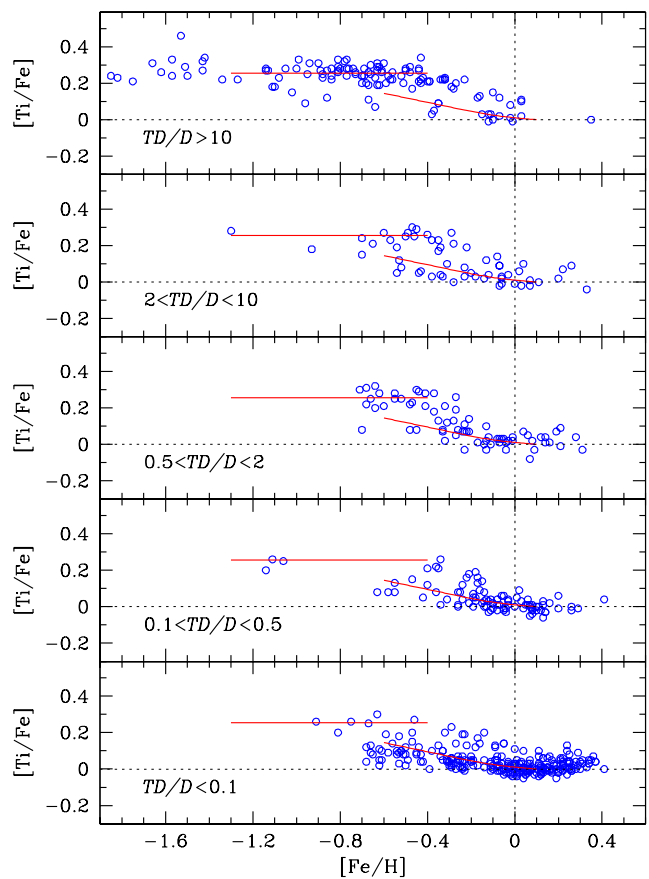}}
\caption{[Ti/Fe] as a function of [Fe/H] selected on
$TD/D$ as indicated in each panel for stars with $\teff>5400$\,K.
To guide the eye, the red lines outline the thick disk abundance plateau 
and the decrease in the thin disk abundance ratio, respectively.
\label{fig:tifetdd}
}
\end{figure}

\begin{figure*}
\resizebox{\hsize}{!}{
\includegraphics[bb=18 160 475 718,clip]{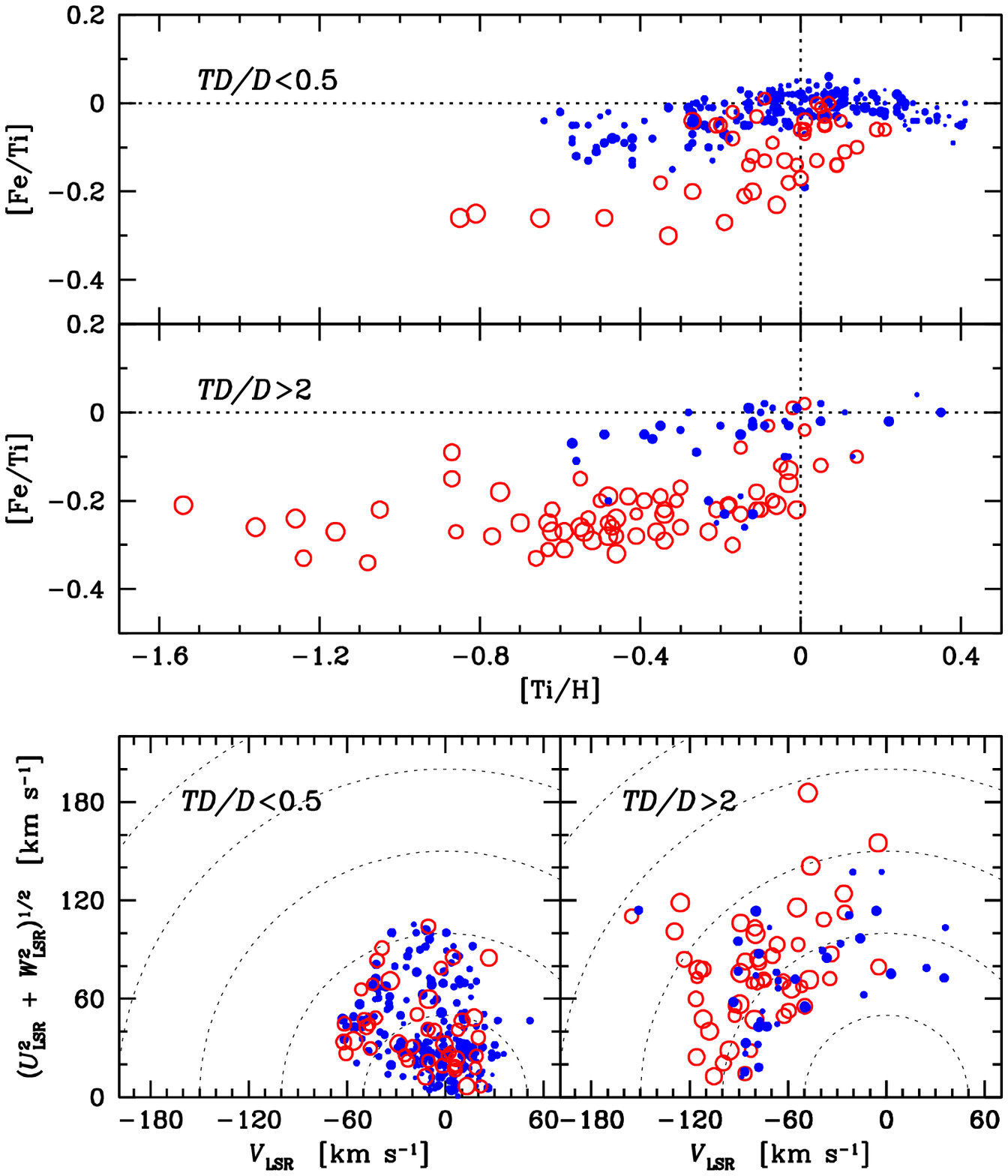}
\includegraphics[bb=18 160 475 718,clip]{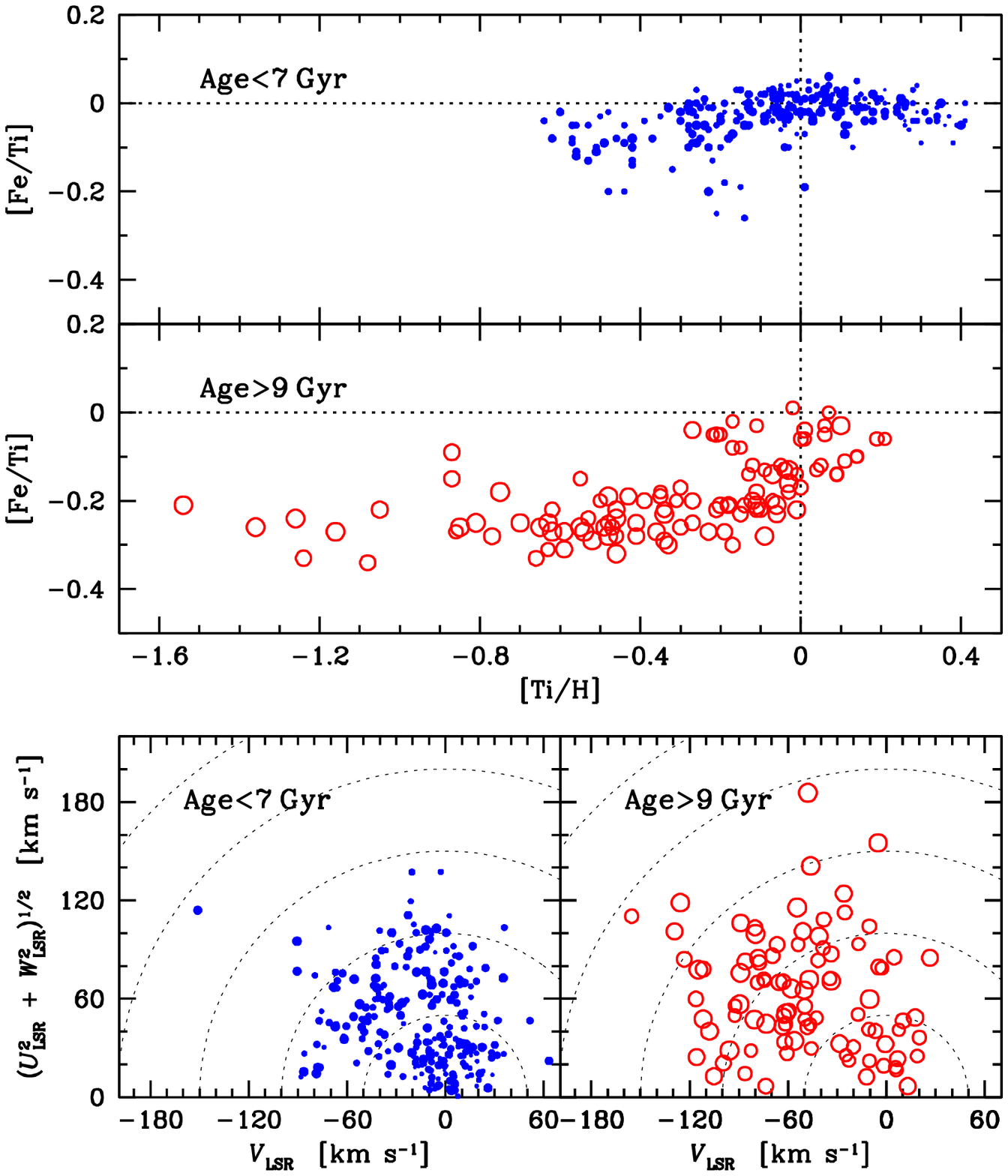}}
\caption{Left-hand side plots show the [Fe/Ti] versus [Ti/H] abundance
trends when using the kinematical criteria as defined in
\cite{bensby2003,bensby2005}. Stars have been colour- and size-coded
depending on their ages. Right-hand side plots shows the abundance
trends when splitting the sample according to their
ages (as indicated). For all plots we have only included stars whose
ages have been better determined than 4\,Gyr (difference between upper
and lower age estimates). \label{fig:confusion}
}
\end{figure*}

To further investigate the mixing of populations when using kinematical 
selection criteria we show in the upper panel on the left-hand side of 
Fig.~\ref{fig:confusion} the [Fe/Ti]-[Ti/H] abundance trends for two 
kinematically selected samples: one where the probabilities of being a 
thin disk star are at least twice that of being a thin disk star 
$(i.e., TD/D<0.5)$; and one where the probabilities of being a thick disk 
star are at least twice that of being a thick disk star $(i.e., TD/D>2)$. 
This time we have coded the sizes of the markers by the ages of the stars and
include only stars with good age estimates ($\rm \Delta Age<4$\,Gyr).  
It is evident that the ``second'', weaker, abundance signature in each 
sample has the same age structure as the main signature in the other sample.
The Toomre diagrams for the two subsamples in the bottom panel 
on the left-hand side of Fig.~\ref{fig:confusion}
shows that the two samples are kinematically very different, with
little overlap. Hence, it is apparent that there are kinematically cold 
stars that are old and $\alpha$-enhanced, as well as kinematically 
hot stars that are young and less $\alpha$-enhanced.

What about stellar age? Could this be a better discriminator between 
the thin and thick disks? The [Fe/Ti]-[Ti/H] abundance trends for two 
samples, one old sample with stars that have estimated ages greater than 
9\,Gyr, and one young sample with stars that have estimated age less than
7\,Gyr, are shown in the upper part on the right-hand side of 
Fig.~\ref{fig:confusion}. Once again we see two very different chemical 
signatures, similar to the ones on the left-hand-side when the selection
of the samples was based on kinematics. However, the difference is 
now that the two abundance trends are somewhat ``cleaner'', with less 
``mirroring" between them. Looking at the Toomre diagrams for these two
age-selected subsamples (bottom panel on the right-hand side of 
Fig.~\ref{fig:confusion}), there is a large kinematical
overlap, i.e. there are many young stars with hot 
kinematics and many old stars with cold kinematics.

In summary, we note that there appears to be no perfect way of 
selecting thin and thick disk stars. While velocities and distances
can be pinpointed to rather high accuracies there seem to be a large 
kinematical overlap between the two populations. Ages on the other hand 
appear to be better, but as good ages are notoriously difficult to obtain, 
there is also a significant overlap (due to the errors).
However, it appears as if stellar ages might be a somewhat better discriminator
when selecting thin and thick disk stars from nearby stellar samples.  
In kinematically selected thin disk samples we are prone to 
be contaminated by the low-velocity tail of the thick disk, and 
especially so at lower metallicities, and for kinematically selected 
thick disk samples we are prone to be contaminated by the high-velocity 
tail of the thin disk, and especially so at higher metallicities.

\section{Discussions}

\subsection{Ages and metallicities}

\subsubsection{Old and metal-rich stars?}

Recent high-resolution spectroscopic studies indicate that 
most stars with thick disk kinematics are older than those with 
thin disk kinematics \citep[e.g.,][]{gratton2000,feltzing2001b,bensby2005}.
However, considerably larger samples available in photometric studies 
such as the GCS indicate the existence of a significant number of 
stars with thin disk kinematics that have high ages ($>10$\,Gyr).
Figure~\ref{fig:agefe}a shows the age-metallicity relation
for our sample, and we also see that our sample possibly
contains such old and metal-rich stars. However, the stars
that have ages greater than 10\,Gyr and metallicities higher than
solar, all have large uncertainties (red small dots in Fig.~\ref{fig:agefe}).
Hence, the parameters for these stars are doubtful and cast doubt on the 
existence of (very) old and metal-rich (super-solar) stars.

\subsubsection{Age-metallicity relations?}

In \cite{bensby_amr} we investigated whether stars with kinematics typical
for the thick disk showed any signs of an age-metallicity
relation. We found, in accordance with other studies 
\cite[e.g.,][]{haywood2006,schuster2006}, that stars with kinematics 
typical of the thick disk show an age-metallicity relation such that more 
metal-rich stars on average are younger than the less metal-poor 
stars in the sample. The stars older than about 8\,Gyr in 
Fig.~\ref{fig:agefe}a show a trend of declining metallicity with age,
consistent with the age-metallicity
relation seen for thick disk stars in the studies mentioned above.
Younger stars do not show this behaviour. Instead there appears to be 
a rather large scatter in age over the whole
metallicity range ($-$0.8 to +0.4\,dex), i.e., no apparent age-metallicity 
relation.

\subsubsection{[$\alpha$/Fe] as a proxy for age?}

\begin{figure}
\resizebox{\hsize}{!}{
\includegraphics{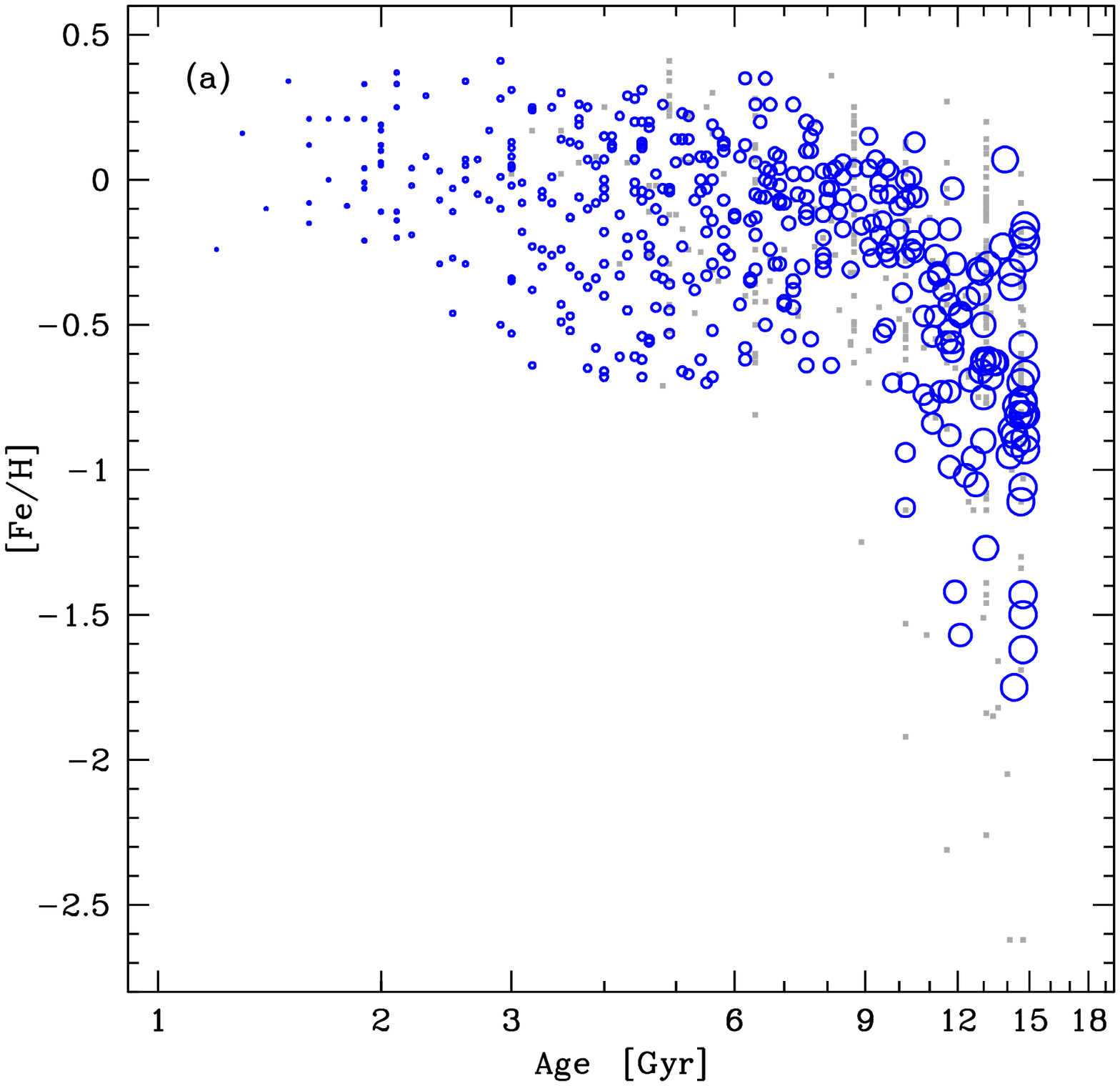}}
\resizebox{\hsize}{!}{
\includegraphics[bb=18 144 592 420,clip]{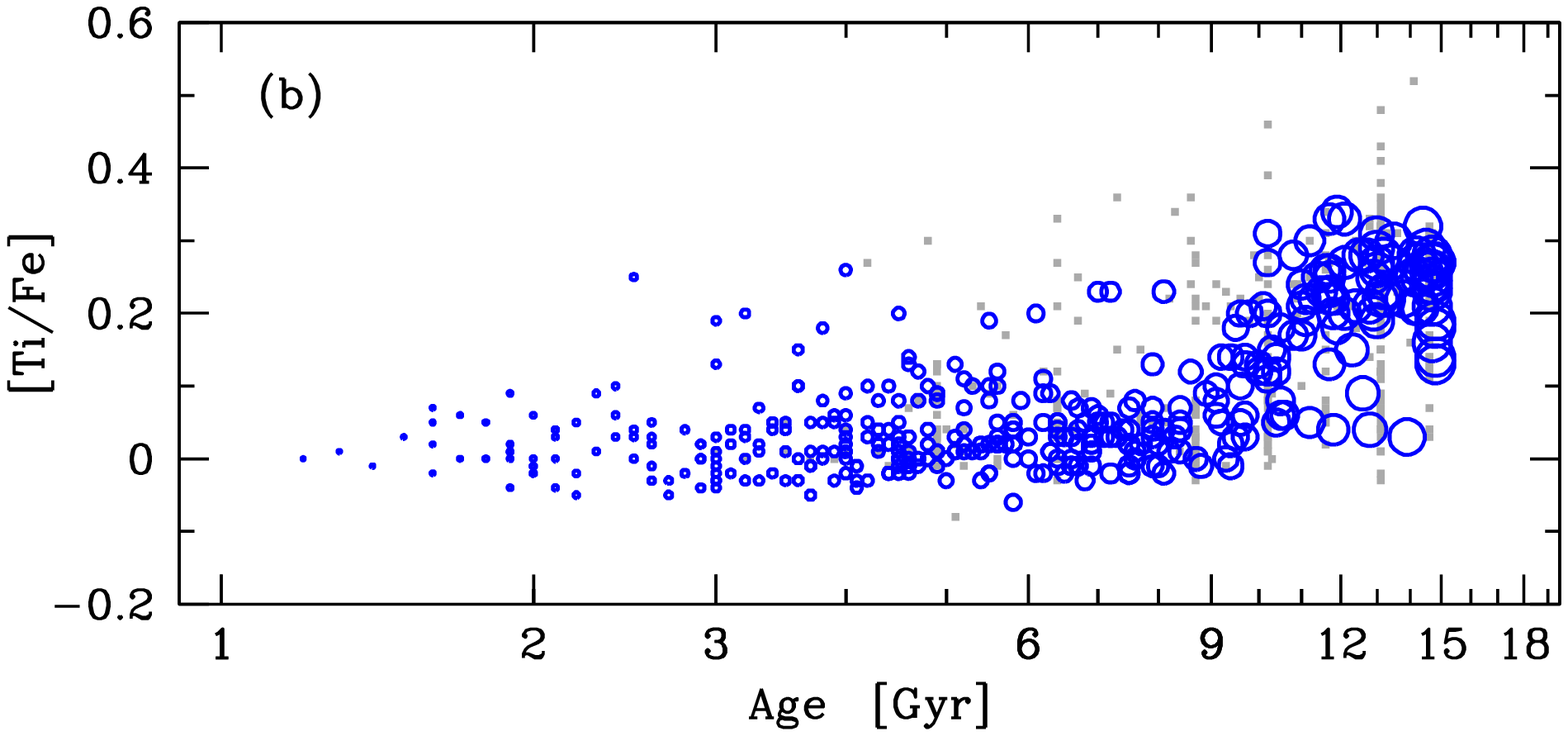}}
\caption{Age-metallicity relation for those stars
that have an age difference between upper and lower estimate of
less than 4\,Gyr. The sizes of the circles have been scaled 
with the ages of the stars. Stars with larger age uncertainties
are shown as small grey dots.
\label{fig:agefe}
}
\end{figure}

Recently, \cite{navarro2011} have argued that it is better to
identify stars with different populations
based on their elemental abundances rather than other
properties such as kinematics.  That a statistical selection based on
kinematics causes overlaps between various abundance trends is evident
from the nature of that selection process (see Sect.~\ref{sec:criteria}), 
and casts doubt on the reality of distinct trends for different stellar
populations. This argument was used, e.g., by \cite{bovy2012} when
they investigated the scale-height of mono-abundance populations
(i.e., stars that fall in a narrow range of elemental abundances,
e.g., [$\alpha$/Fe] and [Fe/H]) in the SEGUE data set. 

To better understand the formation and evolution of the Milky Way, 
it is very desirable to have stellar ages as well as elemental 
abundances. Given the overall structure of the elemental abundance 
patterns and ages observed in the Milky Way \citep[e.g.,][]{edvardsson1993}, 
it has been suggested that the amount of $\alpha$-enhancement in a 
star can be used as a proxy for the age of a star \citep{liu2012,haywood2013}. 
However, age is a very difficult property to derive for most stars 
\citep[e.g.,][]{soderblom2010}. As our sample contains a fair 
portion of turn-off and sub-giant stars we are in a position to 
investigate whether old ages are a common feature for all stars 
with enhanced [$\alpha$/Fe] in the Solar neighbourhood. 
Figure~\ref{fig:agefe}b shows that this is indeed the case for stars 
older than about 8\,Gyr and thus that [Ti/Fe] can be used as a proxy 
for age for stars, in the sense that young and old stellar populations 
can be distinguished. Other studies are also finding that various 
$\alpha$-elements correlate with ages in this sense. For example 
\cite{ramirez2013}, in their Fig. 17, show the same results as our 
Fig.~\ref{fig:agefe}b, but for [O/Fe] as a function of age. 

However, this result is only valid for dwarf stars in the immediate Solar
neighbourhood. We do not know if the same is true elsewhere in the
Galaxy or indeed recoverable for other stellar evolutionary stages.
\cite{bensby2013} provides data for 58 microlensed dwarf and turn-off
stars in the Galactic bulge. These stars, tentatively, show the same 
trend as the stars in the Solar neighbourhood, making it plausible that
the connection between $\alpha$-enhancement and age is a property shared
by many stellar populations in the Galaxy.

\subsubsection{A lower metallicity limit for thin disk}

Thin disk stars with metallicities below $\rm [Fe/H]<-0.7$ are 
apparently not found in spectroscopic studies in the literature 
\citep[see, e.g.][]{fuhrmann2004,reddy2003,soubiran2005}. One of the few
studies that does claim to have thin disk stars at lower metallicities, 
reaching $\rm [Fe/H]\approx-1$, is \cite{mishenina2004}. It is clear, 
however, that those few stars have chemical compositions similar to what 
is found in the thick disk, even though the kinematic properties place them
as thin disk stars. Hence their thin disk status is ambiguous.

Out of the $>14\,000$ stars in the GCS, there are 11\,010 stars 
that are potential thin disk stars according to their kinematics. 
1378 of those stars have 
ages estimated to be older than 7\,Gyr, and 156 stars have 
$\rm [M/H]<-0.7$. Our sample originally contained 27 thin disk
stars with metallicity estimates in the GCS less than $\rm [Fe/H]<-0.7$.
However, nine of these stars could not be analysed due to the fact that they 
were binaries or rotated too fast. Out of the remaining 18 stars, only 
one remained below $\rm [Fe/H]<-0.7$ after the spectroscopic analysis.
Therefore, we believe that $\rm [Fe/H]\approx-0.7$
could be interpreted as a lower metallicity limit for the Galactic 
thin disk.

\subsection{Metal-rich and $\alpha$-enhanced stars}

\begin{figure}
\resizebox{\hsize}{!}{
\includegraphics[bb=18 144 592 477,clip]{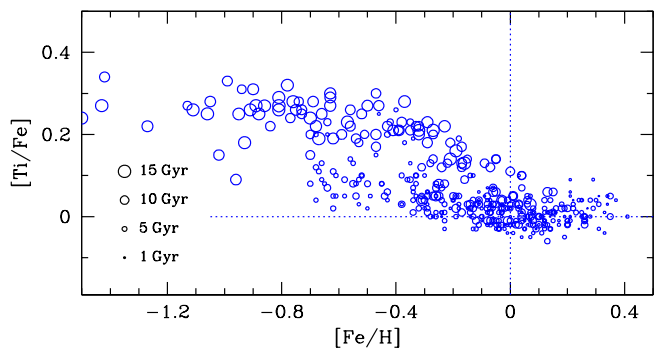}}
\caption{[Ti/Fe] versus [Fe/H] for stars that
have low age uncertainties (the differences between upper and lower age 
estimates are less than 4\,Gyr). The the sizes of the circles are 
scaled with the ages of the stars as indicated in the figure. 
}
\label{fig:tife_vol_sel}
\end{figure}
\begin{figure}
\resizebox{\hsize}{!}{
\includegraphics{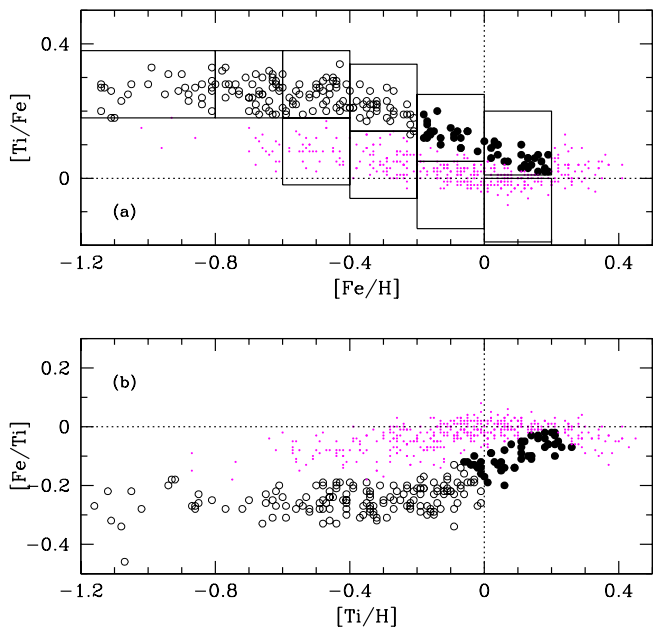}}
\caption{
The solid black circles mark stars that are $\alpha$-enhanced
and metal-rich (HAMR stars); the empty black circles mark stars that are
$\alpha$-enhanced at lower [Fe/H] (a.k.a. potential thick
disk); and the small blue circles mark stars with low or
moderate $\alpha$-enhancement (a.k.a. potential thin disk
stars).
}
\label{fig:hamr1}
\end{figure}

In Fig.~\ref{fig:tife_vol_sel} we show the [Ti/Fe] abundances trends
for all stars in our sample where the upper and lower age estimates differ
by at most 4\,Gyr. 
We find a similar division of the stellar sample as seen by 
\cite{fuhrmann1998,fuhrmann2000unpubl,fuhrmann2004,fuhrmann2008,fuhrmann2011}, 
but now in Ti. We see that the Ti-enhanced stars are 
the oldest stars. However, a major
difference is that we have deliberately searched for metal-rich stars 
with hot kinematics. As a result, we have stars that could be associated
with the thick disk (high $\rm [\alpha/Fe]$ ratios and high ages) 
that are more metal-rich than can be found in Fuhrmann's sample
(which is volume complete for $d<25$\,pc and thus rarer types of stars
may be missing).
There are not many of them, and most of them in our sample
are found outside the 25\,pc sphere within which Fuhrmann's
stars are located. Hence, our sample has the potential to 
trace the thick disk to higher metallicities \citep{bensby2007letter2}.

\begin{figure}
\resizebox{\hsize}{!}{
\includegraphics[bb=18 160 592 440,clip]{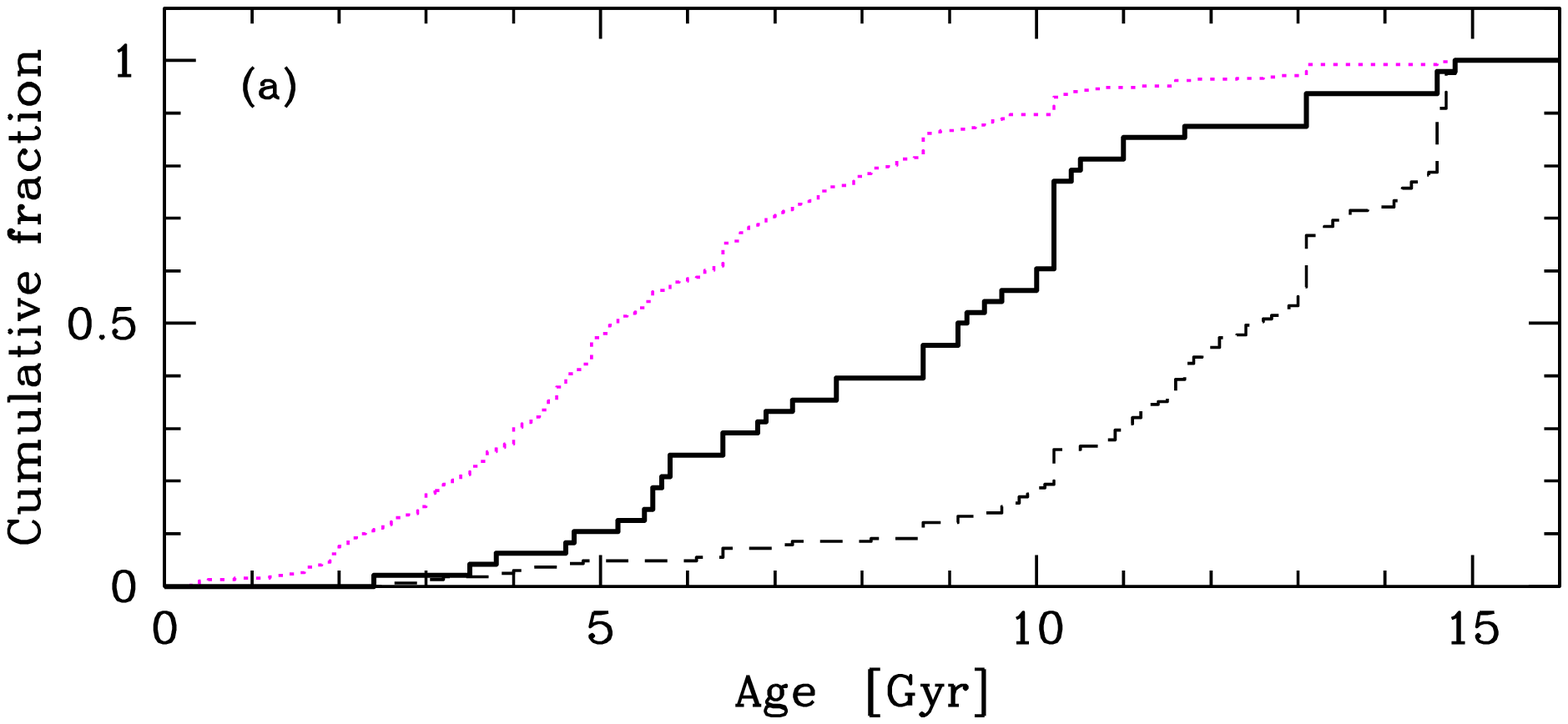}}
\resizebox{\hsize}{!}{
\includegraphics{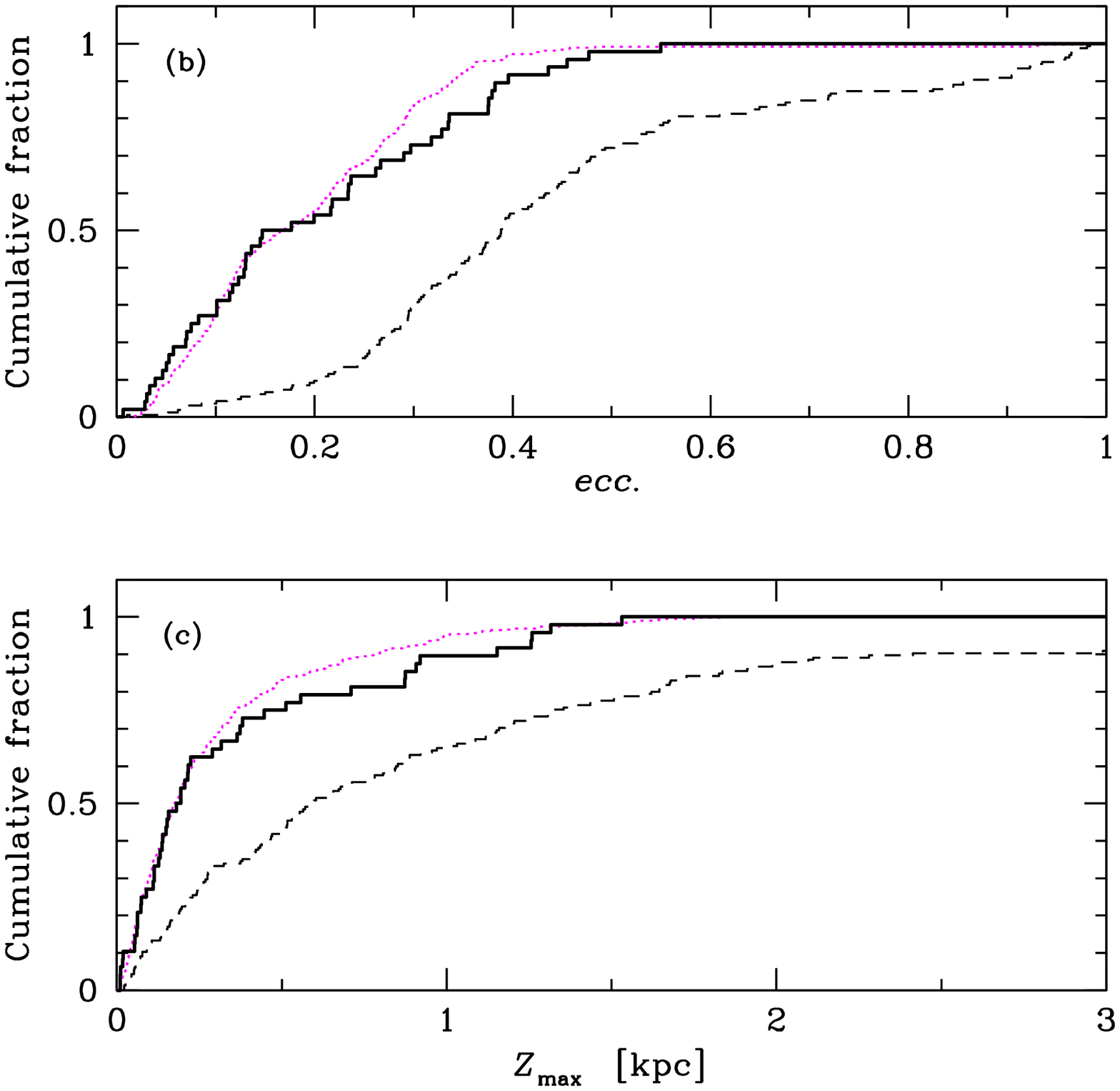}}
\caption{
Cumulative histograms
for the eccentricity, $\zmax$, and age  distributions
for the three different samples in Fig.~\ref{fig:hamr1}a. 
HAMR stars are marked by solid black
lines, potential thick disk stars by dashed black lines, and potential
thin disk stars by dotted blue lines.
}
\label{fig:hamr2}
\end{figure}
\begin{figure}
\resizebox{\hsize}{!}{
\includegraphics{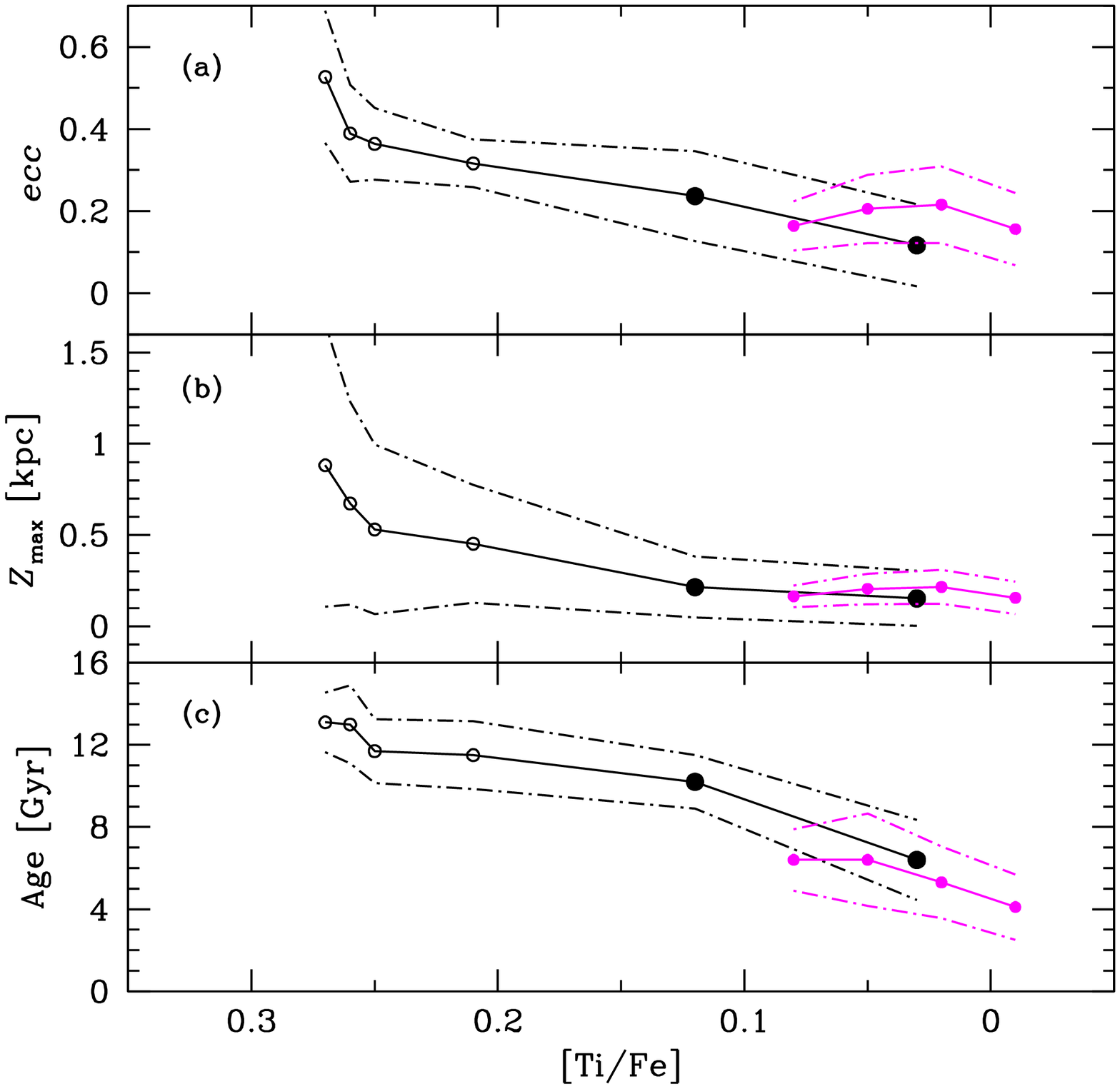}}
\caption{
Median values of the eccentricity, $\zmax$, and age
for the stars in the boxes in Fig.~\ref{fig:hamr1}a.
HAMR stars are marked by solid black
circles, potential thick disk stars by open black circles, 
and potential thin disk stars by solid blue circles.}
\label{fig:hamr3}
\end{figure}

A different aspect of metal-rich and $\alpha$-enhanced stars
was put forward by \cite{adibekyan2011} who claimed a new 
$\alpha$-enhanced and metal-rich population (high-$\alpha$ and 
metal-rich stars, hereafter HAMR stars), 
distinct from both the thin disk and the thick disk.
This HAMR population showed up as stars with [Fe/H] around solar values
that have $\alpha$-enhancement greater than what is seen 
for the bulk of the stars at $\rm [Fe/H]\approx 0$.
These stars were also separated from
the thick disk by a ``gap'' in metallicity
at $\rm [Fe/H]=-0.2$ and $\alpha$-enhancement at $\rm [\alpha/Fe]=+0.17$. 
The kinematical properties
resembled those of the thin disk population, i.e., circular orbits
confined to the Galactic plane.

In our sample we have several stars around solar [Fe/H] that have
higher $\alpha$-enhancements than the bulk of disk stars at
similar metallicities (see, e.g., Figs.~\ref{fig:abundancetrends} 
and \ref{fig:tife_vol_sel}).
In Fig.~\ref{fig:hamr1} we show the abundance trends for Ti
with our HAMR stars marked by larger solid black circles, [Ti/Fe]-enhanced
stars at lower [Fe/H] (typical thick disk stars) by open circles, and 
low-[Ti/Fe] stars (typical thin disk stars) by magenta coloured dots. 
The approximate separation in Fig.~\ref{fig:hamr1} has been done
by eye.
Figures~\ref{fig:hamr2}a-c then show cumulative histograms
of the age, eccentricity, and $\zmax$ distributions for these three
different groups of stars. First we see that the HAMR stars have 
an age distribution in between those of the two disks, and that there might be
``bumps'' around 6-7\,Gyr and 10-12\,Gyr, which are the typical ages
for stars of the thin and thick disks. Looking at the eccentricity and $\zmax$
distributions, it is clear that the HAMR stars are very similar to the
low-$\alpha$ stars associated with the thin disk. 

So, what are these HAMR stars, where do they come from, and
should they be classified as a stellar population of their own?
And if so, is there a metallicity gap between the thick disk and this
newly found HAMR population? To further investigate this, we will divide the sample into ``mono-abundance'' populations according to the boxes
in Fig.~\ref{fig:hamr1}a. Figure~\ref{fig:hamr1}b shows the sample but
with [Ti/H] as the reference element. In Figs.~\ref{fig:hamr3}a-c we then show 
how the median eccentricity, median $\zmax$, and median age varies
with [Ti/Fe] for the stars in the boxes in Fig.~\ref{fig:hamr1}a.
The plots also show the 1-$\sigma$ dispersions around the median.
It is evident that the eccentricity, $\zmax$, and age for the HAMR
stars (black filled circles) follow smoothly upon the downward trend 
with [Ti/Fe] set by the ``thick disk'' stars (open circles). 
We also see that the ``thin disk'' stars (magenta coloured filled circles) 
more or less follow upon the trend set by the thick disk and HAMR stars regarding
eccentricity and $\zmax$. For the ages, there could be a potential
gap around 7-8\,Gyr, indicating that the most metal-rich,
thick disk/HAMR stars are older than the most metal-poor, thin disk stars.

In summary, we cannot claim that the HAMR stars form unique 
population as claimed by \cite{adibekyan2011}. More likely, it may
just be the metal-rich extension of the thick disk. This implies 
that the thick disk potentially reaches metallicities as high as 
$\rm [Fe/H]\approx +0.2$, somewhat higher than what we found in 
\cite{bensby2007letter2}. 
The disparate results between our study and the \cite{adibekyan2012} study
could be due to that both we and them have complex selection functions.
Larger samples with a controlled and well-defined selection function, such as 
for instance the sample from the Gaia-ESO survey \citep{gilmore2012}, 
will reveal the existence or non-existence 
of a unique HAMR stellar population.

\begin{figure}
\resizebox{\hsize}{!}{
\includegraphics{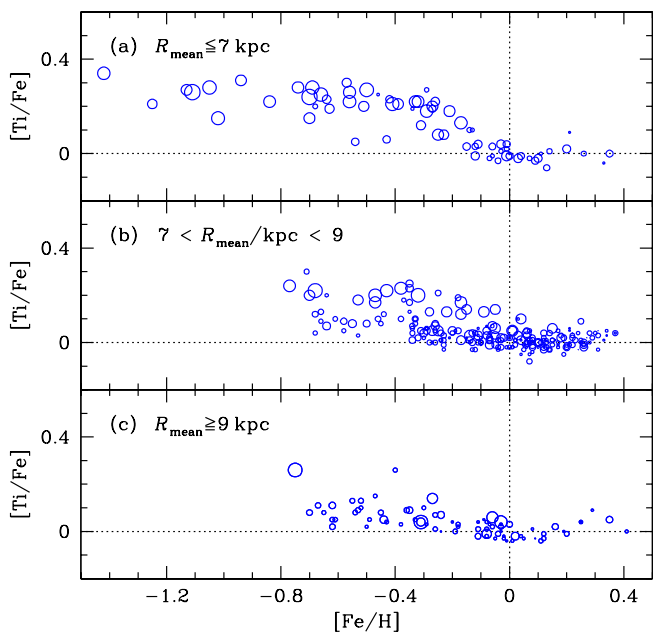}}
\caption{[Ti/Fe]-[Fe/H] abundance trends
for stars with different $R_{\rm mean}$. 
Only stars for which he difference between the upper and
lower age estimates is less than 4\,Gyr are included.
The sizes of the circles have been scaled 
with the ages of the stars.
\label{fig:rmeantife}
}
\end{figure}

\subsection{Radial variation}

\begin{figure}
\resizebox{\hsize}{!}{
\includegraphics{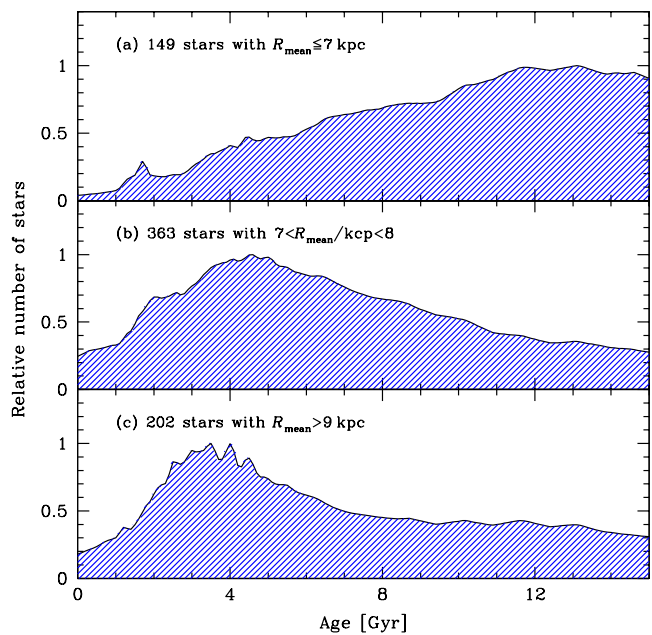}}
\caption{Sums of individual age probability distributions
for stars with different $R_{\rm mean}$. All stars are included.
\label{fig:rmeanage}
}
\end{figure}

The mean of the apo- and pericentric distances of the stellar orbit, 
$\rmean$, can be used as a proxy for the galactocentric radius of the 
birth place for a star \citep[e.g.,][]{grenon1987,edvardsson1993}. 
However, with the recent advancement of the theory of radial migration, 
rearranging the orbits of stars through processes such as churning 
and blurring throughout the history of the Galaxy 
\citep[e.g.,][]{sellwood2002,schonrich2009a,schonrich2009b}, the usage 
of $\rmean$ as a proxy for the birthplace of a star could be dubious. 
We will, however, start by using $\rmean$ as a first approximation. 
Figure~\ref{fig:rmeantife}
shows the $\rm [Fe/Ti]-[Ti/Fe]$ trends for stars with different $\rmean$,
with the sizes of the circles scaled with the ages of the stars.
We see that the sample with $\rmean<7$\,kpc
mainly contains old and $\alpha$-enhanced stars with a small fraction 
of younger and less $\alpha$-enhanced stars. The opposite is true for 
the sample with $\rmean>9$\,kpc, which mainly contains young and 
less $\alpha$-enhanced stars and very few old and $\alpha$-enhanced stars.
The sample with orbits that stay close to that of the Sun ($7<\rmean<9$\,kpc)
contain stars that divide into two trends, one with old and $\alpha$-enhanced
stars and one with young and less $\alpha$-enhanced stars. 
The age distributions in Figs.~\ref{fig:rmeanage} further show that the
sample with $\rmean<7$\,kpc contains stars of all ages with a slight
over-representation of old stars; the sample of stars with orbits close to the 
Sun contains mainly stars with ages less than $\sim 8$\,Gyr and only a few
older stars; while the sample with $\rmean>9$\,kpc contains almost only younger
stars. These findings indicate that the old and $\alpha$-enhanced stars
mainly come from small galactocentric radii while the young and less 
$\alpha$-enhanced stars mainly come from large galactocentric radii.

\begin{figure*}
\resizebox{\hsize}{!}{
\includegraphics{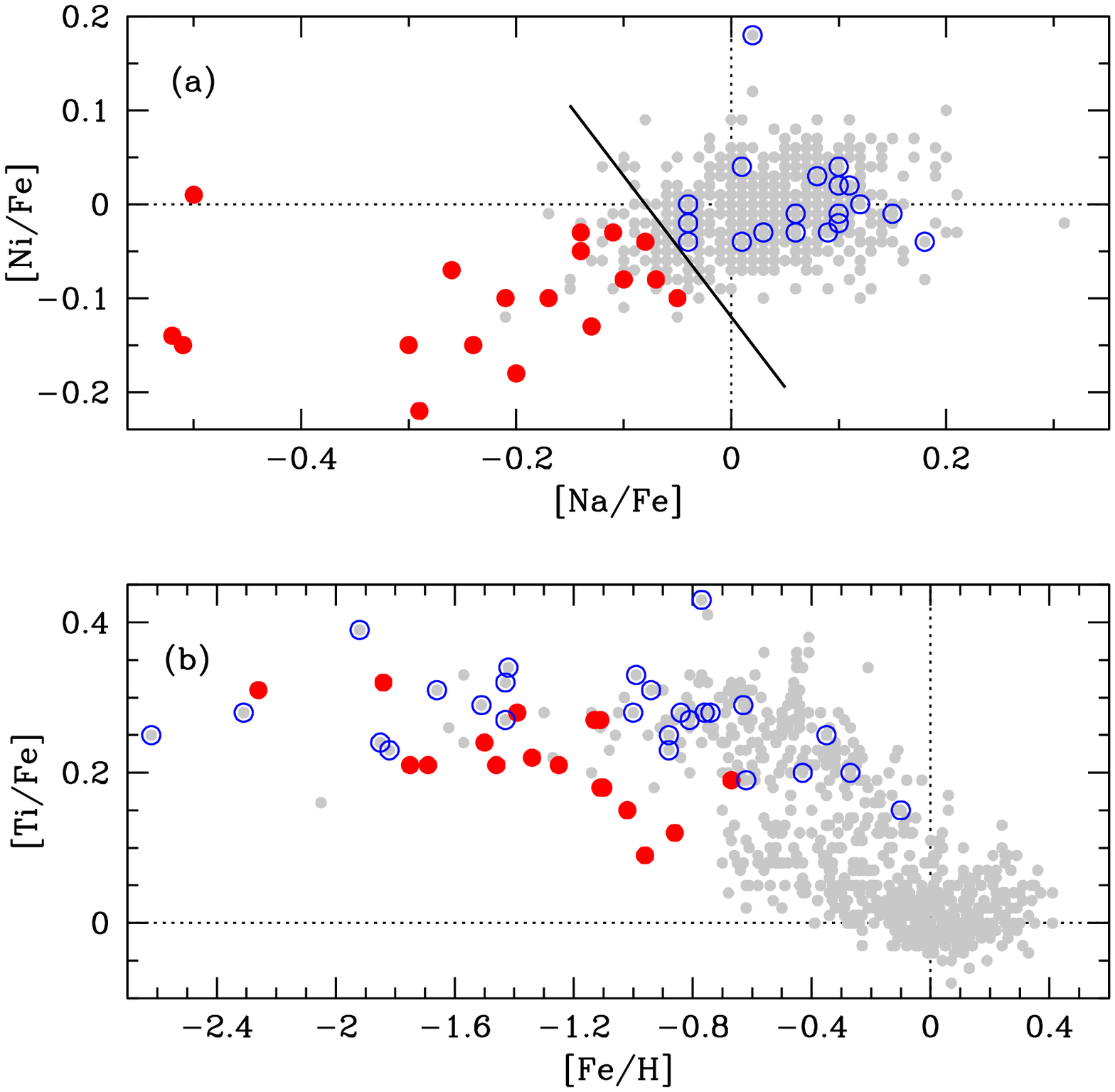}
\includegraphics[bb=18 144 592 718,clip]{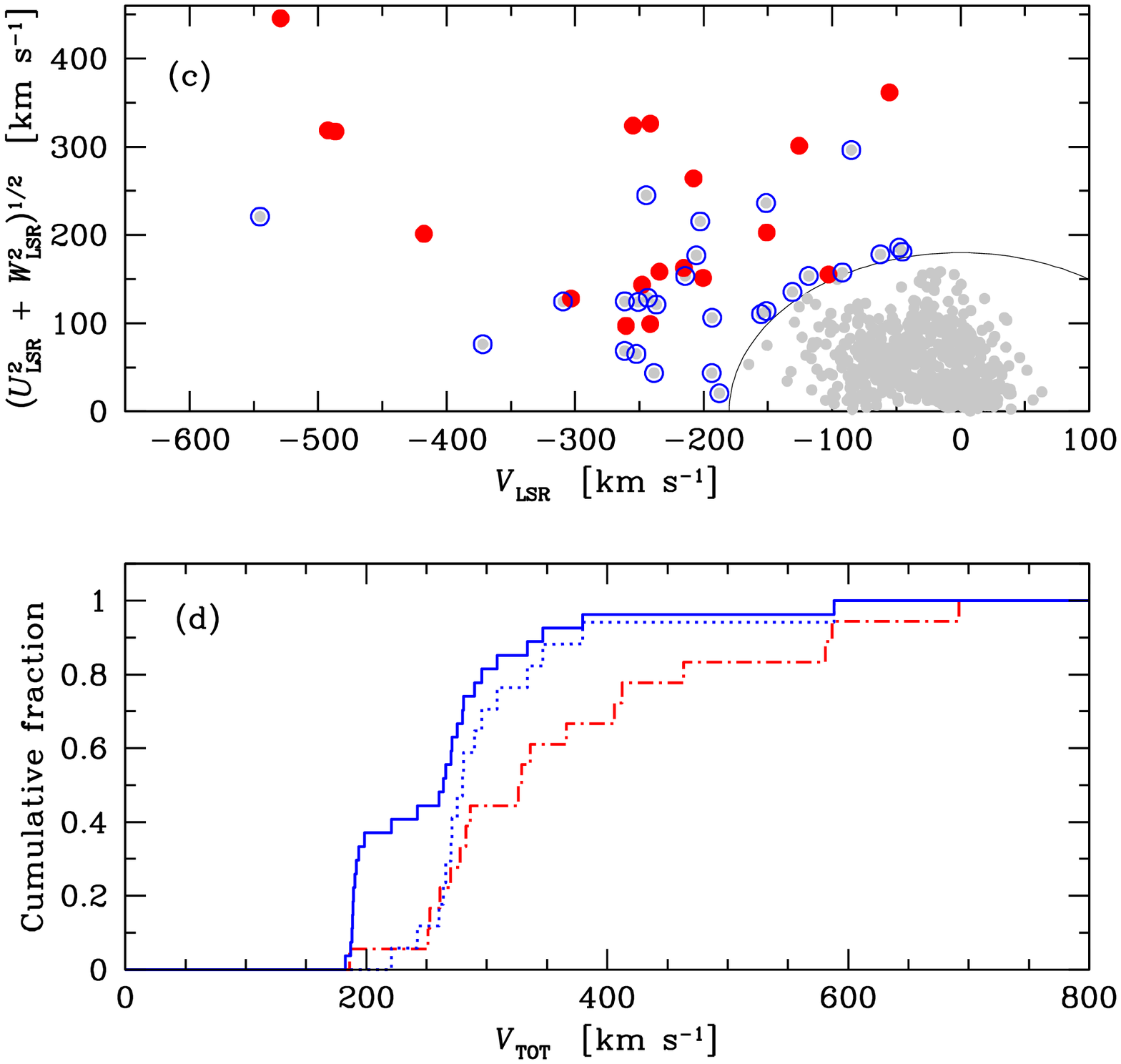}}
\caption{\label{fig:halo}
(a) [Ni/Fe]-[Na/Fe] for the sample where the solid line marks the 
approximate separation between the low-$\alpha$ and high-$\alpha$
halo populations discovered by \cite{nissen2010}. Stars that have
a total space velocity greater than $V_{\rm TOT}>180\,\kms$ have
been marked by open blue circles if located on the upper-right 
side of the dividing line, and by solid red circles if located
on the lower-left side. (b) [Ti/Fe]-[Fe/H] for the sample with the
same coding as in (a). (c) Toomre diagram with the same coding as in (a).
The curved line marks $V_{\rm TOT}=180\,\kms$. (d) Cumulative
distributions of $V_{\rm TOT}$: dashed red line represents
the red stars in (a)-(c), solid blue line the blue stars in (a)-(c),
and dotted blue line the blue stars in (a)-(c) but only including
stars with $V_{\rm TOT}>200\,\kms$.
}
\end{figure*}

To address the question of whether $\rmean$ is a valid proxy for 
stellar  birthplace, we see that the abundance trends for the different 
$\rmean$ bins
in Figs.~\ref{fig:rmeantife}a--c are essentially identical to the ones
found by \cite{bensby2010letter,bensby2011letter}, who studied
44 red giants located {\it in situ} in the inner disk ($R=4$ to 7\,kpc)
and 20 red giants located {\it in situ} in the outer disk ($R=9$ to 12\,kpc). 
Especially the lack of younger stars with low [Ti/Fe] at $\rm [Fe/H]<-0.2$
in Fig.~\ref{fig:rmeantife}a, and the lack of older
$\alpha$-enhanced stars in Fig.~\ref{fig:rmeantife}c, agree very well
with the inner and outer disk {\it in situ} red giant samples of
\cite{bensby2010letter,bensby2011letter}. These similarities
could validate the use of $\rmean$.

In \cite{bensby2011letter} the lack 
of $\alpha$-enhanced stars in the outer disk, even if they were 
located far ($>1$\,kpc) from the plane, was interpreted as being due to the fact 
that the thick disk had a much shorter scale-length than the thin disk. 
Shortly thereafter, \cite{cheng2012_2} used 5650 stars from 
the SEGUE survey and confirmed the short scale-length of the thick disk. 
The local data presented here appears to confirm these conclusions.

\subsection{Kinematics groups and star streams}

\subsubsection{Low-$\alpha$ halo}

The abundance trend plots in Fig.~\ref{fig:abundancetrends}
show a small number of stars with $\rm [Fe/H]\approx -1$
that have lower [Ti/Fe] than the majority of the low
metallicity stars in our sample. These stars are of the same type 
stars as the inner halo stars found in
\citet{nissen1997,nissen2010}. The selection criteria
applied to their halo sample was that the stars should have 
$V_{\rm TOT}>180$\,km\,s$^{-1}$, plus a
metallicity criterion (derived from $uvby$ photometry).  
To their surprise, they found that the
halo stars clearly split into two abundance trends, one with the high
constant $\alpha$-enhancement, and one that shows a straight decline
from $\rm [Fe/H]\approx -1.6$ to $-0.8$. 
Although with some overlap, the low-$\alpha$ stars identified
by \cite{nissen2010} generally had
higher values on their total space velocities than the high-$\alpha$ ones. 
An even better discriminator turned out to be the [Ni/Fe]-[Na/Fe] 
abundance space where there was a very clear distinction, 
essentially without overlap, between high- and low-$\alpha$ stars.

In Fig.~\ref{fig:halo}a we show the [Ni/Fe]-[Na/Fe]
abundance space for our stars and the dividing line found
\cite{nissen2010} is marked out. The stars that have ${\it V}_{\rm TOT}>180\,\kms$
and that fall on either side of the dividing line is marked by solid red
and open blue circles, respectively. Figure~\ref{fig:halo}b then
shows the [Ti/Fe]-[Fe/H] abundance plot with the potential high- and
low-$\alpha$ stars marked. It is clear that the abundance pattern 
observed by \cite{nissen2010} is also present in our sample, a low-$\alpha$
pattern and a high-$\alpha$ pattern the diverge with [Fe/H]. In Fig.~\ref{fig:halo}c
we then show the Toomre diagram with the high- and low-$\alpha$ halo stars specially marked.
Although they appeared to show more different velocity distributions in the
\cite{nissen2010} paper, it appears as if the low-$\alpha$ halo stars in our sample
on average have higher total space velocities. This is further illustrated in
Fig.~\ref{fig:halo}d that shows the cumulative distributions of the total
space velocities for the two samples.  As we have quite a few high-$\alpha$
stars (blue circles) in the Toomre diagram just outside the $V_{\rm TOT}=180\,\kms$
line, we also show the cumulative distributions when restricting to stars
with velocities greater than 200\,$\kms$.

Our results show that with our sample we can, at least tentatively, 
confirm the \cite{nissen2010} finding that in the 
stellar halo, as sampled in the Solar neighbourhood, there exist two 
different elemental abundance trends. The different abundance trends
are most likely indicative of differing origins for these stars, 
and should be more extensively explored with larger and complete samples,
such as, e.g., the sample from the Gaia-ESO survey \citep{gilmore2012}.

\begin{figure}
\resizebox{\hsize}{!}{
\includegraphics{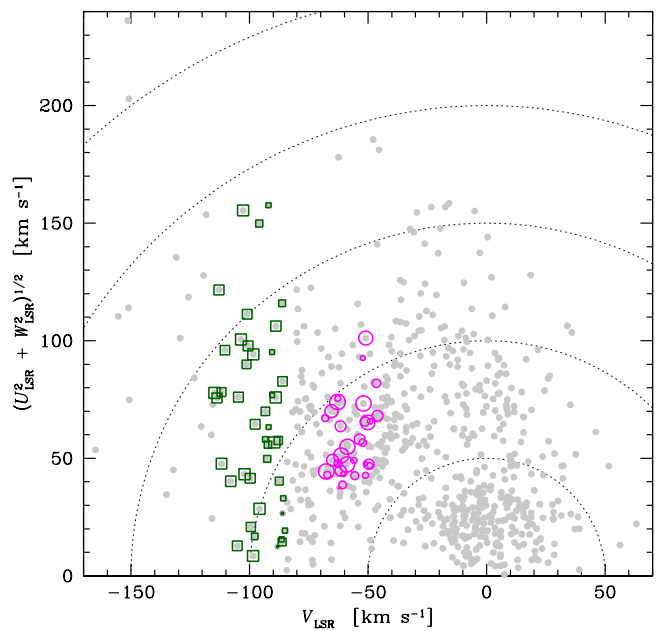}
}
\caption{\label{fig:toomre2}
Stars that have kinematic probabilities of belonging to
the Hercules stream ($Her/TD>2$ and $Her/D>2$)
are marked by larger circles, and candidate Arcturus stream
stars by larger squares. The sizes of the markers have been scaled with the
ages of the stars.
}
\end{figure}
\begin{figure}
\resizebox{\hsize}{!}{
\includegraphics{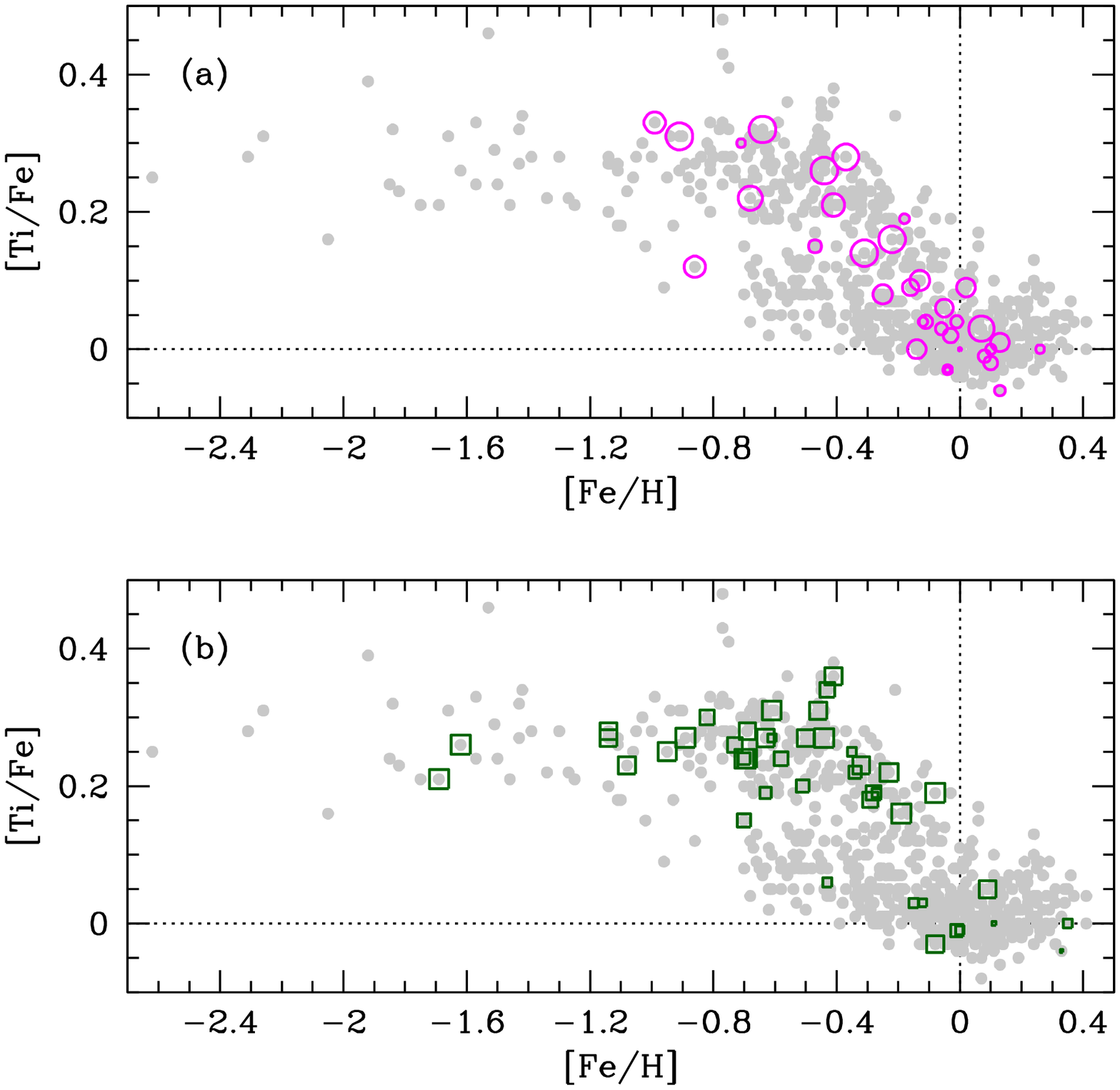}
}
\resizebox{\hsize}{!}{
\includegraphics[bb=18 144 592 430,clip]{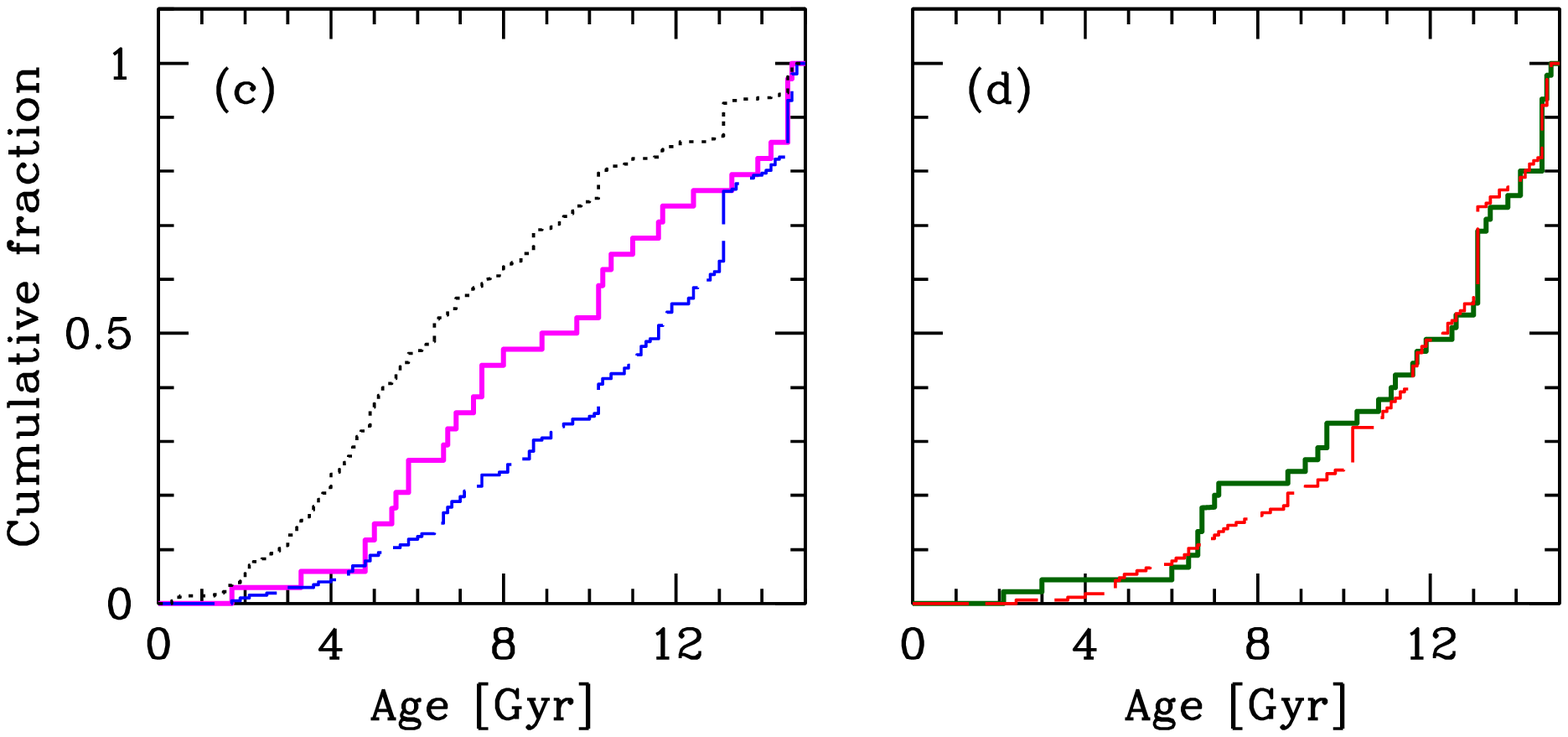}
}
\caption{\label{fig:agehercarc}
(a) and (b) show the [Ti/Fe]-[Fe/H] abundance plots for the
candidate Hercules stream and Arcturus group stars, respectively.
Same markers as in Fig.~\ref{fig:toomre2}. (c) and (d) show the
cumulative age distributions for the candidate Hercules and Arcturus
stars, respectively (coloured as in (a) and (b).
In (c) the dotted black line represents stars with $7<\rmean<9$\,kpc
and the dashed blue line stars with $\rmean<7$\,kpc (compare 
Fig.~\ref{fig:rmeantife}). In (d) the dashed red line represents
stars with $TD/D>10$. 
}
\end{figure}

\subsubsection{The Hercules stream}

One structure of particular interest is the Hercules stream. 
As the Hercules stream has a net velocity drift
away from the Galactic centre, it has been speculated that the Hercules 
stream stars have a dynamical origin in the inner parts of the Galaxy 
where they were kinematically heated by the central bar 
\citep[e.g.,][]{dehnen2000,famaey2005}. In \cite{bensby2007letter}
we found that stars that could kinematically be associated
with the Hercules stream did not have a distinct chemical signature
but showed a mixture of abundances and ages as seen in the thin and 
thick disks. 
As our analysis has been updated since the 2007 paper 
(see Sect.~\ref{sec:analysis}), and also since the comparison in \cite{bensby2007letter}
was made to the smaller thin and thick disk sample of 102 stars in \cite{bensby2005}
we here reproduce the abundance plot comparison in Fig.~\ref{fig:agehercarc}a.
Also, while we in \cite{bensby2007letter} chose to include all stars with
kinematical probabilities $Herc/TD$ and $Herc/D$ to be greater than 1 (one), we
have here required  that these probability ratios should be at least 2.

Our candidate Hercules stream stars are marked in 
Fig.~\ref{fig:toomre2}a by magenta circles, and also in the corresponding
abundance plot in Fig.~\ref{fig:agehercarc}a.
We see that the abundance trend resembles very much the trend
we see for the inner disk stars in Fig.~\ref{fig:rmeantife}a, i.e.,
there are $\alpha$-enhanced and old stars, as well as 
younger and less $\alpha$-enhanced around solar [Fe/H]. 
There is also a lack of low-$\alpha$ stars at lower [Fe/H], in the same way 
as for the inner disk sample.
Figure~\ref{fig:agehercarc}c shows a comparison between the age distributions
of the candidate Hercules stream stars and the ``inner disk sample'' that 
have $\rmean<7$\,kpc (dashed blue line) as well as the stars in our 
sample that have $7<\rmean<9$\,kpc (dotted black line). It is clear that the 
it does not fit either of them, but lies somewhere in between.
As the Hercules stream stars have a net velocity component directed
radially outwards from the Galactic centre, the resemblance  is 
consistent with origins at slightly smaller Galactocentric radii.

\subsubsection{The Arcturus moving group}

\cite{gilmore2002}, and later \cite{wyse2006},
identified a group of stars
lagging behind the local standard of rest (LSR) by $\sim$100\,$\kms$, 
that they claim to be associated with a disrupted satellite that merged 
with the Milky Way 10-12\,Gyr ago. \cite{navarro2004} suggest that these 
stars are the same group of stars that \cite{eggen1971} associated with 
the bright star Arcturus, whose Galactic orbital velocity also lags behind the 
LSR by $\sim$100\,$\kms$.

Both \cite{williams2009} and \cite{ramya2012} observed candidate members
of the Arcturus moving group but could find no clear chemical signature
of those stars compared to the abundance pattern seen in the Solar
neighbourhood, and conclude that it most probably owes its origin to
dynamical perturbations within the Galaxy.

The green squares in Fig.~\ref{fig:toomre2}a with 
$-115<\vlsr<-85\,\kms$ mark the stars in our sample that could
be associated with the Arcturus moving group. The corresponding
abundance plot is shown in
Fig.~\ref{fig:agehercarc}b. We find that a majority of the stars
are $\alpha$-enhanced and older than 10\,Gyr, and a few
have younger ages and are less $\alpha$-enhanced. 
The general appearance is very similar to what wee see in the thick 
disk. The similarities to the thick disk is further
demonstrated in Fig.~\ref{fig:agehercarc}d that compares the age distribution
of the candidate Arcturus group stars (solid green line) with stars that are 
likely thick disk stars ($TD/D>10$, dashed red line). The two distributions
almost identical. A two-sided $KS$-test gives $D=0.09$ and $p=0.91$, showing
that the hypothesis that the two age distributions are the same cannot
be rejected at the 91\,\% level.

The extended range in [Fe/H] and level of $\alpha$-enhancement in the
candidate Arcturus group stars is very different to what is 
seen in dwarf galaxies \citep{venn2004}. Instead these properties,
as well as stellar ages, are very similar to what is seen in the thick disk.
Hence, we conclude that there is no clear signature of an extra-galactic origin,
nor disrupted cluster, for the Arcturus group, confirming
the conclusions by \citealt{williams2009} and \citealt{ramya2012}).
The structure in velocity space associated with the Arcturus group
is more likely a dynamical feature caused within the Milky Way.
Actually, \cite{gardner2010}, and more recently \cite{monari2013}, showed that 
the Galactic long bar produces a kinematic feature in 
velocity space with the same parameters as those of the Arcturus 
moving group. In contrast, the short bar is believed to be responsible 
for the kinematic feature associated with the Hercules stream, assuming
that there are, in fact, two bars.

\section{Summary}

We have conducted a homogeneous detailed elemental abundance study
of 714 F and G dwarf stars in the Solar neighbourhood. The stars in the
sample were selected on basis of their kinematics and metallicities
to trace the age and abundance structures of the Galactic thin and thick 
disks, the stellar halo, as well as kinematic groups such as the 
Hercules stream and the 
Arcturus moving group. Hence, the selection function is very complex 
and the sample should not be used to infer various parameter distributions of 
the two disks. It is, however, well-suited to probe the properties of 
kinematic sub-structures in the Galactic disk and the extremes of the 
thin and thick disks.

The analysis is based on equivalent width measurements in high-resolution
and high signal-to-noise spectra, and 1-D, LTE, plane-parallel 
MARCS model stellar atmospheres. Stellar
parameters and abundances were determined following a purely 
spectroscopic approach, i.e., surface gravity from ionisation balance
between \ion{Fe}{i} and \ion{Fe}{ii}, effective temperature from
excitation balance of \ion{Fe}{i}, and microturbulence from the balance
of \ion{Fe}{i} with reduced line strength. All \ion{Fe}{i} abundances
were corrected for NLTE effects on a line-by-line basis in every step
of the analysis. We note that the excitation and ionisation balance method
appears to fail for stars on the lower main sequence ($\teff\lesssim5600$\,K
and $\log g\gtrsim4.2$), producing an erroneously horizontal main sequence
that is not seen if $\log g$ is determined from Hipparcos parallaxes.
As we wanted to keep the analysis strictly spectroscopic, and 
distance-independent, we applied an empirical correction that was derived
through comparisons to stars that have small uncertainties in their
parallaxes.

In summary, our main findings and conclusions are:
\begin{enumerate}

\item The Solar neighbourhood appears to contain two stellar populations
that have distinct, elemental abundance trends with a gap in 
the $\rm [\alpha/Fe]-[Fe/H]$
plane for metallicities between $\rm -0.7<[Fe/H]<-0.2$. This gap becomes
more prominent if stars that are more susceptible to uncertainties 
($\teff<5400$\,K) are discarded. 

\item The $\alpha$-enhanced population is old and reaches at least
solar metallicities, if not higher. It also shows an age-metallicity
relation, from $\sim$10\,Gyr below $\rm [Fe/H]<-0.4$ to
around 8\,Gyr at $\rm [Fe/H]\approx 0$.

\item The $\alpha$-poor population has a lower metallicity limit
around $\rm [Fe/H]\approx -0.7$. It does not show an age-metallicity
relation, but a wide spread in ages (between 2-7\,Gyr) over the 
whole metallicity range.

\item The $\alpha$-enhanced and metal-rich stars
around solar [Fe/H], claimed by \cite{adibekyan2011} to possibly
be a unique population of its own,
cannot be resolved as a unique
population in our data set. Instead we find that it most likely 
is the metal-rich extension of the thick disk. The status of
the HAMR stars should be further explored with a sample that has
a controlled and well-defined selection function.

\item A majority of the stars that are old and $\alpha$-enhanced
have Galactic orbits with $\rmean<7$\,kpc showing that their
birthplaces are located in the Galactic inner disk, significantly
closer to the Galactic centre than the Sun. The stars with $\rmean>9$\,kpc 
on the other hand, are essentially all young and less $\alpha$-enhanced. 
This finding is consistent with a short scale-length for the thick disk 
that was proposed by \cite{bensby2011letter} and later verified by 
\cite{cheng2012_2} using the Segue G dwarf sample. 

\item Our solar abundances compare within $\pm 0.05$\,dex with those
of nearby and young (thin disk) stars in a narrow metallicity range 
around $\rm [Fe/H]= 0$.
Hence, we cannot claim that the Sun's abundances deviate significantly from
those of other nearby disk stars.

\item Kinematical criteria to select thin and thick disk stars 
are significantly biased. Ages appear
to be a better discriminator, but as ages with small error bars
are notoriously difficult to determine, age criteria also
yield samples with overlap between the two populations, 
although somewhat less than when using kinematical criteria.

\item The Hercules stream does not show
any distinct and/or homogeneous age or abundance patterns. Hence we
confirm previous findings that it
most likely is a feature in velocity space produced by dynamical
interactions with the Galactic (short?) bar.

\item The candidate stars of the Arcturus moving group shows an
abundance pattern that very much resembles what we see in the thick disk,
confirming the conclusion of \cite{ramya2012} that it most likely
is not a disrupted cluster. In addition we show that also the age distribution
is very similar to what is seen in the thick disk. Although an extra-galactic
origin cannot be excluded, our results most likely
points to that also the Arcturus group is a dynamical feature in 
velocity space produced by the Galactic (long?) bar. 

\item
We further find that the standard 1-D, LTE, analysis where surface gravity
is based on ionisation balance of abundances from \ion{Fe}{i} and \ion{Fe}{ii}
lines, and effective temperature is based on
excitation balance of abundances from \ion{Fe}{i} lines, produces an
HR-diagram with where the lower main sequence stars line up horizontally
rather than showing the common steady increase in $\log g$ with decreasing 
$\teff$ on the lower main sequence. We find that this most
likely is an artefact introduced by forcing ionisation balance for these
stars. Where the problem lies, if it is NLTE, 3-D, or both, or something else,
lies beyond the scope of the current paper. Instead we apply an empirical correction
based on a comparison to stellar parameters where $\log g$ is determined 
from stars that have accurate Hipparcos parallaxes with uncertainties less 
than 5\,\%. From that comparison we see that stars in the turn-off,
subgiant, and giant regions of the HR diagram does not seem to be affected.
We also note that the "flat main sequence syndrome" appears to be present in several
other studies that utilise ionisation balance \citep[e.g.,][]{adibekyan2012}, 
while other studies that utilise other methods (Hipparcos parallaxes)
appears not to be hampered by the "flat main sequence syndrome" 
\citep[e.g.,][]{reddy2003,reddy2006}. 

That $\log g$ cannot be readily determined from ionisation balance for lower
main sequence stars may be a problem for studies of distant
stars where accurate distances are rarely available. 
Ages and spectroscopic distances will be severely affected for such stars.
On the other side, studies of distant stars usually utilise turn-off or more 
evolved stars, for which $\log g$:s from ionisation balance appears to be on 
par with $\log g$:s from stars with accurate Hipparcos parallaxes.
\\

\end{enumerate}

The above findings show that the Milky Way indeed appears to have dual 
stellar populations that are chemically distinct, as well as separated
in age. While the Galactic thick disk is more centrally
concentrated than previously thought, the thin disk is the clearly dominant
population in the outer disk, even at large distances from the Galactic plane.
The epoch where we see a separation between the two disks, around 8\,Gyr
ago, coincides with other observational evidence for mergers between the 
Milky Way and another, dwarf galaxy. For instance, \cite{gilmore2002}
and \cite{wyse2006}
claim to have detected debris stars from a major merger $\sim$10\,Gyr
ago, and \cite{deason2013} find that the density profile of the Milky Way
halo is discontinuous, and that this break likely is associated with an early 
(6-9\,Gyr ago) and massive accretion event. 

While this paper has presented the stellar sample, the observations, analysis,
and results, we are 
delving into greater detail on the dichotomy of the Milky Way stellar disk,
and possible formation scenarios for the thick disk (Feltzing et al., 2013, in preparation).

\begin{acknowledgement}

We would like to thank Bengt Gustafsson, Bengt 
Edvardsson, Kjell Eriksson, and Martin Asplund for access to the 
MARCS model atmosphere software and their suite of stellar abundance 
programs. We also thank Giovanni Carraro who kindly provided the 
{\sc grinton} integrator to calculate Galactic orbits.
T.B. was funded by grant No. 621-2009-3911 from The Swedish 
Research Council. S.F. was partly funded by grants No. 621-2011-5042,
621-2008-4095, 621-2005-3181, and 621-2002-3611 from 
The Swedish Research Council, and 2005-2009 S.F. was a Royal Swedish Academy 
of Sciences Research Fellow supported by a grant from the Knut and Alice 
Wallenberg Foundation. This work was also supported by the 
National Science Foundation, grant AST-0448900 to M.S.O. 
This research has made use of the SIMBAD database,
operated at CDS, Strasbourg, France.

\end{acknowledgement}

\bibliographystyle{aa}
\bibliography{referenser}

\begin{thebibliography}{130}
\expandafter\ifx\csname natexlab\endcsname\relax\def\natexlab#1{#1}\fi

\bibitem[{{Abazajian} {et~al.}(2009){Abazajian}, {Adelman-McCarthy},
  {Ag{\"u}eros}, {Allam}, {Allende Prieto}, {An}, {Anderson}, {Anderson},
  {Annis}, {Bahcall}, \& et~al.}]{abazajian2009}
{Abazajian}, K.~N., {Adelman-McCarthy}, J.~K., {Ag{\"u}eros}, M.~A., {et~al.}
  2009, \apjs, 182, 543

\bibitem[{{Adibekyan} {et~al.}(2011){Adibekyan}, {Santos}, {Sousa}, \&
  {Israelian}}]{adibekyan2011}
{Adibekyan}, V.~Z., {Santos}, N.~C., {Sousa}, S.~G., \& {Israelian}, G. 2011,
  \aap, 535, L11

\bibitem[{{Adibekyan} {et~al.}(2012){Adibekyan}, {Sousa}, {Santos}, {Delgado
  Mena}, {Gonz{\'a}lez Hern{\'a}ndez}, {Israelian}, {Mayor}, \&
  {Khachatryan}}]{adibekyan2012}
{Adibekyan}, V.~Z., {Sousa}, S.~G., {Santos}, N.~C., {et~al.} 2012, \aap, 545,
  A32

\bibitem[{{Allen} \& {Santillan}(1991)}]{allen1991b}
{Allen}, C. \& {Santillan}, A. 1991, Revista Mexicana de Astronomia y
  Astrofisica, 22, 255

\bibitem[{{Allende Prieto} {et~al.}(2008){Allende Prieto}, {Majewski},
  {Schiavon}, {Cunha}, {Frinchaboy}, {Holtzman}, {Johnston}, {Shetrone},
  {Skrutskie}, {Smith}, \& {Wilson}}]{allendeprieto2008}
{Allende Prieto}, C., {Majewski}, S.~R., {Schiavon}, R., {et~al.} 2008,
  Astronomische Nachrichten, 329, 1018

\bibitem[{{Antoja} {et~al.}(2012){Antoja}, {Helmi}, {Bienayme},
  {Bland-Hawthorn}, {Famaey}, {Freeman}, {Gibson}, {Gilmore}, {Grebel},
  {Minchev}, {Munari}, {Navarro}, {Parker}, {Reid}, {Seabroke}, {Siebert},
  {Siviero}, {Steinmetz}, {Williams}, {Wyse}, \& {Zwitter}}]{antoja2012}
{Antoja}, T., {Helmi}, A., {Bienayme}, O., {et~al.} 2012, \mnras, 426, L1

\bibitem[{{Arifyanto} \& {Fuchs}(2006)}]{arifyanto2006}
{Arifyanto}, M.~I. \& {Fuchs}, B. 2006, \aap, 449, 533

\bibitem[{{Asplund} {et~al.}(2005){Asplund}, {Grevesse}, \&
  {Sauval}}]{asplundgrevessesauval2005}
{Asplund}, M., {Grevesse}, N., \& {Sauval}, A.~J. 2005, in ASP Conf. Ser. 336:
  Cosmic Abundances as Records of Stellar Evolution and Nucleosynthesis, 25

\bibitem[{{Asplund} {et~al.}(2009){Asplund}, {Grevesse}, {Sauval}, \&
  {Scott}}]{asplund2009}
{Asplund}, M., {Grevesse}, N., {Sauval}, A.~J., \& {Scott}, P. 2009, \araa, 47,
  481

\bibitem[{{Asplund} {et~al.}(1997){Asplund}, {Gustafsson}, {Kiselman}, \&
  {Eriksson}}]{asplund1997}
{Asplund}, M., {Gustafsson}, B., {Kiselman}, D., \& {Eriksson}, K. 1997, \aap,
  318, 521

\bibitem[{{Bagnulo} {et~al.}(2003){Bagnulo}, {Jehin}, {Ledoux}, {Cabanac},
  {Melo}, {Gilmozzi}, \& {The ESO Paranal Science Operations Team}}]{bagnulo}
{Bagnulo}, S., {Jehin}, E., {Ledoux}, C., {et~al.} 2003, The Messenger, 114, 10

\bibitem[{{Barbier-Brossat} \& {Figon}(2000)}]{barbierbrossat2000}
{Barbier-Brossat}, M. \& {Figon}, P. 2000, \aaps, 142, 217

\bibitem[{{Barbier-Brossat} {et~al.}(1994){Barbier-Brossat}, {Petit}, \&
  {Figon}}]{barbierbrossat1994}
{Barbier-Brossat}, M., {Petit}, M., \& {Figon}, P. 1994, \aaps, 108, 603

\bibitem[{{Barklem} \& {Aspelund-Johansson}(2005)}]{barklem2005}
{Barklem}, P.~S. \& {Aspelund-Johansson}, J. 2005, \aap, 435, 373

\bibitem[{{Barklem} \& {O'Mara}(2001)}]{barklem2001}
{Barklem}, P.~S. \& {O'Mara}, B.~J. 2001, Journal of Physics B Atomic Molecular
  Physics, 34, 4785

\bibitem[{{Bedin} {et~al.}(2006){Bedin}, {Piotto}, {Carraro}, {King}, \&
  {Anderson}}]{bedin2006}
{Bedin}, L.~R., {Piotto}, G., {Carraro}, G., {King}, I.~R., \& {Anderson}, J.
  2006, \aap, 460, L27

\bibitem[{{Bensby} {et~al.}(2011{\natexlab{a}}){Bensby}, {Ad{\'e}n},
  {Mel{\'e}ndez}, {Gould}, {Feltzing}, {Asplund}, {Johnson}, {Lucatello},
  {Yee}, {Ram{\'{\i}}rez}, {Cohen}, {Thompson}, {Bond}, {Gal-Yam}, {Han},
  {Sumi}, {Suzuki}, {Wada}, {Miyake}, {Furusawa}, {Ohmori}, {Saito},
  {Tristram}, \& {Bennett}}]{bensby2011}
{Bensby}, T., {Ad{\'e}n}, D., {Mel{\'e}ndez}, J., {et~al.} 2011{\natexlab{a}},
  \aap, 533, A134

\bibitem[{{Bensby} {et~al.}(2010){Bensby}, {Alves-Brito}, {Oey}, {Yong}, \&
  {Mel{\'e}ndez}}]{bensby2010letter}
{Bensby}, T., {Alves-Brito}, A., {Oey}, M.~S., {Yong}, D., \& {Mel{\'e}ndez},
  J. 2010, \aap, 516, L13

\bibitem[{{Bensby} {et~al.}(2011{\natexlab{b}}){Bensby}, {Alves-Brito}, {Oey},
  {Yong}, \& {Mel{\'e}ndez}}]{bensby2011letter}
{Bensby}, T., {Alves-Brito}, A., {Oey}, M.~S., {Yong}, D., \& {Mel{\'e}ndez},
  J. 2011{\natexlab{b}}, \apjl, 735, L46

\bibitem[{{Bensby} \& {Feltzing}(2006)}]{bensby2006}
{Bensby}, T. \& {Feltzing}, S. 2006, \mnras, 367, 1181

\bibitem[{{Bensby} \& {Feltzing}(2010)}]{bensby2010rio}
{Bensby}, T. \& {Feltzing}, S. 2010, in IAU Symposium, Vol. 265, IAU Symposium,
  ed. {K.~Cunha, M.~Spite, \& B.~Barbuy}, 300--303

\bibitem[{{Bensby} {et~al.}(2003){Bensby}, {Feltzing}, \& {Lundstr{\"
  o}m}}]{bensby2003}
{Bensby}, T., {Feltzing}, S., \& {Lundstr{\" o}m}, I. 2003, \aap, 410, 527

\bibitem[{{Bensby} {et~al.}(2004{\natexlab{a}}){Bensby}, {Feltzing}, \&
  {Lundstr{\" o}m}}]{bensby_amr}
{Bensby}, T., {Feltzing}, S., \& {Lundstr{\" o}m}, I. 2004{\natexlab{a}}, \aap,
  421, 969

\bibitem[{{Bensby} {et~al.}(2004{\natexlab{b}}){Bensby}, {Feltzing}, \&
  {Lundstr{\" o}m}}]{bensby2004}
{Bensby}, T., {Feltzing}, S., \& {Lundstr{\" o}m}, I. 2004{\natexlab{b}}, \aap,
  415, 155

\bibitem[{{Bensby} {et~al.}(2005){Bensby}, {Feltzing}, {Lundstr{\" o}m}, \&
  {Ilyin}}]{bensby2005}
{Bensby}, T., {Feltzing}, S., {Lundstr{\" o}m}, I., \& {Ilyin}, I. 2005, \aap,
  433, 185

\bibitem[{{Bensby} {et~al.}(2007{\natexlab{a}}){Bensby}, {Oey}, {Feltzing}, \&
  {Gustafsson}}]{bensby2007letter}
{Bensby}, T., {Oey}, M.~S., {Feltzing}, S., \& {Gustafsson}, B.
  2007{\natexlab{a}}, \apjl, 655, L89

\bibitem[{{Bensby} {et~al.}(2013){Bensby}, {Yee}, {Feltzing}, {Johnson},
  {Gould}, {Cohen}, {Asplund}, {Mel{\'e}ndez}, {Lucatello}, {Han}, {Thompson},
  {Gal-Yam}, {Udalski}, {Bennett}, {Bond}, {Kohei}, {Sumi}, {Suzuki}, {Suzuki},
  {Takino}, {Tristram}, {Yamai}, \& {Yonehara}}]{bensby2013}
{Bensby}, T., {Yee}, J.~C., {Feltzing}, S., {et~al.} 2013, \aap, 549, A147

\bibitem[{{Bensby} {et~al.}(2007{\natexlab{b}}){Bensby}, {Zenn}, {Oey}, \&
  {Feltzing}}]{bensby2007letter2}
{Bensby}, T., {Zenn}, A.~R., {Oey}, M.~S., \& {Feltzing}, S.
  2007{\natexlab{b}}, \apjl, 663, L13

\bibitem[{{Bernstein} {et~al.}(2003){Bernstein}, {Shectman}, {Gunnels},
  {Mochnacki}, \& {Athey}}]{bernstein2003}
{Bernstein}, R., {Shectman}, S.~A., {Gunnels}, S.~M., {Mochnacki}, S., \&
  {Athey}, A.~E. 2003, in Proceedings of the SPIE, Volume 4841, ed. M.~{Iye} \&
  A.~F.~M. {Moorwood}, 1694--1704

\bibitem[{{Binney}(2010)}]{binney2010}
{Binney}, J. 2010, \mnras, 401, 2318

\bibitem[{{Binney}(2012)}]{binney2012}
{Binney}, J. 2012, \mnras, 426, 1328

\bibitem[{{Bovy} {et~al.}(2012){Bovy}, {Rix}, \& {Hogg}}]{bovy2012}
{Bovy}, J., {Rix}, H.-W., \& {Hogg}, D.~W. 2012, \apj, 751, 131

\bibitem[{{Carraro} {et~al.}(2002){Carraro}, {Girardi}, \&
  {Marigo}}]{carraro2002}
{Carraro}, G., {Girardi}, L., \& {Marigo}, P. 2002, \mnras, 332, 705

\bibitem[{{Casagrande} {et~al.}(2010){Casagrande}, {Ram{\'{\i}}rez},
  {Mel{\'e}ndez}, {Bessell}, \& {Asplund}}]{casagrande2010}
{Casagrande}, L., {Ram{\'{\i}}rez}, I., {Mel{\'e}ndez}, J., {Bessell}, M., \&
  {Asplund}, M. 2010, \aap, 512, A54

\bibitem[{{Casagrande} {et~al.}(2011){Casagrande}, {Sch{\"o}nrich}, {Asplund},
  {Cassisi}, {Ramirez}, {Melendez}, {Bensby}, \& {Feltzing}}]{casagrande2011}
{Casagrande}, L., {Sch{\"o}nrich}, R., {Asplund}, M., {et~al.} 2011, \aap, 530,
  A138

\bibitem[{{Chen} {et~al.}(2000){Chen}, {Nissen}, {Zhao}, {Zhang}, \&
  {Benoni}}]{chen2000}
{Chen}, Y.~Q., {Nissen}, P.~E., {Zhao}, G., {Zhang}, H.~W., \& {Benoni}, T.
  2000, \aaps, 141, 491

\bibitem[{{Cheng} {et~al.}(2012){Cheng}, {Rockosi}, {Morrison}, {Lee}, {Beers},
  {Bizyaev}, {Harding}, {Malanushenko}, {Malanushenko}, {Oravetz}, {Pan},
  {Schlesinger}, {Schneider}, {Simmons}, \& {Weaver}}]{cheng2012_2}
{Cheng}, J.~Y., {Rockosi}, C.~M., {Morrison}, H.~L., {et~al.} 2012, \apj, 752,
  51

\bibitem[{{Comer{\'o}n} {et~al.}(2011){Comer{\'o}n}, {Elmegreen}, {Knapen},
  {Salo}, {Laurikainen}, {Laine}, {Athanassoula}, {Bosma}, {Sheth}, {Regan},
  {Hinz}, {Gil de Paz}, {Men{\'e}ndez-Delmestre}, {Mizusawa},
  {Mu{\~n}oz-Mateos}, {Seibert}, {Kim}, {Elmegreen}, {Gadotti}, {Ho},
  {Holwerda}, {Lappalainen}, {Schinnerer}, \& {Skibba}}]{comeron2011}
{Comer{\'o}n}, S., {Elmegreen}, B.~G., {Knapen}, J.~H., {et~al.} 2011, \apj,
  741, 28

\bibitem[{{Deason} {et~al.}(2013){Deason}, {Belokurov}, {Evans}, \&
  {Johnston}}]{deason2013}
{Deason}, A.~J., {Belokurov}, V., {Evans}, N.~W., \& {Johnston}, K.~V. 2013,
  \apj, 763, 113

\bibitem[{{Dehnen}(2000)}]{dehnen2000}
{Dehnen}, W. 2000, \aj, 119, 800

\bibitem[{{Dekker} {et~al.}(2000){Dekker}, {D'Odorico}, {Kaufer}, {Delabre}, \&
  {Kotzlowski}}]{dekker2000}
{Dekker}, H., {D'Odorico}, S., {Kaufer}, A., {Delabre}, B., \& {Kotzlowski}, H.
  2000, in Proc. SPIE Vol. 4008, p. 534-545, Optical and IR Telescope
  Instrumentation and Detectors, Masanori Iye; Alan F. Moorwood; Eds., ed.
  M.~{Iye} \& A.~F. {Moorwood}, 534--545

\bibitem[{{Demarque} {et~al.}(2004){Demarque}, {Woo}, {Kim}, \&
  {Yi}}]{demarque2004}
{Demarque}, P., {Woo}, J.-H., {Kim}, Y.-C., \& {Yi}, S.~K. 2004, \apjs, 155,
  667

\bibitem[{{Edvardsson} {et~al.}(1993){Edvardsson}, {Andersen}, {Gustafsson},
  {Lambert}, {Nissen}, \& {Tomkin}}]{edvardsson1993}
{Edvardsson}, B., {Andersen}, J., {Gustafsson}, B., {et~al.} 1993, \aap, 275,
  101

\bibitem[{{Eggen}(1971)}]{eggen1971}
{Eggen}, O.~J. 1971, \pasp, 83, 271

\bibitem[{{Epstein} {et~al.}(2010){Epstein}, {Johnson}, {Dong}, {Udalski},
  {Gould}, \& {Becker}}]{epstein2010}
{Epstein}, C.~R., {Johnson}, J.~A., {Dong}, S., {et~al.} 2010, \apj, 709, 447

\bibitem[{{ESA}(1997)}]{esa1997}
{ESA}. 1997, {The HIPPARCOS and TYCHO catalogues, ESA SP Series vol no: 1200,
  Noordwijk, Netherlands}

\bibitem[{{Famaey} {et~al.}(2005){Famaey}, {Jorissen}, {Luri}, {Mayor}, {Udry},
  {Dejonghe}, \& {Turon}}]{famaey2005}
{Famaey}, B., {Jorissen}, A., {Luri}, X., {et~al.} 2005, \aap, 430, 165

\bibitem[{{Feltzing} \& {Bensby}(2008)}]{feltzing2008uppsala}
{Feltzing}, S. \& {Bensby}, T. 2008, Physica Scripta Volume T, 133, 014031

\bibitem[{{Feltzing} {et~al.}(2007){Feltzing}, {Fohlman}, \&
  {Bensby}}]{feltzing2007}
{Feltzing}, S., {Fohlman}, M., \& {Bensby}, T. 2007, \aap, 467, 665

\bibitem[{{Feltzing} \& {Gonzalez}(2001)}]{feltzing2001b}
{Feltzing}, S. \& {Gonzalez}, G. 2001, \aap, 367, 253

\bibitem[{{Feltzing} \& {Gustafsson}(1998)}]{feltzing1998}
{Feltzing}, S. \& {Gustafsson}, B. 1998, \aaps, 129, 237

\bibitem[{{Feltzing} {et~al.}(2001){Feltzing}, {Holmberg}, \&
  {Hurley}}]{feltzing2001}
{Feltzing}, S., {Holmberg}, J., \& {Hurley}, J.~R. 2001, \aap, 377, 911

\bibitem[{{Freeman} \& {Bland-Hawthorn}(2002)}]{freeman2002}
{Freeman}, K. \& {Bland-Hawthorn}, J. 2002, \araa, 40, 487

\bibitem[{{Fuhrmann}(1998)}]{fuhrmann1998}
{Fuhrmann}, K. 1998, \aap, 338, 161

\bibitem[{{Fuhrmann}(2000)}]{fuhrmann2000unpubl}
{Fuhrmann}, K. 2000, unpublished

\bibitem[{{Fuhrmann}(2004)}]{fuhrmann2004}
{Fuhrmann}, K. 2004, Astronomische Nachrichten, 325, 3

\bibitem[{{Fuhrmann}(2008)}]{fuhrmann2008}
{Fuhrmann}, K. 2008, \mnras, 384, 173

\bibitem[{{Fuhrmann}(2011)}]{fuhrmann2011}
{Fuhrmann}, K. 2011, \mnras, 414, 2893

\bibitem[{{Gardner} \& {Flynn}(2010)}]{gardner2010}
{Gardner}, E. \& {Flynn}, C. 2010, \mnras, 405, 545

\bibitem[{{Gilmore} {et~al.}(2012){Gilmore}, {Randich}, {Asplund}, {Binney},
  {Bonifacio}, {Drew}, {Feltzing}, {Ferguson}, {Jeffries}, {Micela},
  {Negueruela}, {Prusti}, {Rix}, {Vallenari}, {Alfaro}, {Allende-Prieto},
  {Babusiaux}, {Bensby}, {Blomme}, {Bragaglia}, {Flaccomio}, {Francois},
  {Irwin}, {Koposov}, {Korn}, {Lanzafame}, {Pancino}, {Paunzen},
  {Recio-Blanco}, {Sacco}, {Smiljanic}, {van Eck}, \& {Walton}}]{gilmore2012}
{Gilmore}, G., {Randich}, S., {Asplund}, M., {et~al.} 2012, The Messenger, 147,
  25

\bibitem[{{Gilmore} {et~al.}(2002){Gilmore}, {Wyse}, \& {Norris}}]{gilmore2002}
{Gilmore}, G., {Wyse}, R.~F.~G., \& {Norris}, J.~E. 2002, \apjl, 574, L39

\bibitem[{{Gratton} {et~al.}(2000){Gratton}, {Carretta}, {Matteucci}, \&
  {Sneden}}]{gratton2000}
{Gratton}, R.~G., {Carretta}, E., {Matteucci}, F., \& {Sneden}, C. 2000, \aap,
  358, 671

\bibitem[{{Grenon}(1987)}]{grenon1987}
{Grenon}, M. 1987, Journal of Astrophysics and Astronomy, 8, 123

\bibitem[{{Gustafsson} {et~al.}(1975){Gustafsson}, {Bell}, {Eriksson}, \&
  {Nordlund}}]{gustafsson1975}
{Gustafsson}, B., {Bell}, R.~A., {Eriksson}, K., \& {Nordlund}, A. 1975, \aap,
  42, 407

\bibitem[{{Gustafsson} {et~al.}(2008){Gustafsson}, {Edvardsson}, {Eriksson},
  {J{\o}rgensen}, {Nordlund}, \& {Plez}}]{gustafsson2008}
{Gustafsson}, B., {Edvardsson}, B., {Eriksson}, K., {et~al.} 2008, \aap, 486,
  951

\bibitem[{{Haywood}(2006)}]{haywood2006}
{Haywood}, M. 2006, \mnras, 371, 1760

\bibitem[{{Haywood} {et~al.}(2013){Haywood}, {Di Matteo}, {Lehnert}, {Katz}, \&
  {Gomez}}]{haywood2013}
{Haywood}, M., {Di Matteo}, P., {Lehnert}, M., {Katz}, D., \& {Gomez}, A. 2013,
  arXiv:1305.4663 [astro-ph.GA]

\bibitem[{{Helmi} {et~al.}(2006){Helmi}, {Navarro}, {Nordstr{\"o}m},
  {Holmberg}, {Abadi}, \& {Steinmetz}}]{helmi2006}
{Helmi}, A., {Navarro}, J.~F., {Nordstr{\"o}m}, B., {et~al.} 2006, \mnras, 365,
  1309

\bibitem[{{H{\o}g} {et~al.}(2000){H{\o}g}, {Fabricius}, {Makarov}, {Urban},
  {Corbin}, {Wycoff}, {Bastian}, {Schwekendiek}, \& {Wicenec}}]{hoeg2000}
{H{\o}g}, E., {Fabricius}, C., {Makarov}, V.~V., {et~al.} 2000, \aap, 355, L27

\bibitem[{{Humphreys} \& {Larsen}(1995)}]{humphreys1995}
{Humphreys}, R.~M. \& {Larsen}, J.~A. 1995, \aj, 110, 2183

\bibitem[{{Ilyin}(2000)}]{ilyin2000}
{Ilyin}, I.~V. 2000, Ph.D.~Thesis, University of Oulu

\bibitem[{{Joshi}(2007)}]{joshi2007}
{Joshi}, Y.~C. 2007, \mnras, 378, 768

\bibitem[{{Kaufer} {et~al.}(1999){Kaufer}, {Stahl}, {Tubbesing}, {Norregaard},
  {Avila}, {Francois}, {Pasquini}, \& {Pizzella}}]{kaufer1999}
{Kaufer}, A., {Stahl}, O., {Tubbesing}, S., {et~al.} 1999, The Messenger, 95, 8

\bibitem[{{Kiselman}(1993)}]{kiselman1993}
{Kiselman}, D. 1993, \aap, 275, 269

\bibitem[{{Kobayashi} {et~al.}(2006){Kobayashi}, {Umeda}, {Nomoto}, {Tominaga},
  \& {Ohkubo}}]{kobayashi2006}
{Kobayashi}, C., {Umeda}, H., {Nomoto}, K., {Tominaga}, N., \& {Ohkubo}, T.
  2006, \apj, 653, 1145

\bibitem[{{Korotin} {et~al.}(2011){Korotin}, {Mishenina}, {Gorbaneva}, \&
  {Soubiran}}]{korotin2011}
{Korotin}, S., {Mishenina}, T., {Gorbaneva}, T., \& {Soubiran}, C. 2011,
  \mnras, 415, 2093

\bibitem[{{Kupka} {et~al.}(1999){Kupka}, {Piskunov}, {Ryabchikova}, {Stempels},
  \& {Weiss}}]{vald_3}
{Kupka}, F., {Piskunov}, N., {Ryabchikova}, T.~A., {Stempels}, H.~C., \&
  {Weiss}, W.~W. 1999, \aaps, 138, 119

\bibitem[{{Lambert}(1989)}]{lambert1989}
{Lambert}, D.~L. 1989, in American Institute of Physics Conference Series, Vol.
  183, Cosmic Abundances of Matter, ed. C.~J. {Waddington}, 168--199

\bibitem[{{Lee} {et~al.}(2011){Lee}, {Beers}, {An}, {Ivezi{\'c}}, {Just},
  {Rockosi}, {Morrison}, {Johnson}, {Sch{\"o}nrich}, {Bird}, {Yanny},
  {Harding}, \& {Rocha-Pinto}}]{lee2011}
{Lee}, Y.~S., {Beers}, T.~C., {An}, D., {et~al.} 2011, \apj, 738, 187

\bibitem[{{Lind} {et~al.}(2011){Lind}, {Asplund}, {Barklem}, \&
  {Belyaev}}]{lind2011}
{Lind}, K., {Asplund}, M., {Barklem}, P.~S., \& {Belyaev}, A.~K. 2011, \aap,
  528, A103

\bibitem[{{Lind} {et~al.}(2012){Lind}, {Bergemann}, \& {Asplund}}]{lind2012}
{Lind}, K., {Bergemann}, M., \& {Asplund}, M. 2012, \mnras, 427, 50

\bibitem[{{Lindegren} \& {Feltzing}(2013)}]{lindegren2013}
{Lindegren}, L. \& {Feltzing}, S. 2013, \aap, 553, A94

\bibitem[{{Liu} \& {van de Ven}(2012)}]{liu2012}
{Liu}, C. \& {van de Ven}, G. 2012, \mnras, 425, 2144

\bibitem[{{Mashonkina} \& {Gehren}(2001)}]{mashonkina2001}
{Mashonkina}, L. \& {Gehren}, T. 2001, \aap, 376, 232

\bibitem[{{Mayor} {et~al.}(2003){Mayor}, {Pepe}, {Queloz}, {Bouchy},
  {Rupprecht}, {Lo Curto}, {Avila}, {Benz}, {Bertaux}, {Bonfils}, {Dall},
  {Dekker}, {Delabre}, {Eckert}, {Fleury}, {Gilliotte}, {Gojak}, {Guzman},
  {Kohler}, {Lizon}, {Longinotti}, {Lovis}, {Megevand}, {Pasquini}, {Reyes},
  {Sivan}, {Sosnowska}, {Soto}, {Udry}, {van Kesteren}, {Weber}, \&
  {Weilenmann}}]{mayor2003}
{Mayor}, M., {Pepe}, F., {Queloz}, D., {et~al.} 2003, The Messenger, 114, 20

\bibitem[{{Mel{\'e}ndez} {et~al.}(2009){Mel{\'e}ndez}, {Asplund}, {Gustafsson},
  \& {Yong}}]{melendez2009sun}
{Mel{\'e}ndez}, J., {Asplund}, M., {Gustafsson}, B., \& {Yong}, D. 2009, \apjl,
  704, L66

\bibitem[{{Mel{\'e}ndez} \& {Barbuy}(2009)}]{melendez2009}
{Mel{\'e}ndez}, J. \& {Barbuy}, B. 2009, \aap, 497, 611

\bibitem[{{Mel{\'e}ndez} {et~al.}(2012){Mel{\'e}ndez}, {Bergemann}, {Cohen},
  {Endl}, {Karakas}, {Ram{\'{\i}}rez}, {Cochran}, {Yong}, {MacQueen},
  {Kobayashi}, \& {Asplund}}]{melendez2012}
{Mel{\'e}ndez}, J., {Bergemann}, M., {Cohen}, J.~G., {et~al.} 2012, \aap, 543,
  A29

\bibitem[{{Minchev} {et~al.}(2012){Minchev}, {Famaey}, {Quillen}, {Dehnen},
  {Martig}, \& {Siebert}}]{minchev2012}
{Minchev}, I., {Famaey}, B., {Quillen}, A.~C., {et~al.} 2012, \aap, 548, A127

\bibitem[{{Mishenina} {et~al.}(2004){Mishenina}, {Soubiran}, {Kovtyukh}, \&
  {Korotin}}]{mishenina2004}
{Mishenina}, T.~V., {Soubiran}, C., {Kovtyukh}, V.~V., \& {Korotin}, S.~A.
  2004, \aap, 418, 551

\bibitem[{{Mitschang} {et~al.}(2013){Mitschang}, {De Silva}, {Zucker},
  {Anguiano}, {Bensby}, \& {Feltzing}}]{mitschang2013}
{Mitschang}, A., {De Silva}, G., {Zucker}, D., {et~al.} 2013, arXiv:1312.1759
  [astro-ph.GA]

\bibitem[{{Monari} {et~al.}(2013){Monari}, {Antoja}, \& {Helmi}}]{monari2013}
{Monari}, G., {Antoja}, T., \& {Helmi}, A. 2013, arXiv:1306.2632 [astro-ph.GA]

\bibitem[{{Navarro} {et~al.}(2011){Navarro}, {Abadi}, {Venn}, {Freeman}, \&
  {Anguiano}}]{navarro2011}
{Navarro}, J.~F., {Abadi}, M.~G., {Venn}, K.~A., {Freeman}, K.~C., \&
  {Anguiano}, B. 2011, \mnras, 412, 1203

\bibitem[{{Navarro} {et~al.}(2004){Navarro}, {Helmi}, \&
  {Freeman}}]{navarro2004}
{Navarro}, J.~F., {Helmi}, A., \& {Freeman}, K.~C. 2004, \apjl, 601, L43

\bibitem[{{Nave} {et~al.}(1994){Nave}, {Johansson}, {Learner}, {Thorne}, \&
  {Brault}}]{nave1994}
{Nave}, G., {Johansson}, S., {Learner}, R.~C.~M., {Thorne}, A.~P., \& {Brault},
  J.~W. 1994, \apjs, 94, 221

\bibitem[{{Nissen}(2004)}]{nissen2004}
{Nissen}, P.~E. 2004, in Origin and Evolution of the Elements, Carnegie
  Observatories Astrophysics Series, Vol.~4, (Eds.) A. McWilliam and M. Rauch,
  Pasadena: Carnegie Observatories, 156

\bibitem[{{Nissen} \& {Schuster}(1997)}]{nissen1997}
{Nissen}, P.~E. \& {Schuster}, W.~J. 1997, \aap, 326, 751

\bibitem[{{Nissen} \& {Schuster}(2010)}]{nissen2010}
{Nissen}, P.~E. \& {Schuster}, W.~J. 2010, \aap, 511, L10

\bibitem[{{Nordstr{\" o}m} {et~al.}(2004){Nordstr{\" o}m}, {Mayor}, {Andersen},
  {Holmberg}, {Pont}, {J{\o}rgensen}, {Olsen}, {Udry}, \&
  {Mowlavi}}]{nordstrom2004}
{Nordstr{\" o}m}, B., {Mayor}, M., {Andersen}, J., {et~al.} 2004, \aap, 418,
  989

\bibitem[{{Piskunov} {et~al.}(1995){Piskunov}, {Kupka}, {Ryabchikova}, {Weiss},
  \& {Jeffery}}]{vald_1}
{Piskunov}, N.~E., {Kupka}, F., {Ryabchikova}, T.~A., {Weiss}, W.~W., \&
  {Jeffery}, C.~S. 1995, \aaps, 112, 525

\bibitem[{{Piskunov} \& {Valenti}(2002)}]{piskunov2002}
{Piskunov}, N.~E. \& {Valenti}, J.~A. 2002, \aap, 385, 1095

\bibitem[{{Prochaska} {et~al.}(2000){Prochaska}, {Naumov}, {Carney},
  {McWilliam}, \& {Wolfe}}]{prochaska2000}
{Prochaska}, J.~X., {Naumov}, S.~O., {Carney}, B.~W., {McWilliam}, A., \&
  {Wolfe}, A.~M. 2000, \aj, 120, 2513

\bibitem[{{Ram{\'{\i}}rez} {et~al.}(2013){Ram{\'{\i}}rez}, {Allende Prieto}, \&
  {Lambert}}]{ramirez2013}
{Ram{\'{\i}}rez}, I., {Allende Prieto}, C., \& {Lambert}, D.~L. 2013, \apj,
  764, 78

\bibitem[{{Ram{\'{\i}}rez} {et~al.}(2010){Ram{\'{\i}}rez}, {Asplund},
  {Baumann}, {Mel{\'e}ndez}, \& {Bensby}}]{ramirez2010}
{Ram{\'{\i}}rez}, I., {Asplund}, M., {Baumann}, P., {Mel{\'e}ndez}, J., \&
  {Bensby}, T. 2010, \aap, 521, A33

\bibitem[{{Ram{\'{\i}}rez} {et~al.}(2009){Ram{\'{\i}}rez}, {Mel{\'e}ndez}, \&
  {Asplund}}]{ramirez2009}
{Ram{\'{\i}}rez}, I., {Mel{\'e}ndez}, J., \& {Asplund}, M. 2009, \aap, 508, L17

\bibitem[{{Ramya} {et~al.}(2012){Ramya}, {Reddy}, \& {Lambert}}]{ramya2012}
{Ramya}, P., {Reddy}, B.~E., \& {Lambert}, D.~L. 2012, \mnras, 425, 3188

\bibitem[{{Reddy} {et~al.}(2006){Reddy}, {Lambert}, \& {Allende
  Prieto}}]{reddy2006}
{Reddy}, B.~E., {Lambert}, D.~L., \& {Allende Prieto}, C. 2006, \mnras, 367,
  1329

\bibitem[{{Reddy} {et~al.}(2003){Reddy}, {Tomkin}, {Lambert}, \& {Allende
  Prieto}}]{reddy2003}
{Reddy}, B.~E., {Tomkin}, J., {Lambert}, D.~L., \& {Allende Prieto}, C. 2003,
  \mnras, 340, 304

\bibitem[{{Ruchti} {et~al.}(2010){Ruchti}, {Fulbright}, {Wyse}, {Gilmore},
  {Bienaym{\'e}}, {Binney}, {Bland-Hawthorn}, {Campbell}, {Freeman}, {Gibson},
  {Grebel}, {Helmi}, {Munari}, {Navarro}, {Parker}, {Reid}, {Seabroke},
  {Siebert}, {Siviero}, {Steinmetz}, {Watson}, {Williams}, \&
  {Zwitter}}]{ruchti2010}
{Ruchti}, G.~R., {Fulbright}, J.~P., {Wyse}, R.~F.~G., {et~al.} 2010, \apjl,
  721, L92

\bibitem[{{Ryabchikova} {et~al.}(1999){Ryabchikova}, {Piskunov}, {Stempels},
  {Kupka}, \& {Weiss}}]{vald_2}
{Ryabchikova}, T., {Piskunov}, N., {Stempels}, H., {Kupka}, F., \& {Weiss}, W.
  1999, Physica Scripta, T83, 162

\bibitem[{{Sackmann} {et~al.}(1993){Sackmann}, {Boothroyd}, \&
  {Kraemer}}]{sackmann1993}
{Sackmann}, I.-J., {Boothroyd}, A.~I., \& {Kraemer}, K.~E. 1993, \apj, 418, 457

\bibitem[{{Sch{\"o}nrich} \& {Binney}(2009{\natexlab{a}})}]{schonrich2009a}
{Sch{\"o}nrich}, R. \& {Binney}, J. 2009{\natexlab{a}}, \mnras, 396, 203

\bibitem[{{Sch{\"o}nrich} \& {Binney}(2009{\natexlab{b}})}]{schonrich2009b}
{Sch{\"o}nrich}, R. \& {Binney}, J. 2009{\natexlab{b}}, \mnras, 399, 1145

\bibitem[{{Sch{\"o}nrich} {et~al.}(2010){Sch{\"o}nrich}, {Binney}, \&
  {Dehnen}}]{schonrich2010}
{Sch{\"o}nrich}, R., {Binney}, J., \& {Dehnen}, W. 2010, \mnras, 403, 1829

\bibitem[{{Schuster} {et~al.}(2006){Schuster}, {Moitinho}, {M{\'a}rquez},
  {Parrao}, \& {Covarrubias}}]{schuster2006}
{Schuster}, W.~J., {Moitinho}, A., {M{\'a}rquez}, A., {Parrao}, L., \&
  {Covarrubias}, E. 2006, \aap, 445, 939

\bibitem[{{Sellwood} \& {Binney}(2002)}]{sellwood2002}
{Sellwood}, J.~A. \& {Binney}, J.~J. 2002, \mnras, 336, 785

\bibitem[{{Soderblom}(2010)}]{soderblom2010}
{Soderblom}, D.~R. 2010, \araa, 48, 581

\bibitem[{{Soubiran} {et~al.}(2003){Soubiran}, {Bienaym{\' e}}, \&
  {Siebert}}]{soubiran2003}
{Soubiran}, C., {Bienaym{\' e}}, O., \& {Siebert}, A. 2003, \aap, 398, 141

\bibitem[{{Soubiran} \& {Girard}(2005)}]{soubiran2005}
{Soubiran}, C. \& {Girard}, P. 2005, \aap, 438, 139

\bibitem[{{Tautvai{\v s}ien{\.e}} {et~al.}(2001){Tautvai{\v s}ien{\.e}},
  {Edvardsson}, {Tuominen}, \& {Ilyin}}]{tautvaisiene2001}
{Tautvai{\v s}ien{\.e}}, G., {Edvardsson}, B., {Tuominen}, I., \& {Ilyin}, I.
  2001, \aap, 380, 578

\bibitem[{{Th{\'e}venin} \& {Idiart}(1999)}]{thevenin1999}
{Th{\'e}venin}, F. \& {Idiart}, T.~P. 1999, \apj, 521, 753

\bibitem[{{Trevisan} {et~al.}(2011){Trevisan}, {Barbuy}, {Eriksson},
  {Gustafsson}, {Grenon}, \& {Pomp{\'e}ia}}]{trevisan2011}
{Trevisan}, M., {Barbuy}, B., {Eriksson}, K., {et~al.} 2011, \aap, 535, A42

\bibitem[{{Valenti} \& {Fischer}(2005)}]{valenti2005}
{Valenti}, J.~A. \& {Fischer}, D.~A. 2005, \apjs, 159, 141

\bibitem[{{van Leeuwen}(2007)}]{vanleeuwen2007}
{van Leeuwen}, F. 2007, \aap, 474, 653

\bibitem[{{Venn} {et~al.}(2004){Venn}, {Irwin}, {Shetrone}, {Tout}, {Hill}, \&
  {Tolstoy}}]{venn2004}
{Venn}, K.~A., {Irwin}, M., {Shetrone}, M.~D., {et~al.} 2004, \aj, 128, 1177

\bibitem[{{Williams} {et~al.}(2009){Williams}, {Freeman}, {Helmi}, \& {RAVE
  Collaboration}}]{williams2009}
{Williams}, M.~E.~K., {Freeman}, K.~C., {Helmi}, A., \& {RAVE Collaboration}.
  2009, in IAU Symposium, Vol. 254, IAU Symposium, ed. J.~{Andersen},
  {Nordstr{\"o}ara}, B.~{m}, \& J.~{Bland-Hawthorn}, 139--144

\bibitem[{{Wyse} {et~al.}(2006){Wyse}, {Gilmore}, {Norris}, {Wilkinson},
  {Kleyna}, {Koch}, {Evans}, \& {Grebel}}]{wyse2006}
{Wyse}, R.~F.~G., {Gilmore}, G., {Norris}, J.~E., {et~al.} 2006, \apjl, 639,
  L13

\bibitem[{{Yanny} {et~al.}(2009){Yanny}, {Rockosi}, {Newberg}, {Knapp},
  {Adelman-McCarthy}, {Alcorn}, {Allam}, {Allende Prieto}, {An}, {Anderson},
  {Anderson}, {Bailer-Jones}, {Bastian}, {Beers}, {Bell}, {Belokurov},
  {Bizyaev}, {Blythe}, {Bochanski}, {Boroski}, {Brinchmann}, {Brinkmann},
  {Brewington}, {Carey}, {Cudworth}, {Evans}, {Evans}, {Gates}, {G{\"a}nsicke},
  {Gillespie}, {Gilmore}, {Nebot Gomez-Moran}, {Grebel}, {Greenwell}, {Gunn},
  {Jordan}, {Jordan}, {Harding}, {Harris}, {Hendry}, {Holder}, {Ivans},
  {Ivezi{\v c}}, {Jester}, {Johnson}, {Kent}, {Kleinman}, {Kniazev},
  {Krzesinski}, {Kron}, {Kuropatkin}, {Lebedeva}, {Lee}, {French Leger},
  {L{\'e}pine}, {Levine}, {Lin}, {Long}, {Loomis}, {Lupton}, {Malanushenko},
  {Malanushenko}, {Margon}, {Martinez-Delgado}, {McGehee}, {Monet}, {Morrison},
  {Munn}, {Neilsen}, {Nitta}, {Norris}, {Oravetz}, {Owen}, {Padmanabhan},
  {Pan}, {Peterson}, {Pier}, {Platson}, {Re Fiorentin}, {Richards}, {Rix},
  {Schlegel}, {Schneider}, {Schreiber}, {Schwope}, {Sibley}, {Simmons},
  {Snedden}, {Allyn Smith}, {Stark}, {Stauffer}, {Steinmetz}, {Stoughton},
  {SubbaRao}, {Szalay}, {Szkody}, {Thakar}, {Sivarani}, {Tucker}, {Uomoto},
  {Vanden Berk}, {Vidrih}, {Wadadekar}, {Watters}, {Wilhelm}, {Wyse}, {Yarger},
  \& {Zucker}}]{yanni2009}
{Yanny}, B., {Rockosi}, C., {Newberg}, H.~J., {et~al.} 2009, \aj, 137, 4377

\bibitem[{{Yoachim} \& {Dalcanton}(2006)}]{yoachim2006}
{Yoachim}, P. \& {Dalcanton}, J.~J. 2006, \aj, 131, 226

\bibitem[{{Zucker} {et~al.}(2012){Zucker}, {de Silva}, {Freeman},
  {Bland-Hawthorn}, \& {Hermes Team}}]{zucker2012}
{Zucker}, D.~B., {de Silva}, G., {Freeman}, K., {Bland-Hawthorn}, J., \&
  {Hermes Team}. 2012, in Astronomical Society of the Pacific Conference
  Series, Vol. 458, Galactic Archaeology: Near-Field Cosmology and the
  Formation of the Milky Way, ed. W.~{Aoki}, M.~{Ishigaki}, T.~{Suda},
  T.~{Tsujimoto}, \& N.~{Arimoto}, 421

\end{thebibliography}

\begin{appendix}

\section{Kinematical selection criteria}
\label{sec:kincriteria}

The kinematical criteria that we have used as a starting point 
to select candidate thin and thick disk stars assumes that the 
Galactic space velocities  ($U_{\rm LSR}$, $V_{\rm LSR}$, and $W_{\rm LSR}$
of the stellar populations have Gaussian distributions,
\begin{equation}
	f = k
	\cdot \exp\left({- \frac{(U_{\rm LSR}-U_{\rm asym})^{2}}{2\,\sigma_{\rm U}^{2}} -
                           \frac{(V_{\rm LSR}-V_{\rm asym})^{2}}{2\,\sigma_{\rm V}^{2}} -
                           \frac{W^{2}_{\rm LSR}}{2\,\sigma_{\rm W}^{2}}}\right),
\label{eq:probabilities}
\end{equation}
where
\begin{equation}
        k = \frac{1}{(2\pi)^{3/2}\,\sigma_{\rm U}\sigma_{\rm V}\sigma_{\rm W}} 
\label{eq:normalization}
\end{equation}
normalises the expression. $\sigma_{\rm U}$, $\sigma_{\rm V}$, and $\sigma_{\rm W}$ 
are the characteristic velocity dispersions, and $V_{\rm asym}$ is the asymmetric 
drift.  The values for the velocity dispersions, 
rotational lags, and normalisations in the Solar neighbourhood that we 
used are listed in Table~\ref{tab:dispersions}.

To get the 
probability (which we will call D, TD, and H, for the thin disk, thick disk,
and stellar halo, respectively) that a given star belongs to a specific population
the probabilities from Eq.~(\ref{eq:probabilities}) should be multiplied
by the observed 
fractions ($X$) of each population in the Solar neighbourhood. By then dividing 
the thick disk probability (TD)
with the thin disk (D) and halo (H) probabilities, respectively, we get two 
relative probabilities for
the thick disk-to-thin disk (TD/D) and thick disk-to-halo (TD/H) membership, i.e.
\begin{equation}
        {\rm TD/D} = \frac{X_{\rm TD}}{X_{\rm D}}\cdot \frac{f_{\rm TD}}{f_{\rm D}}, \label{eq:tdd}
\end{equation}
and likewise for other probability ratios.

\begin{table}
\centering
\caption{
\label{tab:dispersions}
        Characteristics for stellar populations 
        in the Solar neighbourhood.$^{\dagger}$
        }
\begin{tabular}{lrrrrrl}
\hline \hline\noalign{\smallskip}
        & $\sigma_{\rm U}$
        & $\sigma_{\rm V}$
        & $\sigma_{\rm W}$
        & $U_{\rm asym}$
        & $V_{\rm asym}$ 
        & $X$  \\
\noalign{\smallskip}
        & \multicolumn{5}{c}{-----------~~[km\,s$^{-1}$]~~-----------}     
        & \\
\noalign{\smallskip}
\hline\noalign{\smallskip}
   Thin disk   &  35  & 20 & 16 &    0  &  $-15$ &  0.85  \\
   Thick disk  &  67  & 38 & 35 &    0  &  $-46$ &  0.09  \\
   Halo        & 160  & 90 & 90 &    0  & $-220$ &  0.0015 \\
   Hercules    &  26  &  9 & 17 & $-40$ &  $-50$ &  0.06    \\
   Arcturus    &   ?  &  ? &  ? & ?     & $-100$ &   ?      \\
\hline
\end{tabular}
\flushleft
$^{\dagger}${\scriptsize
		Columns (2)-(4) give the velocity dispersions ($\sigma_{\rm U}$, 
		$\sigma_{\rm V}$, and $\sigma_{\rm W}$) for the different populations
		in col.~(1); cols.~(5)-(6) give the the asymmetric drifts 
		(in $U$ and $V$) relative to the LSR; and col.~(7) gives the 
        normalisation fractions for each population in the Solar neighbourhood
        (in the Galactic plane).
        Values are taken from
        \cite{bensby2005,bensby2007letter} for the thin disk, thick disk, 
        the stellar halo, and the Hercules stream. For the Arcturus moving group
        only the $\vlsr$ velocity is know and is taken from \cite{williams2009}. 
        }
\end{table}

\section{Description of error analysis method}
\label{sec:uncertainties2}

The method is taken from \cite{epstein2010} and is based on the fact 
that the four stellar 
parameters $m_{j} = (\teff,\,\xi_{\rm t},\,\log g,\,\log ({\rm Fe}))$ 
have been determined using four observables, $o_{i}$:
\begin{itemize}
\item $o_{1}$: The first observable is the slope from the linear 
regression when plotting abundances from \ion{Fe}{i} lines versus 
excitation potential. For the best fit of the effective temperature 
($\teff$) this slope should be zero; 
\item $o_{2}$: The second observable is the slope from the linear
regression when plotting abundances from \ion{Fe}{i} lines versus 
reduced line strength ($\log(W/\lambda)$). For the best fit of the 
micro-turbulence parameter ($\xi_{\rm t}$) this slope should be zero;
\item $o_{3}$: The third observable is the
abundances from \ion{Fe}{i} and \ion{Fe}{ii} lines. For a correctly
determined surface gravity, they should be equal; 
\item $o_{4}$: The fourth observable is the difference between
the output abundance from \ion{Fe}{i} lines and the input metallicity
of the stellar model that is used. For the best fit this
difference should be zero.
\end{itemize}

Each observable can be written as a linear combination of deviations
from the best fit model:
\begin{equation}
o_{i} = o_{i}^{0} + \sum_{j=1}^{4}b_{ij}(m_{j}-m_{j}^{0}),
\label{eq:observable}
\end{equation}
where $b_{ij}=\partial o_{i}/\partial m_{j} = \Delta o_{i}/\Delta m_{j}$ 
are the partial derivatives of the observables.
The values for $b_{ij}$ are determined by varying each of the
stellar parameters one at a time by an amount of $\Delta m_{j}$.
We choose to set  $\Delta m_{1}=\pm 100$\,K, $\Delta m_{2}=\pm0.1\,\kms$, 
$\Delta m_{3}=\pm0.1$\,dex, and $\Delta m_{4}=\pm0.1$\,dex. 
Applying these changes in the
stellar parameters, we then calculate four sets of new abundances for all
lines. Compared to the best fit model, we will now see changes in the 
observables $\Delta o_{i} = o_{i} - o_{i}^{0}$ (where $o_{i}^0$ is the
value of the observable from the best fit model). 
Equation~\ref{eq:observable} gives a system of equations to be solved. 
Inverting the $4\times4$ matrix of $b_{ij}$ gives a new $4\times4$
matrix of elements $c_{ik}$.
As each observable $o_{i}$ is associated with an error 
($\sigma_{k}$), the uncertainties in the stellar parameters ($m_{i}$) 
can be then solved as:
\begin{equation}
\label{eq:errors2}
\sigma(m_{i}) = \sqrt{\sum_{k=1}^{4} c_{ik}^2 \sigma_{k}^2}.
\end{equation}
For $o_{1}$, which is the slope of the \ion{Fe}{i} abundances versus
excitation potential that is used for the determination of $\teff$, 
we take $\sigma_{1}$ as the uncertainty of the
linear regression in that plot. For $o_{2}$, which is the slope of the
abundances from \ion{Fe}{i} lines versus reduced line strength,
we take $\sigma_{2}$ as the uncertainty of the linear regression in 
that plot. For $o_{3}$, $\sigma_{3}$ is
connected to the formal errors in the \ion{Fe}{i} and \ion{Fe}{ii}
abundances.
$\sigma_{4}$, associated with the observable for the 
balance between input and output abundances, is similar to $\sigma_{3}$,
but since we only use \ion{Fe}{i} lines to measure $\rm \log(Fe)$, we use
the formal error that we measure for abundances from 
\ion{Fe}{i} lines as $\sigma_{4}$.
The final estimates of the uncertainties in the stellar parameters,
as calculated by Eq.~(\ref{eq:errors2}), are given together with the
best fit values of the stellar parameters in Table~\ref{tab:parameters}.

The measured abundance of an element ($X$) can be written as a
linear combination of deviations from the best fit model
\begin{equation}
X = X_0 + \sum_{j=1}^{4}\kappa_{j}(m_{j}-m_{j}^{0}) = X_0 + \sum_{j=1}^{4}\alpha_{j}(o_{j}-o_{j}^{0}),
\end{equation}
where the partial derivatives
$\kappa_j = \partial X/\partial m{j} = \Delta X / \Delta m_j$
are calculated for all elements ($X$) by changing the stellar
model atmosphere parameters by the same amounts as when determining
$b_{ij}$ above, and $\alpha_j$ is given by
\begin{equation}
\alpha_{j} = \sum_{k=1}^{4} \kappa_{k} \cdot c_{kj}.
\end{equation}
The error in the measured average abundance for an element
then becomes
\begin{equation}
\sigma_{X} = \sqrt{
\sigma_{X_0}^2 + \sum_{k=1}^{4}\alpha_{k}^{2}\cdot\sigma_{k}^{2}}
\end{equation}
where $\sigma_k$ are the uncertainties in the observables
as given above, and $\sigma_{X_0}$ is the formal error of the
measured abundance.
The uncertainty in a measured abundance ratio [$X/Y$] is then
\begin{equation}
\sigma_{XY} = \sqrt{
\sigma_{X}^2 + \sigma_{Y}^2 - 2\sum_{k=1}^{4}\alpha_{k,\,X} \cdot \alpha_{k,\,Y} \cdot \sigma_{k}^{2}}
\end{equation}
Uncertainties in the stellar parameters and in the abundance ratios
([$X$/Fe] and [$X$/Ti]) are given in Table~\ref{tab:parameters}
for all 714 stars.

\section{Description of online tables}

We are providing three online tables. The first
table (Table~\ref{tab:rejected}) lists the stars that were
rejected from further analysis. The reasons are given in the table
but the main causes are that the stars are either spectroscopic binaries
and/or rotated too fast to allow for proper measurements of the equivalent
widths. The next table (Table~\ref{tab:atomdata}) 
gives the atomic data and the equivalent widths
and elemental abundances for individual lines in the Sun.
The third table (Table~\ref{tab:parameters})
gives the results, kinematics, ages, abundance ratios,
and uncertainties for the full sample of 714 stars.
Details on all three tables are given below.

\begin{table}[ht]
\centering
\setlength{\tabcolsep}{1.5mm}
 \caption{
The following stars were observed but rejected from analysis. The
table is only available in the online version of the paper and in electronic
form at the CDS via anonymous ftp to {\tt cdsarc.u-strasbg.fr (130.79.125.5)}
or via {\tt http://cdsweb.u-strasbg.fr/Abstract.html} 
  \label{tab:rejected}
        } 
\scriptsize
\begin{tabular}{cc}
\hline\hline
\noalign{\smallskip}
{HIP} &
Comment \\
\noalign{\smallskip}
\hline
\noalign{\smallskip}
 \vdots   &   \vdots   \\
 6492      &  Spectroscopic binary    \\
\vdots    &    \vdots  \\
\noalign{\smallskip}
\hline
\end{tabular}
\end{table}

\begin{table}[ht]
\centering
\setlength{\tabcolsep}{0.9mm}
 \caption{
Atomic line data. The 
table is only available in the online version of the paper and in electronic 
form at the CDS via anonymous ftp to {\tt cdsarc.u-strasbg.fr (130.79.125.5)}
or via {\tt http://cdsweb.u-strasbg.fr/Abstract.html}
   \label{tab:atomdata}
        }
\scriptsize
\begin{tabular}{lccrccrccll}
\hline\hline
\noalign{\smallskip}
Atom         &
$\lambda$    &
             &
$\log gf$    &
             &
             &
             &
             &
             &
             &
Ref.          \\
             &
[{\AA}]      &
             &
             &
             &
             &
             &
             &
             &
             &
              \\
(1)          &
(2)          &
(3)          &
(4)          &
(5)          &
(6)          &
(7)          &
(8)          &
(9)          &
(10)         &
(11)          \\        
\noalign{\smallskip}
\hline
\noalign{\smallskip}
   $\vdots$ & $\vdots$ & $\vdots$ & $\vdots$ & $\vdots$ & $\vdots$ & $\vdots$ & $\vdots$ & $\vdots$ & $\vdots$ & $\vdots$ \\
  Fe {\sc i}   &  5242.491  &   3.634  &   -0.97   &     1.40 & 5.754E+07 &   86.2 &  1.00 & X X &    a1I       z1H     &  BFL03       \\
   $\vdots$ & $\vdots$ & $\vdots$ & $\vdots$ & $\vdots$ & $\vdots$ & $\vdots$ & $\vdots$ & $\vdots$ & $\vdots$ & $\vdots$ \\
\noalign{\smallskip}
\hline
\end{tabular}
\end{table}

\begin{table}
\centering
\scriptsize
\caption{The online table has 714 stars with the following columns. The
table is only available in the online version of the paper and in electronic
form at the CDS via anonymous ftp to {\tt cdsarc.u-strasbg.fr (130.79.125.5)}
or via {\tt http://cdsweb.u-strasbg.fr/Abstract.html}}  
\label{tab:parameters}
\begin{tabular}{clll}
\hline\hline
\noalign{\smallskip}
  column  & label & unit & comment \\
\noalign{\smallskip}
\hline
\noalign{\smallskip}
(1)  &   HIP   &            &          \\
(2)  &   $\teff$ &     K     &          \\
(3)  &   $\epsilon(\teff)$ & K &          \\
(4)  &   $\log g$ &         &          \\
(5)  &   $\epsilon(\log g)$ & &          \\
(6)  &   $\xi_{\rm t}$    & $\kms$ &          \\
(7)  &   $\epsilon(\xi_{\rm t})$ & $\kms$ &     \\
(8)  &   $\log\varepsilon({\rm \ion{Fe}{i}})$ & &  absolute abundance from \ion{Fe}{i} lines  \\
(9)  &   $\log\varepsilon({\rm \ion{Fe}{ii}})$ & &  absolute abundance from \ion{Fe}{ii} lines  \\
(10) &   [Fe/H]  &    Sun       &          \\
(11) &   [O/Fe]  &    Sun       &          \\
(12) &   [Na/Fe] &    Sun       &          \\
(13) &   [Mg/Fe] &    Sun       &          \\
(14) &   [Al/Fe] &    Sun       &          \\
(15) &   [Si/Fe] &    Sun       &          \\
(16) &   [Ca/Fe] &    Sun       &          \\
(17) &   [Ti/Fe] &    Sun       &          \\
(18) &   [Cr/Fe] &    Sun       &          \\
(19) &   [Ni/Fe] &    Sun       &          \\
(20) &   [Zn/Fe] &    Sun       &          \\
(21) &   [Y/Fe]  &    Sun       &          \\
(22) &   [Ba/Fe] &    Sun       &          \\
(23) &   $N$(\ion{Fe}{i})   &   &  number of lines used  \\
(24) &   $N$(\ion{Fe}{ii})   &   &    \\
(25) &   $N$(\ion{O}{i})   &   &    \\
(26) &   $N$(\ion{Na}{i})   &   &   \\
(27) &   $N$(\ion{Mg}{i})   &   &    \\
(28) &   $N$(\ion{Si}{i})   &   &    \\
(29) &   $N$(\ion{Ca}{i})   &   &    \\
(30) &   $N$(\ion{Ti}{i})   &   &   \\
(31) &   $N$(\ion{Ti}{ii})   &   &    \\
(32) &   $N$(\ion{Cr}{i})   &   &   \\
(33) &   $N$(\ion{Cr}{ii})   &   &    \\
(34) &   $N$(\ion{Ni}{i})   &   &    \\
(35) &   $N$(\ion{Zn}{i})   &   &   \\
(36) &   $N$(\ion{Y}{ii})   &   &    \\
(37) &   $N$(\ion{Ba}{ii})   &   &    \\
(38) &   $\rm\sigma\ion{Fe}{i}$  & &  1-$\sigma$ line-to-line dispersion \\
(39) &   $\rm\sigma\ion{Fe}{ii}$  & &  \\
(40) &   $\rm\sigma\ion{O}{i}$  & &  \\
(41) &   $\rm\sigma\ion{Na}{i}$  & &  \\
(42) &   $\rm\sigma\ion{Mg}{i}$  & &  \\
(43) &   $\rm\sigma\ion{Si}{i}$  & &  \\
(44) &   $\rm\sigma\ion{Ca}{i}$  & &  \\
(45) &   $\rm\sigma\ion{Ti}{i}$  & &  \\
(46) &   $\rm\sigma\ion{Ti}{ii}$  & &  \\
(47) &   $\rm\sigma\ion{Cr}{i}$  & &  \\
(48) &   $\rm\sigma\ion{Cr}{ii}$  & &  \\
(49) &   $\rm\sigma\ion{Ni}{i}$  & &  \\
(50) &   $\rm\sigma\ion{Zn}{i}$  & &  \\
(51) &   $\rm\sigma\ion{Y}{ii}$  & &  \\
(52) &   $\rm\sigma\ion{Ba}{ii}$  & &  \\
(53) &   $\rm\epsilon[Fe/H]$ & &  abundance ratio uncertainty \\
(54) &   $\rm\epsilon[O/Fe]$ & &   \\
(55) &   $\rm\epsilon[Na/Fe]$ & &   \\
(56) &   $\rm\epsilon[Mg/Fe]$ & &   \\
(57) &   $\rm\epsilon[Al/Fe]$ & &   \\
(58) &   $\rm\epsilon[Si/Fe]$ & &   \\
(59) &   $\rm\epsilon[Ca/Fe]$ & &   \\
(60) &   $\rm\epsilon[Ti/Fe]$ & &   \\
(61) &   $\rm\epsilon[Cr/Fe]$ & &   \\
(62) &   $\rm\epsilon[Ni/Fe]$ & &   \\
(63) &   $\rm\epsilon[Zn/Fe]$ & &   \\
(64) &   $\rm\epsilon[Y/Fe]$ & &   \\
(65) &   $\rm\epsilon[Ba/Fe]$ & &   \\
(66) &   $\rm\epsilon[Ti/H]$ & &  abundance ratio uncertainty \\
(67) &   $\rm\epsilon[O/Ti]$ & &   \\
(68) &   $\rm\epsilon[Na/Ti]$ & &   \\
(69) &   $\rm\epsilon[Mg/Ti]$ & &   \\
(70) &   $\rm\epsilon[Al/Ti]$ & &   \\
(71) &   $\rm\epsilon[Si/Ti]$ & &   \\
(72) &   $\rm\epsilon[Ca/Ti]$ & &   \\
(73) &   $\rm\epsilon[Cr/Ti]$ & &   \\
(74) &   $\rm\epsilon[Ni/Ti]$ & &   \\
(75) &   $\rm\epsilon[Zn/Ti]$ & &   \\
(76) &   $\rm\epsilon[Y/Ti]$ & &   \\
(77) &   $\rm\epsilon[Ba/Ti]$ & &   \\
\noalign{\smallskip}
\hline
\end{tabular}
\end{table}
\setcounter{table}{2}

\begin{table}
\centering
\scriptsize
\caption{\it continued}  
\begin{tabular}{clll}
\hline\hline
\noalign{\smallskip}
  column  & label & unit & comment \\
\noalign{\smallskip}
\hline
\noalign{\smallskip}
(78)  &   Age   &   Gyr         &  Best age\\
(79)  &   Agel  &   Gyr         &  lower limit on age \\
(80)  &   Ageu  &   Gyr         &  upper limit on age \\
(81)  &   mass  &   sun         &  Best mass \\
(82)  &   massl &   sun         &  lower limit on mass \\
(83)  &   massu &   sun         &  upper limit on mass \\
(84)  &   $d$   &   kpc         &          \\
(85)  &   $l$   &   deg         &          \\
(86)  &   $b$   &   deg         &          \\
(87)  &   $X$   &   kpc         &          \\
(88)  &   $Y$   &   kpc         &          \\
(89)  &   $Z$   &   kpc         &          \\
(90)  &   $R_{\rm min}$     &  kpc          &          \\
(91)  &   $R_{\rm max}$     &  kpc          &          \\
(92)  &   $\rmean$     &   kpc         &          \\
(93)  &   $\zmax$     &    kpc        &          \\
(94)  &   $e$        &            &          \\
(95)  &   $L_{\rm Z}$     &            &          \\
(96)  &   $E/E_{\rm LSR}$ &        & total energy normalised to the LSR \\
(97)  &   $\ulsr$      &   $\kms$         &          \\
(98)  &   $\vlsr$      &   $\kms$         &          \\
(99)  &   $\wlsr$      &   $\kms$         &          \\
(100) &   $TD/D$        &            &          \\
(101) &   $TD/H$        &            &          \\
(102) &   $Her/TD$        &            &          \\
(103) &   $Her/D$        &            &          \\
(104) &   $\teff$ (u.i.b) &     K     & un-corrected ion.bal temp         \\
(105) &   $\log g$ (u.i.b) &         & un-corrected ion.bal grav         \\
(106) &   $\teff$ (hip) &     K     & hipparcos temp         \\
(107) &   $\log g$ (hip) &         & hipparcos grav         \\
\noalign{\smallskip}
\hline
\end{tabular}
\end{table}

\end{appendix}

\end{document}